\documentclass[longauth]{aa}  
\usepackage{sidecap}
\usepackage{sidecap}

\usepackage{graphicx}
\usepackage{txfonts}

\usepackage[acronym,toc]{glossaries}
\usepackage{hyperref}
\usepackage{amsmath}
\usepackage{bbold}
\usepackage{natbib}
\usepackage{tikz}
\usepackage{booktabs}
\usepackage{tabularx}
\usepackage[utf8]{inputenc}
\usepackage[normalem]{ulem}
\usepackage{placeins}
\usepackage{multirow}

\usetikzlibrary{shapes.geometric,arrows.meta,backgrounds,calc,automata}

\hypersetup{
	colorlinks=true,
	filecolor=magenta,
	urlcolor=blue,
}
\glsdisablehyper

\newcommand{\tpa}{PTF10tpa}
\newcommand{\hb}{\href{https://wis-tns.org/object/2010hb}{SN~2010hb}}
\newcommand{\wmf}{PTF10wmf}
\newcommand{\xlr}{PTF10xlr}
\newcommand{\jc}{\href{https://wis-tns.org/object/2010jc}{SN~2010jc}}
\newcommand{\icthree}{\href{https://wis-tns.org/object/2011az}{SN~2011az}}
\newcommand{\ngctwo}{\href{https://wis-tns.org/object/2011fd}{SN~2011fd}}
\newcommand{\ngcfour}{\href{https://wis-tns.org/object/2011fv}{SN~2011fv}}
\newcommand{\pgc}{\href{https://wis-tns.org/object/2011fy}{SN~2011fy}}
\newcommand{\cer}{LSQ12cer}
\newcommand{\css}{\href{https://wis-tns.org/object/2012ch}{SN~2012ch}}
\newcommand{\icone}{\href{https://wis-tns.org/object/2012eh}{SN~2012eh}}
\newcommand{\ljg}{PTF12ljg}
\newcommand{\hi}{\href{https://wis-tns.org/object/2012hi}{SN~2012hi}}
\newcommand{\hnj}{LSQ12hnj}
\newcommand{\ugc}{\href{https://wis-tns.org/object/2013bm}{SN~2013bm}}
\newcommand{\bjx}{iPTF13bjx}
\newcommand{\zw}{\href{https://www.wis-tns.org/object/2012ho}{SN~2012ho}}
\newcommand{\icthreefive}{\href{https://www.wis-tns.org/object/2012fs}{SN~2012fs}}
\newcommand{\ds}{\href{https://www.wis-tns.org/object/2013ds}{SN~2013ds}}
\newcommand{\fvq}{LSQ12fvq}

\newcommand{\sem}{\href{https://www.wis-tns.org/object/1999em}{SN~1999em}}
\newcommand{\gi}{\href{https://www.wis-tns.org/object/1999gi}{SN~1999gi}}

\newcommand{\hn}{\href{https://www.wis-tns.org/object/2003hn}{SN~2003hn}}
\newcommand{\fourdj}{\href{https://www.wis-tns.org/object/2004dj}{SN~2004dj}}
\newcommand{\et}{\href{https://www.wis-tns.org/object/2004et}{SN~2004et}}
\newcommand{\ay}{\href{https://www.wis-tns.org/object/2005ay}{SN~2005ay}}
\newcommand{\cs}{\href{https://www.wis-tns.org/object/2005cs}{SN~2005cs}}
\newcommand{\bk}{\href{https://www.wis-tns.org/object/2008bk}{SN~2008bk}}
\newcommand{\ib}{\href{https://www.wis-tns.org/object/2009ib}{SN~2009ib}}
\newcommand{\aw}{\href{https://www.wis-tns.org/object/2012aw}{SN~2012aw}}
\newcommand{\ej}{\href{https://www.wis-tns.org/object/2012ej}{SN~2013ej}}
\newcommand{\eaw}{\href{https://www.wis-tns.org/object/2017eaw}{SN~2017eaw}}
\newcommand{\aoq}{\href{https://www.wis-tns.org/object/2018aoq}{SN~2018aoq}}

\newcommand{\ixf}{\href{https://www.wis-tns.org/object/2023ixf}{SN~2023ixf}}

\newacronym{lmc}{LMC}{Large Magellanic Cloud}
\newacronym{trgb}{TRGB}{Tip of the Red Giant Branch}
\newacronym{flrw}{FLRW}{Friedmann-Lema\^{i}tre-Robertson-Walker}
\newacronym{holicow}{H0LiCOW}{$H_0$ Lenses in COSMOGRAIL's Wellspring}
\newacronym{snfactory}{SNfactory}{Nearby Supernova Factory}
\newacronym{adH0cc}{adH0cc}{accurate determination of $H_0$ with core-collapse supernovae}
\newacronym[plural={SNe~II}, \glsshortpluralkey={SNe~II}, longplural={Type~II supernovae}]{sneII}{SN~II}{Type~II supernova}
\newacronym[plural={SNe~II-P}, \glsshortpluralkey={SNe~\mbox{II-P}}, longplural={Type~\mbox{II-P} supernovae }]{sneIIp}{SN~\mbox{II-P}}{Type~\mbox{II-P} supernova}
\newacronym[plural={SNe~Ia}, \glsshortpluralkey={SNe~Ia}, longplural={Type~Ia supernovae}]{sneIa}{SN~Ia}{Type~Ia supernova}
\newacronym[plural={SNe}, \glsshortpluralkey={SNe}, longplural={supernovae}]{sne}{SN}{supernova}
\newacronym{bao}{BAO}{baryon acoustic oscillations}
\newacronym{ptf}{PTF}{Palomar Transient Factory}
\newacronym{lsq}{LSQ}{La Silla-QUEST}
\newacronym{crts}{CRTS}{Catalina Real-Time Transient Survey}
\newacronym{issp}{ISSP}{Italian Supernovae Search Project}
\newacronym{iptf}{iPTF}{intermediate Palomar Transient Factory}
\newacronym{scm}{SCM}{standardized candle method}
\newacronym{lcdm}{$\Lambda$CDM}{$\Lambda$ cold dark matter}
\newacronym{bbn}{BBN}{Big Bang nucleosynthesis}
\newacronym{kde}{KDE}{kernel-density estimate}
\newacronym{ppf}{PPF}{percent-point function}
\newacronym{csp}{CSP-I}{Carnegie Supernova Project-I}
\newacronym{sdss}{SDSS}{Sloan Digital Sky Survey}
\newacronym[plural={ToEs},firstplural={times of explosion (ToEs)}]{toe}{ToE}{time of explosion}
\newacronym{cmb}{CMB}{Cosmic Microwave Background}
\newacronym{mcmc}{MCMC}{Markov chain Monte Carlo}
\newacronym{csm}{CSM}{circumstellar medium}
\newacronym{zam}{ZAM}{zero age mass}
\newacronym{epm}{EPM}{expanding photosphere method}
\newacronym{pmm}{PMM}{photospheric magnitude method}
\newacronym{pcm}{PCM}{photometric color method}
\newacronym{ned}{NED}{NASA/IPAC Extragalactic Database}
\newacronym{loss}{LOSS}{Lick Observatory Supernova Search}
\newacronym{kait}{KAIT}{Katzman Automatic Imaging Telescope}
\newacronym{prompt}{PROMPT}{Panchromatic Robotic Optical Monitoring and Polarimetry Telescopes}
\newacronym{ntt}{NTT}{New Technology Telescope}
\newacronym{ctio}{CTIO}{Cerro Tololo Inter-American Observatory}
\newacronym{lco}{LCO}{Las Cumbres Observatory}
\newacronym{osc}{OSC}{Open Supernova Catalogue}
\newacronym{pcc}{PCC}{Pearson correlation coefficient}
\newacronym{aic}{AIC}{Akaike information criterion}
\newacronym{bic}{BIC}{Bayesian information criterion}
\newacronym{sndb}{UCB-SNDB}{The UC Berkeley SNDB}
\newglossaryentry{bessell12}{
	name={Bessell-12},
	description={System of revised UBVRI photonic passbands by \cite{Bessell2012}}}
 \newglossaryentry{H0}{
  name = $H_0$ ,
  description = The Hubble-Lemaître constant}

\newcommand{\vhb}{\ensuremath{v_{\mathrm{H}\beta}}}

\newsavebox{\modelbox}

\graphicspath{{./}{figures/}}

\makeglossaries

\begin{document}

    \title{An improved standardization framework for Type II-P supernovae using a hierarchical Bayesian approach}
   \titlerunning{An improved Type II-P SNe standardization framework}
   \authorrunning{Holas et al.}

   \author{A.~Holas\inst{\ref{aff:hits}}
          \and
          S.~Taubenberger\inst{\ref{aff:mpa}}
          \and
          C.~Vogl\inst{\ref{aff:mpa}\and\ref{aff:origins}}
          \and
          D.~Rubin\inst{\ref{aff:hawaii2}\and\ref{aff:lbnl}}
          \and
          M.~Vincenzi\inst{\ref{aff:lbnl}\and\ref{aff:oxford}}
          \and
          W.~Hillebrandt\inst{\ref{aff:mpa}\and\ref{aff:origins}}
          \and
          G.~Aldering\inst{\ref{aff:lbnl}}
          \and
          P.~Antilogus\inst{\ref{aff:cnrs}}
          \and
          C.~Aragon\inst{\ref{aff:lbnl}\and\ref{aff:coe}}
          \and
          S.~Bailey\inst{\ref{aff:lbnl}}
          \and
          C.~Baltay\inst{\ref{aff:yale}}
          \and
          S.~Benitez-Herrera\inst{\ref{aff:esa}}
          \and
          S.~Bongard\inst{\ref{aff:cnrs}}
          \and
          K.~Boone\inst{\ref{aff:lbnl}\and\ref{aff:ucb}\and\ref{aff:dirac}}
          \and
          C.~Buton\inst{\ref{aff:cnrs2}}
          \and
          G.~Csörnyei\inst{\ref{aff:eso}}
          \and
          Y.~Copin\inst{\ref{aff:cnrs2}}
          \and
          S.~Dixon\inst{\ref{aff:lbnl}\and\ref{aff:ucb}}
          \and
          M.~Fink\inst{\ref{aff:mpa}\and\ref{aff:wuerzburg}}
          \and
          D.~Fouchez\inst{\ref{aff:cnrs3}}
          \and
          E.~Gangler\inst{\ref{aff:cnrs2}\and\ref{aff:cnrs4}}
          \and
          R.~Gupta\inst{\ref{aff:lbnl}}
          \and
          B.~Hayden\inst{\ref{aff:lbnl}\and\ref{aff:stsi}}
          \and
          E.~E.~O.~Ishida\inst{\ref{aff:cnrs4}}
          \and
          A.~G.~Kim\inst{\ref{aff:lbnl}}
          \and
          M.~Klauser\inst{\ref{aff:mpa}}
          \and
          M.~Kowalski\inst{\ref{aff:berlin}\and\ref{aff:desy}}
          \and
          M.~Kromer\inst{\ref{aff:hits}}
          \and
          D.~K.~K\"{u}sters\inst{\ref{aff:ucb}\and\ref{aff:desy}}
          \and
          J.~Nordin\inst{\ref{aff:lbnl}\and\ref{aff:berlin}}
          \and
          R.~Pain\inst{\ref{aff:cnrs}}
          \and
          E.~Pecontal\inst{\ref{aff:lyon}}
          \and
          R. Pereira\inst{\ref{aff:cnrs2}}
          \and
          S.~Perlmutter\inst{\ref{aff:lbnl}\and\ref{aff:ucb}}
          \and 
          K.~A.~Ponder\inst{\ref{aff:ucb}}
          \and
          D.~Rabinowitz\inst{\ref{aff:yale}}
          \and
          M.~Rigault\inst{\ref{aff:cnrs2}}
          \and
          K.~Runge\inst{\ref{aff:lbnl}}
          \and
          M.~Sasdelli\inst{\ref{aff:adelaide}}
          \and
          C.~Saunders\inst{\ref{aff:lbnl}\and\ref{aff:ucb}\and\ref{aff:sorbonne}\and\ref{aff:princeton}}
          \and
          G.~Smadja\inst{\ref{aff:cnrs2}}
          \and
          N.~Suzuki\inst{\ref{aff:lbnl}\and\ref{aff:kavli},\ref{aff:florida}}
          \and
          C.~Tao\inst{\ref{aff:bejing},\ref{aff:cnrs3}}
          \and
          R.~C.~Thomas\inst{\ref{aff:lbnl},\ref{aff:lbnl2}}
          }
          
   \institute{Heidelberg Institute for Theoretical Studies, Schloss-Wolfsbrunnenweg 35, 69118 Heidelberg, Germany\label{aff:hits}\\
        \email{alexander.holas@mailbox.org}
        \and             
        Max-Planck-Institut für Astrophysik, Karl-Schwarzschild-Str. 1, D-85748 Garching, Germany\label{aff:mpa}
        \and
        Exzellenzcluster ORIGINS, Boltzmannstr. 2, 85748 Garching, Germany\label{aff:origins}
        \and
        European Southern Observatory, Karl-Scharzschild-Str. 2, Garching 85748, Germany\label{aff:eso}
        \and
        Institute for Astronomy, University of Hawaii, 2680 Woodlawn Drive, Honolulu, HI 96822, USA\label{aff:hawaii}
        \and
        Department of Physics and Astronomy, University of Hawaii, 2505 Correa Rd, Honolulu, HI, 96822, USA\label{aff:hawaii2}
        \and
        Physics Division, Lawrence Berkeley National Laboratory, 1 Cyclotron Road, Berkeley, CA, 94720, USA\label{aff:lbnl}
        \and
        Department of Physics, University of Oxford, Denys Wilkinson Building, Keble Road, Oxford OX1 3RH, United Kingdom\label{aff:oxford}
        \and
        Laboratoire de Physique Nucl\'eaire et des Hautes Energies, CNRS/IN2P3, Sorbonne Universit\'e, Universit\'e de Paris, 4 place Jussieu, 75005 Paris, France\label{aff:cnrs}
        \and
        College of Engineering, University of Washington 371 Loew Hall, Seattle, WA, 98195, USA\label{aff:coe}
        \and
        Department of Physics, Yale University, New Haven, CT, 06250-8121, USA\label{aff:yale}
        \and
        European Space Astronomy Center (ESAC), European Space Agency, Camino Bajo del Castillo s/n, 28692 Villanueva de la Cañada, Madrid, Spain\label{aff:esa}
        \and
        Department of Physics, University of California Berkeley, 366 LeConte Hall MC 7300, Berkeley, CA, 94720-7300, USA\label{aff:ucb}
        \and
        DIRAC Institute, Department of Astronomy, University of Washington, 3910 15th Ave NE, Seattle, WA 98195, USA\label{aff:dirac}
        \and
        Univ Lyon, Universit\'e Claude Bernard Lyon~1, CNRS/IN2P3, IP2I Lyon, F-69622, Villeurbanne, France\label{aff:cnrs2}
        \and
        Institut f\"ur Theoretische Physik und Astrophysik, Universit\"at W\"urzburg, Emil-Fischer-Stra{\ss}e 31, 97074 W\"urzburg, Germany\label{aff:wuerzburg}
        \and
        Aix Marseille Univ, CNRS/IN2P3, CPPM, Marseille, France\label{aff:cnrs3}
        \and
        Universit\'e Clermont Auvergne, CNRS/IN2P3, Laboratoire de Physique de Clermont, F-63000 Clermont-Ferrand, France\label{aff:cnrs4}
        \and
        Space Telescope Science Institute, 3700 San Martin Drive Baltimore, MD, 21218, USA\label{aff:stsi}
        \and
        Institut f\"ur Physik,  Humboldt-Universitat zu Berlin, Newtonstr. 15, 12489 Berlin, Germany\label{aff:berlin}
        \and
        DESY, D-15735 Zeuthen, Germany\label{aff:desy}
        \and
        Centre de Recherche Astronomique de Lyon, Universit\'e Lyon 1, 9 Avenue Charles Andr\'e, 69561 Saint Genis Laval Cedex, France\label{aff:lyon}
        \and
        The University of Adelaide, Adelaide SA 5000, Australia\label{aff:adelaide}
        \and
        Sorbonne Universit\'es, Institut Lagrange de Paris (ILP), 98 bis Boulevard Arago, 75014 Paris, France\label{aff:sorbonne}
        \and
        Princeton University, Department of Astrophysics, 4 Ivy Lane, Princeton, NJ, 08544, USA\label{aff:princeton}
        \and
        Kavli Institute for the Physics and Mathematics of the Universe, The University of Tokyo Institutes for Advanced Study, The University of Tokyo, 5-1-5 Kashiwanoha, Kashiwa, Chiba 277-8583, Japan\label{aff:kavli}
        \and
        Department of Physics, Florida State University, 77 Chieftan Way, Tallahassee, FL 32306-4350, USA\label{aff:florida}
        \and
        Tsinghua Center for Astrophysics, Tsinghua University, Beijing 100084, China\label{aff:bejing}
        \and
        Computational Cosmology Center, Computational Research Division, Lawrence Berkeley National Laboratory, 1 Cyclotron Road, Berkeley, CA, 94720, USA\label{aff:lbnl2}
        }
         
   \date{Received November 21, 2024; accepted TBD}

\abstract{
We use \glspl{sneIIp} observed as part of the \gls{snfactory} experiment as standardizable candles to develop an improved standardization framework.
These consist of spectrophotometric measurements of low-redshift
supernovae in the wavelength range $3200 < \lambda < 10000$\,\AA{} using a two-channel
integral field optical spectrograph, featuring host galaxy
subtraction through follow-up observations and an accurate flux
calibration. The quality and homogeneous nature of the measured dataset makes it
unique and well suited for use in the SN~II-P \gls{scm}.
In this work, we develop an improved method to determine the expansion velocities of \glspl{sneIIp} based on Gaussian Process regression. We extend the conventional \gls{scm}
by an additional correction term containing the H$_\alpha$ absorption-to-emission ratio and a hierarchical Bayesian framework.
We find that the additional correction improves the quality of the standardization, particularly around $30$ to $40$ days after the explosion.
Here, we are also able to reduce the scatter of the Hubble-flow \glspl{sneIIp} from $0.28$\,mag down to $0.18$\,mag compared to previous work.
As an example application of this methodology we measure $H_0$ using a pre-existing calibrator sample of SNe~II having Cepheid- or \gls{trgb}-based distance moduli. We find a value of $H_0 = 70.6^{+4.5}_{-4.3}$\,km\,s$^{-1}$\,Mpc$^{-1}$, including only statistical uncertainty, corresponding to a precision of $6.2$\,\%.
We find variations of $1.4$\,km\,s$^{-1}$\,Mpc$^{-1}$ between hierarchical model variations --- well within our quoted statistical uncertainty.
In contrast, without the additional correction term, we obtain $H_0 = 73.0^{+6.9}_{-6.1}$\,km\,s$^{-1}$\,Mpc$^{-1}$ --- a reduced precision of $8.9$\,\% --- highlighting the improved precision of our new methodology. We identify and discuss inconsistencies between the measured parent populations of the calibrator and Hubble-flow standardization characteristics --- present in this and literature samples --- that are not explained by a simple selection on magnitude and which will need to be improved upon as \gls{sneIIp} samples grow.
\glsresetall
}

   \keywords{instrumentation: spectrographs -- supernovae: general -- cosmology: distance scale}

   \maketitle

\section{Introduction} \label{sec:intro}

The discovery of the expansion of the universe by Georges Lemaître
\citep{Lemaitre1927a} and Edwin Hubble \citep{Hubble1929} ushered in the era of modern cosmology. The discovery of the accelerating expansion of the universe 
\citep{Riess1998,Schmidt1998,Perlmutter1999} showed us evidence of a more complex universe, and now there is evidence that the expansion may be accelerating in an unanticipated manner \citep{rubin2023a,desi2025a,desi2025c}.
Despite being the oldest of these, and having improved tremendously in precision and accuracy, not all estimates of the rate of expansion, the Hubble-Lemaître constant $H_0$, agree, prompting studies like ours.

As a practical matter, measuring $H_0$ locally, in the manner of \cite{Hubble1929}, rests on a ``distance ladder'', with the first rung being a geometrical distance to a celestial source, the second measuring similar ``standard candle'' or ``standard ruler'' sources in other galaxies, such as the brightness of the \gls{trgb} \citep{Madore2008a,Jang2017b,Jang2017c,Freedman2019,Yuan2019}, the period-luminosity relation of Cepheid variable stars \citep{freedman2001a, Riess2009, Riess2016, Riess2020} and masers \citep{Pesce2020}.
Finally, the third rung using yet brighter ``standardizable'' candles or rulers that can be found in galaxies in the so-called Hubble flow, where gravity-induced peculiar velocities are subdominant to the cosmological expansion velocity. 
This three-rung distance ladder approach is necessitated by the current lack of geometric sources or second-rung sources directly measurable in the Hubble-flow. Such local measurements occur late in the history of the Universe and are therefore often called ``late time'' measurements of $H_0$.

Alternatively, ``early time'' measurements of $H_0$ use \gls{bao}, imprinted on the \gls{cmb} during recombination \citep{peebles1970,Bennett2003a,Spergel2007,Collaboration2018}. For \gls{cmb} measurements this \gls{bao} scale can be calculated from first principles. The \gls{bao} scale is also detected in the clustering of galaxies \citep{eisenstein2005}, enabling a Hubble diagram to be constructed from the era of recombination down to lower redshift \citep{beutler2011a,aubourg2015,ross2015a,alam2017a,Collaboration2018,Alam2021a}.

Somewhat in between are the measurements of $H_0$ from gravitational lens time delays \citep{Shajib2019,Wong2019}, or model-dependent methods like the expanding photosphere method \citep[EPM; ][]{Kirshner1974, vogl2024a}. These do not depend on standard candles or recombination theory {\it per se}, but do rely on simple dark matter profiles for the lens and can be susceptible to the so-called mass-sheet degeneracy.
\glsunset{epm}
\glsunset{sne}

Depending on the method, a discrepancy between late- and early-time $H_0$ measurements ranging up to 6\,$\sigma$ is found \cite[e.g.][]{diValentino2021}. The \gls{cmb}/\gls{bao} early-time approach occupies the lower end of the range, at $\sim68$\,km\,s$^{-1}$\,Mpc$^{-1}$. The most often compared late time measurements range from $H_0=69.8\pm1.9$\,km\,s$^{-1}$\,Mpc$^{-1}$ \citep{Freedman2019} using \gls{trgb} distances for the second rung and \glspl{sneIa} for the third rung,\footnote{More recently updated to
$70.4\pm1.9$\,km\,s$^{-1}$\,Mpc$^{-1}$ in \citet{Freedman2025}} up to  $H_0=73.04\pm1.04$\,km\,s$^{-1}$\,Mpc$^{-1}$ \citep{Riess2022a} using Cepheids for the second rung and \glspl{sneIa} for the third rung. This disagreement --- between early time and some late-time measurements, or between different late-time measurements --- is often referred to as ``the Hubble tension.''
Whether caused by yet unknown physics or unaccounted systematic errors in the measurements and techniques, the Hubble tension is at the center of various cosmological investigations.

Another type of exploding star can be used to measure distances between the second and third rungs of the distance ladder: \glspl{sneII}. Compared to \gls{sneIa}, they have the advantage that they are more frequent
\citep{Li2011,Graur2017} and the physics behind them is better understood. The
drawback is that they are less luminous than \glspl{sneIa}
\citep{Richardson2014} and thus limited to lower redshifts for the same
observational conditions.

Over the years several methods have been developed to determine distances from \glspl{sneII}:
\begin{enumerate}
\item The \gls{epm}, which uses the angular size of the emitting regions and the surface brightness to infer the luminosity distance of the \gls{sne} \citep{Kirshner1974,vogl2024a}.
\item The \gls{scm}, a method based on the observation that more luminous \glspl{sne} exhibit higher expansion velocities \citep{Hamuy2001,Hamuy2002,Jaeger2022a}.
\item The \gls{pmm}, generalizing the \gls{scm} by using multiple phases at once \citep{Rodriguez2014,Rodriguez2019}.
\item The \gls{pcm}, which only relies on photometry and uses the color and slopes of light curves to obtain standardized magnitudes \citep{Jaeger2015,Jaeger2020a}.
\end{enumerate}
For a comprehensive review of various \glspl{sneII} distance determination methods, see e.g. \cite{jaeger2023a}.

Unfortunately, with current methodologies \glspl{sneII} cannot yet compete with
the $\sim 1.4$\,\% precision in $H_0$ \citep{Riess2022a} of a Cepheid-calibrated
\gls{sneIa} distance ladder.
At the time of writing, the most precise \glspl{sneII} measurement reaches a precision of $2.5$\,\% \citep{vogl2024a} using a tailored \gls{epm} approach, and the most precise \gls{scm} measurement achieves a $5$\,\% precision \citep{Jaeger2022a}.
As a direct consequence, there are few surveys dedicated to the observation of \glspl{sneII} specifically for the purpose of measuring $H_0$.

In this work, we apply the \gls{scm} to a sample of \glspl{sneIIp} observed by
the Nearby Supernova Factory \citep[SNfactory; ][]{Aldering2002} -- one of the few \glspl{sneIIp} data sets collected
with the explicit purpose of determining $H_0$. It features accurately calibrated spectra including a host galaxy subtraction obtained through follow-up observations.
\glsunset{snfactory}
Its quality and homogeneous nature makes it unique and well suited
for use in the \gls{scm} and it will help to reduce the negative influence of
systematic data calibration issues, which potentially contribute to the Hubble
tension. Most importantly, exchanging \glspl{sneIa} in a Cepheid- or \gls{trgb}-
calibrated distance ladder with \glspl{sneIIp} could eventually provide evidence
that the use of \glspl{sneIa} is not the origin of the Hubble tension. 

To this end, we revisit and revise most steps in the \gls{scm}, with the aim of
increasing the precision of the \gls{scm} and making \glspl{sneII} a more viable
alternative to \gls{sneIa}-based measurements. With that in mind, we extend
the classically used \gls{scm} (e.g.
\citealt{Nugent2006,Gall2018,Jaeger2020b,Jaeger2022a}) by an additional correction
involving the H$_\alpha$ absorption-to-emission ratio, $a/e$. Moreover, we bring the
\gls{scm} closer to the \gls{sneIa} standardization technique by
implementing a hierarchical Bayesian framework similar to the UNITY framework of
\citet{Rubin2015a,rubin2023a}.
In this work, we focus primarily on studying the impact of this more sophisticated standardization approach; obtaining a value for $H_0$ is a secondary objective.
In Section~\ref{sec:data} we elaborate on the data sets utilized, i.e., the
\gls{snfactory} sample and our calibrator sample. The methods used, such as the
spectral line fitting, phase interpolation, and the \gls{scm} itself, are
described in Sections~\ref{sec:methods} and \ref{sec:scm_methods}. Following this, we present and discuss
the results of this analysis in Sections~\ref{sec:results} to \ref{sec:djcomp_head}. We conclude our
findings in Section~\ref{sec:conclusion}.

\section{Data} \label{sec:data}
The data sample analyzed in this work consists of two distinct sub-samples:
a Hubble-flow sample of $21$ \glspl{sne} (listed in Table~\ref{tab:analysissample} in Appendix~\ref{app:sample_prop}) observed by the \gls{snfactory} \citep{Aldering2002} in the form of 158 flux-calibrated spectra (thereby eliminating the need for separate optical imaging), as well as a calibrator sample composed of $12$ \glspl{sne} (listed in Table~\ref{tab:calibratordist}) with known distances, whose data has been collected from various archival sources.

\subsection{\gls{snfactory} data sample}\label{sec:data_snfactory}
Here we describe the initial selection of the SNfactory Hubble-flow SN~II-P sample and the follow-up data and its processing. We then discuss the phase determination from the spectra and external photometry.

\subsubsection{SN selection}

While primarily focused on SNe~Ia, \gls{snfactory} also observed several \glspl{sneIIp}, $21$ of which make up the Hubble-flow sample of this work. These targets were discovered by various transient surveys such as the Catalina Real-Time Transient Survey (CRTS; \citealt{Drake2009}), the Palomar Transient Factory (PTF; \citealt{Law2009}), the intermediate Palomar Transient Factory (iPTF; \citealt{Kulkarni2013}), La Silla-QUEST (LSQ; \citealt{Baltay2013}), and the Italian Supernovae Search Project (ISSP), as well as amateur astronomers: Tom Boles, Maurice Gavin, Ron Arbour, Koichi Itagaki.
\begin{figure}
    \includegraphics[]{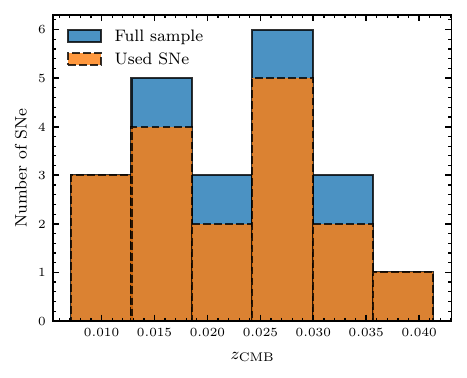}
    \caption{CMB redshift distribution of the $21$ \glspl{sne} in the Hubble-flow sample. The exact $z_\mathrm{CMB}$ values can be found in Table~\ref{tab:analysissample}.}
    \label{fig:redshiftdistr}
\end{figure}
The \gls{cmb} redshift distribution of the \glspl{sne} is illustrated in Figure~\ref{fig:redshiftdistr}, ranging from $0.01$ up to around $0.04$.

The various discovery sources have different survey depths, with the amateur astronomer surveys being comparatively shallow, the ISSP having intermediate depth, while PTF, iPTF and LSQ have a deeper survey depth. In this sense, our sample is most like the Carnegie Supernova Project (CSP) subsample that makes up about 50\% of the total \cite{Jaeger2020b,Jaeger2022a} Hubble-flow sample. About half of our \glspl{sne} were spectroscopically typed before we started their follow-up (using spectrographs on Keck, VLT, NTT and Asiago), while roughly half were first classified using SNIFS. We do not observe strong survey-dependent trends in the observed quantities, such as I-band magnitude or H$_\beta$ velocity versus redshift within our sample. Rather, the \glspl{sne} from different survey sources are mixed throughout the sample redshift range. For example, both the highest- and lowest-redshift \glspl{sneII} were found by the ISSP survey and screened by SNIFS. Likewise, the most distant amateur discovery in our sample (\hi), which happened to be typed by the smallest of the spectroscopic follow-up telescopes (Asiago), is at $80\%$ of the distance to the most distant deeper-survey SN (\fvq), typed at the NTT.
Due to this strong mixing, when selection effects are accounted for in the \gls{scm} hierarchical Bayesian model we will assume a survey depth function spanning a wide magnitude range for the \gls{snfactory} sample.
In this context we note that since convolution is commutative, a Gaussian distribution of search depths for our \gls{sne} sample, each represented by the CDF of a Gaussian as in UNITY \citep{rubin2023a}, would produce a CDF equal to that of another, broader, Gaussian.

\subsubsection{Follow-up data and processing}
The \gls{snfactory} spectrophotometry is obtained using our custom-built integral field spectrograph SNIFS \citep{Lantz2004}, following the TIGER concept \citep{Bacon1995,Bacon2001}. SNIFS produces flux-calibrated spectrophotometry of medium resolution ($\sim 3$\,\AA) in a wavelength range of $3200$-$10000$\,\AA{} \citep{rubin2022a}.
The instrument is mounted on the University of Hawaii $2.2$\,m telescope on Mauna Kea. It covers a total field of view of $6^{\prime\prime}\times 6^{\prime\prime}$ through a microlens array consisting of $225$ ($15\times15$) individual lenses, each producing its own spectrum.
The light is divided by a dichroic into a blue and red channels at $5100$\,\AA{} and then recorded by CCDs, each composed of $2048\times4096$ pixels of $15$\,\textmu{}m.
The wavelength calibration of those spectra is done through various arc lamp spectra and the recorded data are then processed by an automated pipeline \citep{Aldering2006,Scalzo2010}.
After removal of the host galaxy emission by use of a spectrophotometric observation of the \gls{sne} location obtained after the \gls{sne} has faded, the spectra are extracted from the ($x, y, \lambda$) data cubes \citep{Bongard2011} and flux calibrated, as well as corrected for atmospheric extinction by observations of various spectroscopic standard stars \citep{Buton2013,rubin2022a}.
SNIFS is also equipped with a parallel imager and a guider, both consisting of a $2048\times 4096$ CCD with $15$\,\textmu{}m pixels. The imager monitors the sky around the field of view of the main spectrograph through a multiple-bandpass filter without reimaging optics, allowing one to estimate the relative atmospheric extinction. This can be used to determine the cloud absorption scale factor needed to achieve flux calibration even in non-photometric nights, resulting in spectra suitable for extracting photometry.

Due to the temporal coverage required to monitor the plateau phase, a fraction of our \gls{sneIIp} follow-up spectra were obtained under conditions of fairly bright Moon.
When scattered off clouds, this light resulted in some failures of the standard \gls{snfactory} processing of the imaging channel.
To ameliorate this situation, software was developed to perform the photometry more robustly.
This includes fitting field stars with a Moffat profile.
Altogether, this reduces the scatter in the lightcurves and avoids catastrophic failures.

We apply host galaxy subtraction to $11$ of our \gls{snfactory} \glspl{sne}, based on galaxy templates taken between one and two years after the respective \gls{sne} was discovered.
The subtraction was performed whenever the subtraction had a non-negligible effect on the flux ($\gtrsim 1\%$).
In the other cases, we skipped the template subtraction when deemed unnecessary due to the small host galaxy contribution.
A detailed list of which \glspl{sne} have undergone host galaxy template subtraction can be found in the last column of Table~\ref{tab:analysissample}.

These \gls{snfactory} spectra are redshift corrected and corrected for Milky Way extinction using the extinction maps of \citet{Schlafly2011} and a \citet{fitzpatrick2019analysis} extinction law through the built-in dust extinction correction of the \textsc{Python} package \textsc{specutils}\footnote{\url{https://github.com/astropy/specutils}}.
The necessary inputs (i.e., redshifts and foreground galactic extinction, $A_V$) have been taken from the NED\footnote{The NASA/IPAC Extragalactic Database (NED) is operated by the Jet Propulsion Laboratory, California Institute of Technology, under contract with the National Aeronautics and Space Administration.} entry of the respective host galaxy. Here, we assume $R_V = 3.1$.
We correct for peculiar velocities using 2M++-SDSS maps \citep{said2020a,peterson2022a,carr2022a}, using the \textsc{Python} package \textsc{pvhub}\footnote{\url{https://github.com/KSaid-1/pvhub}}.
The corrected redshifts are summarized in Table~\ref{tab:pvred}.

From the full Hubble-flow sample of $21$ \glspl{sne}, we exclude three objects from
the standardization: \tpa{}, \jc{}, and \ds. \tpa{} is excluded
due to the badly constrained \gls{toe} and the fact that the first spectrum is
older than $30$ days, making any extrapolation to this phase uncertain.
\jc{} is excluded since it seems to be a luminous
low-velocity \gls{sne} similar to the objects investigated by
\citet{Rodriguez2020}. Its H$_\beta$ velocities are significantly below the
other \glspl{sne} in the sample and thus does not follow the desired relation of
more luminous objects having higher velocity ejecta.
\ds{} only has three early
time spectra with only one spectrum in the plateau phase, preventing interpolation.
A brief overview of the \glspl{sne} of the full Hubble-flow sample can be found in Table~\ref{tab:analysissample}, together with their \glspl{toe}. Except for \bjx, where the \gls{toe} was determined by \citet{Vogl2020} based on an exponential light curve fit, we determine all \glspl{toe} used in this work through either a spectral phase fitting technique or through available optical photometry.

\subsubsection{Spectral phase fitting}
Due to the lack of optical photometry sufficiently constraining the \gls{toe} for most \glspl{sne} in the Hubble-flow sample, we primarily determine the \gls{toe} through a spectral phase fitting technique. Here, we first retrieve the spectral phase of all spectra using the Supernova Identification (SNID) code of \citet{Blondin2007} based on the cross-correlation technique of \citet{Tonry1979} through the \textsc{Python} interface \textsc{pySNID}\footnote{\url{https://github.com/benstahl92/pySNID}}.
In this process we only use the high-quality \gls{sneII} templates of \citet{Guti_rrez_2017} since they have more consistent and reliable phase estimates than SNID's built-in \gls{sne} templates.
Here, we calculate the age and its associated uncertainty by taking the mean and standard deviation of all template ages that fulfill the criterion $r\mathrm{lap} > 0.75 \times r\mathrm{lap}_\mathrm{best}$, where $r\mathrm{lap}_\mathrm{best}$ refers to the $r\mathrm{lap}$ value of the best match, $r\mathrm{lap}$ being a SNID internal match quality measure. 
Before moving on to the next step, we inspect the matches and remove any unsuited results, e.g. where only poor matches have been found, or spectra older than roughly $50$ days after the first spectrum, as the quality of the phase matching deteriorates at later phases and seems to deviate from the expected linear relation.

After inspection we correct the observed phases of the remaining spectra for cosmic time dilation and use them together with their SNID phases in a linear fit where the slope is set to unity, as one would expect the SNID phases to be identical with the observed phases plus an offset. The fitting is done using the nested sampling algorithm MLFriends \citep{Buchner2016,Buchner2017} using the \textsc{Python} package \textsc{UltraNest}\footnote{\url{https://johannesbuchner.github.io/UltraNest}} \citep{Buchner2021}, whereby the axes consists of the observed phases and the SNID phases.
We then use this fit as a first estimate of the \gls{toe}, based on which we reject all spectra later than $45$ days after the explosion for the final \gls{toe} determination (based on the same argument as the previous rejection). A second fit then yields the final \gls{toe} estimate. An illustration of this process can be seen in Figure~\ref{fig:toefit}. For illustrations of other \glspl{sne}, see Appendix~\ref{sec:app_interp}. The end result is stored as a Gaussian \gls{kde} built up from all samples using Silverman's rule \citep{silverman1986}, which we use as a statistical prior in the subsequent analysis.
\begin{figure}
	\includegraphics[]{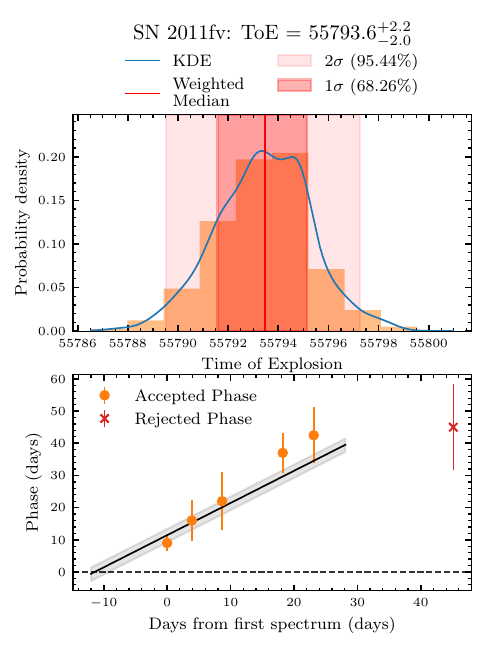}
	\caption{Illustration of the spectral phase fitting procedure. Lower: The individual phases obtained through SNID are plotted, both rejected and accepted phases. Here the rejected phase was excluded from the final \gls{toe} determination as it was determined to be later than $45$ days after the initial \gls{toe} fit. Upper: Depiction of the Gaussian \gls{kde}. The histogram is included for easy visualization of the sampled \glspl{toe}.}
	\label{fig:toefit}
\end{figure}

\subsubsection{Light curve constraints}
We cross-check the results obtained through the SNID spectral phase fitting against available non-detections, discoveries and photometry, either from PTF for the PTF targets, ASRAS\footnote{\url{https://www.rochesterastronomy.org/snimages}} for the LSQ targets or CBAT\footnote{\url{http://www.cbat.eps.harvard.edu}} telegrams for the remaining \glspl{sne}.
As long as the constraints from the non-detections and photometry are not in direct disagreement with the phase fitting results, the result obtained through the fit is used as \gls{toe} for the analysis.
For \wmf{} the available photometry is in disagreement with the fit result, i.e., the first photometry is one day before the estimated \gls{toe}. 
For \hi{} and \hnj{} the last non-detections are roughly one week after the SNID-based phase fitting \gls{toe}.
While this is not formally in disagreement with the SNID estimate, we discard the phase fitting result.
In the case of \hi, the phase matching results are generally poorly constrained, thus suggesting that the \gls{toe} constrained by its light curve is more accurate.
As for \hnj, the first available \gls{snfactory} spectrum is likely younger than the estimated $14$ days, still being a blue and featureless spectrum and being more in line with a \gls{toe} estimated by taking the midpoint between the last non-detection and discovery date (this technique is used by e.g. \citealt{Guti_rrez_2017} as well).
Lastly, the SNID-based \gls{toe} determined for \xlr{} has rather large uncertainties, whereas the available non-detection and discovery date constrain the \gls{toe} rather well. Since both agree within the uncertainties, we opt for the more precise option, i.e. its light curve constraint.
Hence, for these four \glspl{sne} we estimate the \gls{toe} by using this midpoint technique. The resulting \gls{kde} is a Gaussian probability density function with a standard deviation equal to half the interval between detection and non-detection.
This \gls{kde} is also used as a statistical prior in the following analysis.

\begin{table}
    \centering
	\caption{Redshifts of the \gls{snfactory} sample, corrected for peculiar velocities using the 2M++-SDSS map.}
	\label{tab:pvred}
	\begingroup
	\setlength{\tabcolsep}{6pt}
	\renewcommand{\arraystretch}{1.25}
	\begin{tabular}{l|c|l|c}
		\hline\hline
		\gls{sne} name & $z_\mathrm{pv}$ & \gls{sne} name & $z_\mathrm{pv}$ \\
		\hline
		\tpa     & $0.03291$  & \icone       & $0.02174$ \\
		\hb      & $0.01548$  & \icthreefive & $0.01443$ \\
		\wmf     & $0.02730$  & \fvq         & $0.03460$ \\
		\xlr     & $0.01450$  & \ljg         & $0.03291$ \\
		\jc      & $0.02376$  & \hi          & $0.02727$ \\
		\icthree & $0.01448$  & \zw          & $0.00925$ \\
		\ngctwo  & $0.00764$  & \hnj         & $0.01499$ \\
		\ngcfour & $0.04246$  & \ugc         & $0.02704$ \\
		\pgc     & $0.02411$  & \bjx         & $0.02898$ \\
		\cer     & $0.02431$  & \ds          & $0.02513$ \\
		\css     & $0.00968$  & & \\
		\hline
	\end{tabular}
    \endgroup
\end{table}

\subsection{The calibrator sample}\label{sec:data_calib}
To set the scale for our \glspl{sneIIp} Hubble-Lema\^itre diagram, we require a calibrator sample for which there is a distance modulus for the host galaxy as well as optical spectra and photometry of the hosted \glspl{sneIIp} having good phase coverage.
Such distance moduli comprise the ``second rung'' of the distance ladder, and can be obtained using techniques such as the period-luminosity relation for Cepheid stars or \gls{trgb}.
In order to ease comparison with past \gls{scm} work, we focus on the distances collected in Table~1 of \cite{Jaeger2022a}. According to \cite{Jaeger2020b}, these distance moduli have a correlated uncertainty due to the first rung calibration that amounts to 1.2\,km\,s$^{-1}$\,Mpc$^{-1}$ on $H_0$. As will be seen, this is subdominant to the other statistical errors in our analysis, so we simply accept the \cite{Jaeger2020b} value and do not consider it further.

Of these, we do not use SN~2014bc and SN~2022yyz due to their lack of sufficient phase coverage.
Furthermore we exclude \ay{} as we found some inconsistencies in the distance used by \cite{Jaeger2022a}; their cited distance to NGC~3938, the host of \ay{}, refers to the distance to NGC~3982 determined by \cite{Riess2016}.
While both galaxies are part of the Ursa Major groups, they are located in the South and North galaxy group, respectively, and as such their distances may be sufficiently different to impact the \gls{scm}.
Since we cannot establish the extent of this distance error, we exclude \ay{} from our calibrator sample.
To compensate for these cuts, we add \hn{} and \fourdj{}, the only other \glspl{sneIIp} with sufficient phase coverage and available distance moduli at the time when the calibrator sample was set\footnote{Some candidates had to be excluded as extreme outliers, e.g. \ixf{} due to strong CSM interaction and extremely blue color \citep{bostroem2023a}.}.

For many of the \glspl{sne} in \cite{Jaeger2022a} there exist additional distance moduli.
These include the \cite{Freedman2019} \gls{trgb} distance to NGC~1448; \gls{trgb} distances to NGC~2403 from \cite{dalcanton2009a}, \cite{radburn-smith2011a} and \cite{tully2009a}, along with a Cepheid distance from \cite{freedman2001a}; a Cepheid distance to NGC~7783 from \cite{zgirski2017a}; for NGC~3351 the \gls{trgb} distance from \cite{tully2009a} and the Cepheid distance from \cite{freedman2001a}; the \cite{Csornyei2023b} Cepheid distance to NGC~5194; and \gls{trgb} distances to NGC~1559 from \cite{li2024a}.
Although there are some disagreements when there are multiple distance moduli for the same host, they are subdominant to the scatter we find for the \gls{scm}.
\begin{table*}
	\centering
	\caption{Overview of the calibrator sample.}
	\label{tab:calibratordist}
	\begingroup
	\setlength{\tabcolsep}{6pt}
	\renewcommand{\arraystretch}{1.25}
	\begin{tabular}{l|c|c|c|c|c|c}
        \hline\hline
		\gls{sne} name & Host galaxy & $\mu$ (mag) & Calibrator & Distance reference & \gls{toe} (MJD) & \gls{toe} reference \\
        \hline
		\sem & \href{https://ned.ipac.caltech.edu/byname?objname=ngc+1637&hconst=67.8&omegam=0.308&omegav=0.692&wmap=4&corr_z=1}{NGC 1637} & $30.26\pm 0.09$ & Cepheids & 1 (updated from 2) & $51476.5\pm 1.0$ & 3 \\
		\gi & \href{https://ned.ipac.caltech.edu/byname?objname=NGC+3184&hconst=67.8&omegam=0.308&omegav=0.692&wmap=4&corr_z=1}{NGC 3184} & $30.64\pm 0.11$ & Cepheids & 1 (updated from 4) & $51517.8\pm 3.0$ & 3 \\
		\hn & \href{https://ned.ipac.caltech.edu/byname?objname=ngc+1448&hconst=67.8&omegam=0.308&omegav=0.692&wmap=4&corr_z=1}{NGC 1448} & $31.29\pm 0.04$ & Cepheids & 5 & $52863.7\pm 3.8$ & SNID \\
		\fourdj & \href{https://ned.ipac.caltech.edu/byname?objname=NGC+2403&hconst=67.8&omegam=0.308&omegav=0.692&wmap=4&corr_z=1}{NGC 2403} & $27.43\pm 0.15$ & Cepheids & 6 & $53187.0\pm 4.0$ & 7 \\
		\et & \href{https://ned.ipac.caltech.edu/byname?objname=ngc\%206946&hconst=67.8&omegam=0.308&omegav=0.692&wmap=4&corr_z=1}{NGC 6946} & $29.21\pm 0.16$ & \gls{trgb} & From EDD, 8 & $53271.0\pm 1.0$ & 9 \\
		\cs & \href{https://ned.ipac.caltech.edu/byname?objname=NGC+5194&hconst=67.8&omegam=0.308&omegav=0.692&wmap=4&corr_z=1}{NGC 5194} & $29.62\pm 0.09$ & \gls{trgb} & 1 (updated from 10) & $53548.5\pm 0.5$ & 3 \\
		\bk & \href{https://ned.ipac.caltech.edu/byname?objname=NGC+7793&hconst=67.8&omegam=0.308&omegav=0.692&wmap=4&corr_z=1}{NGC 7793} & $27.80\pm 0.08$ & \gls{trgb} & From EDD, 8 & $54542.9\pm 6.0$ & 11 \\
		\ib & \href{https://ned.ipac.caltech.edu/byname?objname=NGC+1559&hconst=67.8&omegam=0.308&omegav=0.692&wmap=4&corr_z=1}{NGC 1559} & $31.49\pm 0.06$ & Cepheids & 5 & $55041.3\pm 3.1$ & 3 \\
		\aw & \href{https://ned.ipac.caltech.edu/byname?objname=NGC+3351&hconst=67.8&omegam=0.308&omegav=0.692&wmap=4&corr_z=1}{NGC 3351} & $29.82\pm 0.09$ & Cepheids & 1 (updated from 12) & $56002.1\pm 1.0$ & 13 \\
		\ej & \href{https://ned.ipac.caltech.edu/byname?objname=NGC+628&hconst=67.8 &omegam=0.308&omegav=0.692&wmap=4&corr_z=1}{NGC 628} & $29.90\pm 0.08$ & \gls{trgb} &1 (updated from 10) & $56496.9\pm 1.0$ & 13 \\
        \eaw & \href{https://ned.ipac.caltech.edu/byname?objname=ngc\%206946&hconst=67.8&omegam=0.308&omegav=0.692&wmap=4&corr_z=1}{NGC 6946} & $29.21\pm 0.16$ & \gls{trgb} & From EDD, 8 & $57885.2\pm 0.1$ & 14 \\
        \aoq & \href{https://ned.ipac.caltech.edu/byname?objname=NGC+4151&hconst=67.8&omegam=0.308&omegav=0.692&wmap=4&corr_z=1}{NGC 4151} & $31.04\pm 0.07$ &  Cepheids & 15 & $58208.5\pm 1.0$ & ASRAS \\
		\hline
	\end{tabular}
	\endgroup
 \tablebib{
 (1)~\citet{Jaeger2020b};
 (2)~\citet{Leonard2003};
 (3)~\citet{Takats2015};
 (4)~\citet{Leonard2002};
 (5)~\citet{Riess2022a};
 (6)~\citet{Saha2006a};
 (7)~\citet{silverman2017a};
 (8)~\citet{anand2021a};
 (9)~\citet{Faran2014};
 (10)~\citet{McQuinn2017};
 (11)~\citet{Jaeger2017};
 (12)~\citet{Kanbur2003};
 (13)~\citet{deJaeger2019};
 (14)~\citet{Dyk2019a};
 (15)~\citet{Yuan2020a} 
 }
\end{table*}

Table~\ref{tab:calibratordist} also contains the adopted \gls{toe} for each calibrator \gls{sne}.\footnote{\citealt{Jaeger2020b,Jaeger2022a} do not specify which \gls{toe} they use in their analysis and it is not guaranteed that the adopted \glspl{toe} are identical.}
The majority of these values were taken from literature (where they were predominantly determined using the previously mentioned midpoint technique), with the exception of \hn{} and \aoq. In the case of \hn{} we find the available value of $52866.5\pm 10$ \citep{Gutierrez2017} to be insufficiently constraining, and therefore we use the more precise value yielded by the SNID phase fitting method.
The resultant fit is illustrated in Figure~\ref{fig:app_03hn_toe}.
As for \aoq, we use the non-detection and discovery date as reported on ASRAS due to the lack of a literature value and apply the midpoint method to obtain the \gls{toe}.

\subsubsection{Calibrator photometry}\label{sec:calibphot}
We collect the photometry from various sources (for details and references, see Appendix~\ref{sec:appendix_calib}) resulting in several photometric systems. Most of the \gls{sne} photometry has been taken in the \gls{kait} \citep{Filippenko2001} photometric system \citep{Ganeshalingam2010}. Some of these \glspl{sne} also have photometry available in the Nickel photometric system \citep{Stahl2019}, but we find there to be a small systematic difference between the Nickel and \gls{kait} photometry, even after correcting it to the revised UBVRI system defined by \citet[henceforth referred to as the \gls{bessell12} system; see below for details on the correction]{Bessell2012}. Since the \gls{kait} photometry was sufficiently well sampled and the Nickel photometry only constitutes a small fraction of the overall photometry data, we decide to drop the Nickel photometry to avoid potential issues caused by this difference.

We apply the AKS correction as described by \cite{Jaeger2015,Jaeger2017} to all calibrator \glspl{sne}.
This correction consists of a so-called S-correction \citep{Stritzinger2002} to account for the difference with respect to the \gls{bessell12} system, a Milky Way extinction correction (AvG correction) and a K-correction \citep{Hamuy1993, Kim1996}. Here, we use a modified version of the approach described by \cite{Jaeger2017} (using large parts of the codebase\footnote{\url{https://github.com/tdejaeger/Astronomy}} provided therein) with two main differences.
First, instead of using the same model spectra from \cite{Dessart2013} for every \gls{sne} to calculate the AKS correction, we use the observed spectra of each \gls{sne} with sufficient wavelength coverage for at least the VRI bands since we only utilize the V and I band for the \gls{scm}. In cases where the spectra do not have sufficient spectral coverage in the red, we extend the wavelength range using the model spectra of \cite{Dessart2013} (see Section~\ref{sec:calibspectra} for details).
Second, instead of calculating the AKS correction at each photometric phase
using the spectrum closest in time, we interpolate the photometry (using the
method described in Section~\ref{sec:phot_interp}) to the spectral phases for
which the correction is calculated. As for the photometry, these AKS values are then analogous to the
photometry interpolated\footnote{Here the hyperparameter optimization is done
using the L-BFGS-B method \citep{Byrd1995,Zhu1997}} back to the photometric
phases and used to correct the observed photometry. The results obtained through
this method are in qualitatively good agreement with the results obtained with the
unmodified \cite{Jaeger2017} approach, although we observe some deviations
especially in the bluer bands at later phases. This is most likely caused by the
blue regions of the later spectra being quite feature rich, which is where the
model spectra of \cite{Dessart2013} differ the most from the observed spectra.
The uncertainties arising from the AKS correction are propagated by adding them in quadrature to the original photometry uncertainties.

\subsubsection{Calibrator spectra}\label{sec:calibspectra}
All spectra of the calibrator \glspl{sne} (for details such as the data source, see Appendix~\ref{sec:appendix_calib}) are redshift corrected and corrected for Milky Way extinction in the same fashion as the \gls{snfactory} spectra, see Section~\ref{sec:data_snfactory}.

For spectra that are used in the AKS correction described in Section~\ref{sec:calibphot} we remove any prominent H$_\alpha$ host galaxy emission lines by using a Piecewise Cubic Hermite Interpolating Polynomial interpolator \citep{Fritsch1984}. Additionally we remove the two most noticeable telluric features at around $6900$\,\AA{} and $7600$\,\AA{} by linear interpolation.
This has been done less out of concern for the spectral analysis (see Section~\ref{sec:spec_fit}), but for a more accurate photometry correction, as, e.g., the \gls{kait} filter transmission curves already contain the telluric absorption lines.
For the spectra of \et{}, we apply the flux correction of \citet{Csornyei2023a}.

Furthermore, we extend the wavelength range of spectra that are used in the AKS correction but have insufficient wavelength coverage in the red. For this we first find a match between the spectrum and the models of \cite{Dessart2013} using SNID.
In cases where SNID struggles to find convincing matches (particularly in the early phases) we find matches using the \textsc{template\_comparison} module of the \textsc{specutils} package according to the lowest returned $\chi^2$ value.
Here, we give preference to spectra that, after visual inspection, yield a better flux fidelity rather than matching features, because these regions are only relevant for the AKS correction.
We then stitch these matches to observed spectra avoiding features in order not to create any offset in the overall continuum flux.
During the stitching process, we resample the wavelength resolution of the model spectra to match the observed spectra.
In some cases of very early blue and featureless spectra, the \cite{Dessart2013} spectra did not yield any suitable matches, and we simply fit a blackbody function to the spectra and use this blackbody spectrum in the stitching procedure.

\section{Data analysis}\label{sec:methods}
In this section, the methodology used for data analysis is described.
All methods included in this section are part of the \textsc{Sccala}\footnote{\url{https://github.com/AlexHls/Sccala}} toolkit and have been applied using it.

\subsection{Synthetic photometry}
For the \gls{snfactory} \glspl{sne}, we obtain synthetic photometry in the \gls{bessell12} system, with the photonic response curves taken from Table~1 therein.
Synthetic magnitudes have been calculated using the STIS005 Vega reference spectrum\footnote{Downloaded from \url{https://archive.stsci.edu/hlsps/reference-atlases/cdbs/calspec/alpha_lyr_stis_005.fits}, last accessed 24.06.2021}.
In addition to the uncertainty originating from pixel-to-pixel variations, the uncertainties include contributions from the gray multi-filter ratios (MFR's) that are used to calibrate the spectra on non-photometric nights.
In addition to the MFR uncertainties, there is a second completely correlated uncertainty, which is also present in photometric nights.
This uncertainty is dubbed ``repeatability'' and has been empirically determined to be around 0.014 mag \citep{rubin2022a}.
The uncertainties of the individual spectra are then combined using inverse-variance weighting to obtain the gray uncertainty in magnitudes of the combined spectrum.

\subsection{Spectral line fitting}\label{sec:spec_fit}
An integral part of the \gls{scm} is the analysis of spectral line features. In particular, the H$_\alpha$ ($\lambda_\mathrm{rest}^\mathrm{air} = 6563$\,\AA) and H$_\beta$ ($\lambda_\mathrm{rest}^\mathrm{air} = 4861$\,\AA) features are of interest here.
Here, the shift of the H$_\beta$ absorption minimum from the rest wavelength is used to determine $v_{\mathrm{H}\beta}$ using the relativistic Doppler shift formula.
In contrast, the fluxes at the H$_\alpha$ absorption minimum and emission maximum are used to obtain a value for $a/e$ \footnote{Our definition differs from those of \citet{Patat1994} and \citet{Gutierrez2014}, who used ratios of pseudo equivalent widths}:
\begin{equation}
    \frac{a}{e} = \frac{f_\mathrm{absorption, min}}{f_\mathrm{emission, max}}.
\end{equation}
$v_{\mathrm{H}\beta}$ is commonly measured by generating template matches, e.g. through SNID with spectra where $v_{\mathrm{H}\beta}$ is known, or by manually fitting a Gaussian on a linear background through, e.g., IRAF \citep{Tody1986}.
As our dataset has sufficiently high signal-to-noise ratio to allow for a direct fit of the respective feature, we generally follow the latter approach. We go, however, beyond fitting a Gaussian to the spectra, since we observe a significant variation in, e.g., the resulting H$_\beta$ velocities depending on how the fit is done. Specifically, we find rather large differences depending on the degree of the background polynomial or whether a skewed or non-skewed Gaussian is used. Since none of these choices can {\it a priori} be assumed to be more correct than others, we take a different approach to avoid this issue altogether.

\begin{figure*}[!hbt]
    \sidecaption
    \includegraphics[width=12cm]{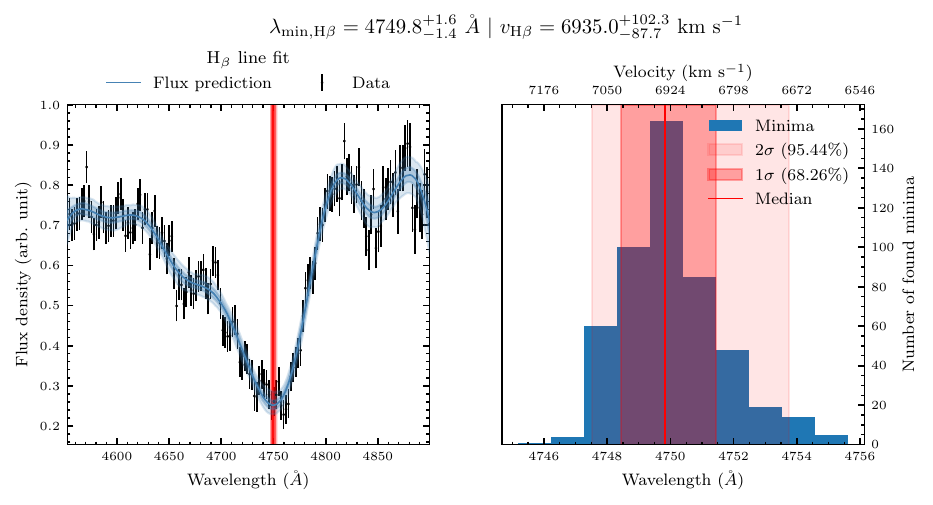}
    \caption{Left: Gaussian Process fit of an H$_\beta$ feature. A
secondary feature can be seen around $4850$\,\AA, as well as a shoulder at
roughly $4650$\,\AA. Both of those are fitted quite well, whereas they would be
problematic in a classical Gaussian fit. The
vertical red lines correspond to the found minima of the predicted curves
(depicted in blue). Right: Histogram of the found absorption minima.}
    \label{fig:hbetafit}
\end{figure*}

\begin{figure*}[!hbt]
    \sidecaption
    \includegraphics[width=12cm]{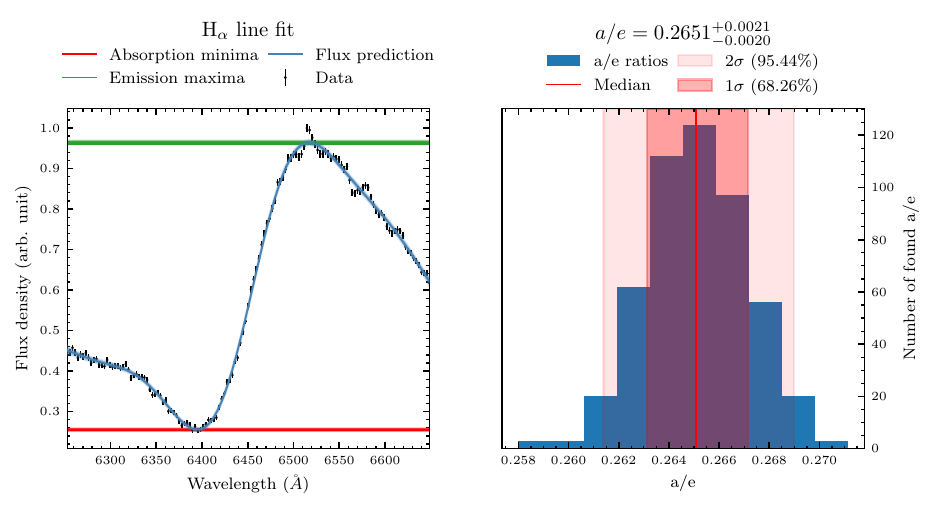}
    \caption{Left: Gaussian Process fit of an H$_\alpha$ feature. While
the choice of the boundaries for a Gaussian fit can be quite arbitrary due to
the asymmetrical shape of a P-Cygni feature, the Gaussian Process fit
simultaneously fits the flux minimum and maximum from which the $a/e$ ratio is
calculated. The horizontal red and green lines correspond to the found minima
and maxima, respectively, of the predicted curves (depicted in blue).
Right: Histogram of the found $a/e$ ratios.}
    \label{fig:halphafit}
\end{figure*}

In this work, we analyze the line features by fitting a non-parametric curve to the feature using Gaussian Process regression (see, e.g., \citealt{rasmussen2005}) utilizing the fast and flexible \textsc{Python} library \textsc{George} \citep{Ambikasaran2015a}.
For this procedure, we have implemented an automatic pipeline in the \textsc{Sccala} toolkit:
First, the input spectrum is cut down to the region around the feature of interest, i.e. either the H$_\beta$ absorption feature or the H$_\alpha$ P-Cygni feature.
This means that the wavelength range is trimmed until the desired absorption minimum and emission maximum are the dominant features in the selection window, ensuring that the correct extrema are fitted.
Second, the normalized flux is modeled using a Gaussian Process with a squared exponential kernel, whereby the hyperparameters are marginalized over using a \gls{mcmc} Ensemble sampler \citep{Goodman2010} through the \textsc{Python} library \textsc{emcee} \citep{ForemanMackey2013}.
In the case of the Hubble-flow sample, we add a second exponential kernel to account for the observed correlated noise originating from the spectral extraction, which is characterized by significantly smaller length scales than the actual feature. Hereby the marginalization step occurs for both kernels simultaneously.
Third, $10\,000$ random samples of the hyperparameter vectors are chosen, from which fitted curves are predicted on a linearly spaced grid with ten times the resolution of the original spectrum.
In the cases where two kernels were used (i.e. for the Hubble-flow sample spectra), the prediction is done using only the first kernel, thus removing the correlated noise. From these fitted curves, the wavelength of the minimum flux is selected in the case of the H$_\beta$ feature and the minimum and maximum flux in the case of the H$_\alpha$ line.
These values are then converted into velocities and $a/e$ ratios, respectively.
Lastly, we manually inspect all the fits and reject inadequate fits, e.g. because the feature has not yet formed in early spectra or because no feature could be detected.
Examples of successful fits can be seen in Figures \ref{fig:hbetafit} and \ref{fig:halphafit} for H$_\beta$ and H$_\alpha$ features, respectively.

\subsection{Phase interpolation}
For the application of the \gls{scm}, SN characteristcs are needed from the same phase for all SNe, which requires interpolation of both the photometry and the spectral line fit values. In this regard, we note that for our SNfactory sample, both photometry and spectral characteristics are simultaneously available at each observed spectral phase.

\subsubsection{Velocity and $a/e$ interpolation}
While in the literature (e.g., \citealt{Hamuy2001,Poznanski2009,Jaeger2017}) the H$_\beta$ velocities conventionally are interpolated using a power-law of the form $v(t)=A\cdot t^\gamma$, we use an interpolation approach based on Gaussian Process regression.
We opted for this method as we found our data to show significant deviations from a power-law (e.g. plateau or linear phases).
Another reason for using Gaussian Process regression was that there exists no established parametric relation for the $a/e$ values.
We implement the Gaussian Process regression again through \textsc{George} in combination with a hyperparameter marginalization through \textsc{emcee}.
In order to avoid problems caused by very early time spectra, we only consider data points within the phase range of interest; in most cases from around $\sim 10$ to $90$~days after explosion.
We additionally take the uncertainty of the ToE into account by drawing random values from the ToE KDE (see Section~\ref{sec:data_snfactory}) and calculating the interpolated values with respect to the resulting phases.
This interpolation is illustrated in Figure~\ref{fig:velocity_fit}.
All \gls{snfactory} \glspl{sne} interpolations are illustrated in Appendix~\ref{sec:app_interp}.
\begin{figure*}
    \sidecaption
    \includegraphics[width=12cm]{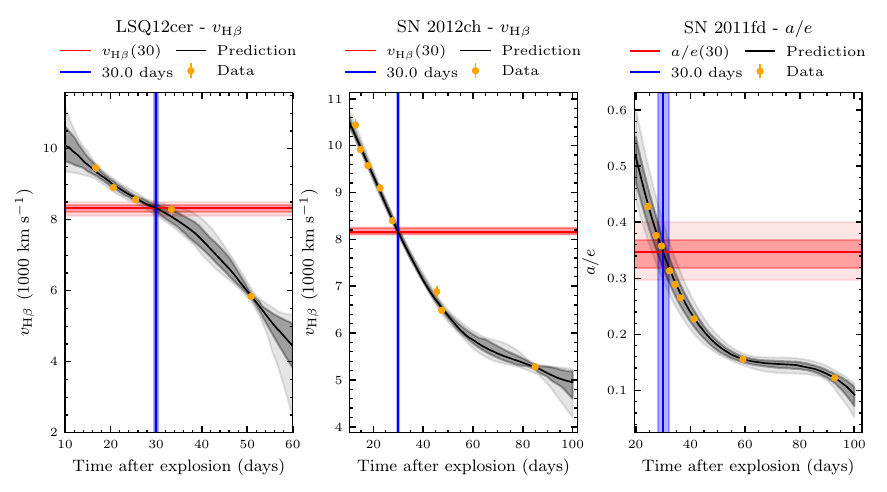}
    \caption{Interpolation of the $v_{\mathrm{H}\beta}$ and $a/e$ values for three different SNe. The shaded regions of the red and blue lines indicate the $1\,\sigma$ and $2\,\sigma$ uncertainty regions of the interpolated values and calculated location of day $30$, respectively. Here, day $30$ is chosen to illustrate how the ToE uncertainty is propagated to the interpolated values. The black line represents the prediction from the Gaussian Process regression, and its shaded regions the $1\,\sigma$ and $2\,\sigma$ uncertainty regions, respectively.}
    \label{fig:velocity_fit}
\end{figure*}

\subsubsection{Photometry}\label{sec:phot_interp}
We interpolate the photometry analogously to the velocities and $a/e$ values, i.e. using Gaussian Process regression.
Here, the main difference is that only data during the plateau phase are used, that is, from around $20$ to $60$ days after explosion.

\subsection{Standardization characteristics consistency}\label{sec:consistency}
With the \gls{sne} standardization characteristic measurements now in hand, we can examine them for their suitability for standardization. This is in the spirit of similar efforts to align the standardization characteristics when measuring $H_0$ using \gls{sneIa}, though even there it has proven challenging \citep[e.g.,][]{Martins2025}.

We begin by comparing our unstandardized luminosity distributions with those from the ZTF volume-limited \gls{sneIIp} sample \citep{das2025a, das2026a}.
Since our luminosities are in I band and phase of $+30$~day while those of ZTF are in r band at peak, we convert our data to r band using the transformation relation of \cite{Jordi2006} and the color curve of \citet{das2025a}.
We then use the luminosity-dependent decline rate formula in \citep{das2026a} to convert to peak.
A Kolmogorov-Smirnov (K-S) test of the resulting $M_r$ magnitudes shows that both our Hubble-flow and calibrator samples agree with the ZTF volume-limited \gls{sneIIp} sample, with K-S probabilities of $0.64$ and $0.18$, respectively.
This may seem surprising at first, since the searches that fed the selection of our Hubble-flow sample do suffer magnitude limits.
But those limits occur fainter than the peak of the ZTF luminosity function, where the numbers of SNe are already following off naturally.
This comports with the K-S probability for agreement of the unstandardized luminosities between our Hubble-flow and calibrator samples of $0.19$.

\begin{figure}
\centering
    \includegraphics[width=\columnwidth]{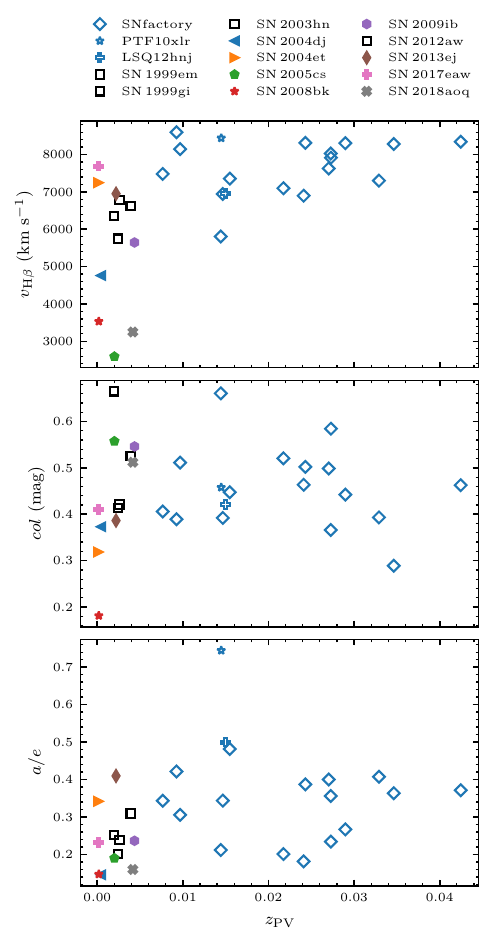}
    \caption{Measured $v_{\mathrm{H}\beta}$, $col$ and $a/e$ values for our HF and calibrator
 sub-samples, as listed in Table~\ref{tab:scmdata}, versus their cosmological redshift . Specific calibrator SNe discussed in the text are denoted with solid color markers to make them easier to locate on the plot. It should be noted that \xlr{} and \hnj{} are part of the SNfactory sample as well.}
    \label{fig:vel_distr}
 \end{figure} 
Despite the good agreement with the intrinsic \gls{sneIIp} luminosity function, and unstandardized luminosities between the Hubble-flow and calibrators, it appears that the \vhb{} and $a/e$ distributions are inconsistent with having been drawn from the same parent population.
For example, our Hubble-flow sample and calibrator \vhb\ values have a probability of only $0.0009$ that they are drawn from the same parent population.
This is evident in Figure~\ref{fig:vel_distr}, where it can be seen that primarily lower \vhb\ \glspl{sneIIp} are absent from our Hubble-flow sample. 
Such \glspl{sneIIp} are intrinsically fainter, but the surveys we sourced for candidates are sufficiently deep to have detected these at the lower end of our redshift range. We note that Table~8 of \citet{Gutierrez2017} indicates a mean \vhb\ $\sim 7100$\,km\,s$^{-1}$ interpolated to a phase of $+30$\,d --- in good agreement with our Hubble flow sample but well above the mean \vhb\ for our calibrator sample.
The reason for the scarcity of low-\vhb\ events in our Hubble-flow sample and the ZTF sample is presently unknown.

There is also disagreement between the $a/e$ distributions, with a K-S probability of 0.01.
It is very likely that \glspl{sneIIp} with obvious \gls{csm} interaction would not have been selected for distance measurements of their Cepheid or TRGBs, since such distance measurements are very expensive.
This could have eliminated large $a/e$ \glspl{sneIIp} from the calibrator sample, and that is what is seen in Figure~\ref{fig:vel_distr}.
Even after removing our two Hubble-flow \glspl{sneIIp} showing clear signs of \gls{csm} interaction, the K-S probability of agreement is only $0.02$.
However, it is plausible that even more exacting limits on signs of \gls{csm} interaction were enforced, given that the calibrator sample is seen to skew to low $a/e$.
Colloquial evidence from VLT spectroscopic follow-up of \glspl{sneIIp} suggests that hints of \gls{csm} interaction are present in roughly half the sample (Hillebrandt, priv. comm.), so this explanation appears plausible.

On the other hand, as is evident in Figure~\ref{fig:vel_distr}, the $col$ standardization characteristics agree very well between samples.
Here, we find a K-S probability of $0.71$.

Despite these differences, and the knowledge that \vhb\ and $a/e$ correlate with luminosity, the luminosity agreement discussed above indicates that these differences do not appear to be strongly driven by selection by the parent \gls{sne} discovery surveys or our selection of which candidates to initially or subsequently follow.

We do not know what impact the additional selection step for distance measurement with Cepheids or TRGB might have.
Actively star-forming galaxies are preferred for Cepheid searches, and often it was required to have hosted a \gls{sneIa}.
It is perhaps telling that while all of our calibrator \glspl{sneIIp} were hosted by NGC galaxies only $23\%$ of $z < 0.01$ \glspl{sneIIp} in the TNS were. 

Interestingly, past SCM analyses exhibit similar differences, which have seemingly passed unrecognized.
For instance, in our literature comparison in Section~\ref{sec:djcomp} we find that the Hubble-flow and calibrator \vhb\ distributions in \citet{Jaeger2022a} have only 0.018 probability of agreement.

Because of these differences, in what follows we will show alternative analyses that remove \glspl{sneIIp} at the extrema --- those with low \vhb\ and high $a/e$ --- such that the Hubble-flow and calibrator standardization characteristics agree better.
We will also show an alternate analysis that allows the standardization coefficients to differ between the Hubble-flow and calibrator \glspl{sne}.

Finally, before proceeding it is worth recalling the role that the standardization characteristic population distribution functions play in the hierarchical Bayesian model we will introduce next.
First, the latent population characteristics distributions are used to include magnitude selection effects and thus remove Malmquist bias.
As we will show, because our standardization dispersion is low, our Malmquist bias correction is small.
The second role of the latent population characteristics distribution is to coax the latent standardization characteristic values of poorly-measured data towards that population distribution.
However, as can be seen in Table~\ref{tab:scmdata}, all of our \glspl{sneIIp} are well-measured compared to the range of the latent population characteristics distributions.
So this benefit of including latent population characteristics distributions should be modest.

\section{Standardizable candle method}\label{sec:scm_methods}
In this section, the standardization model is described.
The models outlined in this section are also part of the \textsc{Sccala}\footnote{\url{https://github.com/AlexHls/Sccala}} toolkit and have been applied using it.
For a full description of the models, see Appendix~\ref{sec:app_model_expl}.

\subsection{$H_0$-free \gls{scm}}\label{sec:h0freescm}
The \gls{scm} is based on the empirical correlation between the \gls{sneIIp} luminosity and the photospheric expansion velocity during the plateau phase \citep{Hamuy2001,Hamuy2001a}, but it has been modified over the years.
In particular, a color term ostensibly to deal with extinction has been added by \cite{Nugent2006}.
We refer to this form of the \gls{scm} including a velocity and color correction as the classical \gls{scm}. In this work, we extend the classical \gls{scm} by an additional correction term that includes the $a/e$ ratio (see Section~\ref{sec:results} for an investigation of this new correction term).
Using these three corrections, the true apparent magnitudes can be modeled as:
\begin{equation}
\begin{split}
    m_I^{\mathrm{true}} & = \mathcal{M}_I - \alpha \cdot \log_{10}\left(\frac{v_{\mathrm{H}\beta}}{\langle v_{\mathrm{H}\beta}\rangle}\right)\\
    & + \beta \cdot (col - \langle col \rangle) + \gamma \cdot \left(\frac{a}{e} - \langle \frac{a}{e}\rangle\right)\\
    & + 5\log_{10} \left(\mathcal{D}_L(z_\mathrm{CMB})\right).
\end{split}
    \label{eq:h0free_mag}
\end{equation}
Here, $\langle \cdot \rangle$ refers to the unweighted mean of the respective quantity and the color $col$ is defined as $col = m_V - m_I$.
The logarithmic form of the velocity term is physically motivated by the square-root relation between velocity and kinetic energy combined with the power-law relation between kinetic energy and luminosity, and the use of the logarithm of the luminosity.
Similarly, $col$ is a logarithmic quantity that is good at capturing both temperature and dust reddening. 
The linear form for $a/e$ is arbitrary.

The coefficients $\alpha$, $\beta$, $\gamma$ and $\mathcal{M}_I$ are free parameters, where $\mathcal{M}_I$ is the so-called Hubble-free absolute magnitude defined by
\begin{equation}
    \mathcal{M}_I = M_I - 5 \log_{10}\left(\frac{H_0}{\mathrm{km} \,\,\mathrm{s}^{-1}\,\, \mathrm{Mpc}^{-1}}\right) + 25.\label{eq:hubble-free-abs-mag}
\end{equation}
Last but not least, $\mathcal{D}_L$ is the Hubble-free luminosity distance given by
\begin{equation}
    \mathcal{D}_L(z_\mathrm{CMB}) = H_0 \cdot d_L(z_\mathrm{CMB}),
\end{equation}
where $d_L$ is the luminosity distance.
In our case, $z_\mathrm{CMB}$ is replaced by the peculiar velocity corrected redshift $z_\mathrm{pv}$ from Table~\ref{tab:pvred}, but this method in principle also works with the uncorrected redshift.
Since the redshifts of our \glspl{sne} are rather small ($\lesssim 0.04$), we use the kinematic expansion of the luminosity distance
\begin{equation}
    d_L \approx \frac{cz}{H_0}\left(1 + \frac{(1-q_0)z}{2} - \frac{(1-q_0-3q_0^2+j_0)z^2}{6}\right)
\end{equation}
as defined by e.g. \citet{Riess2004}, with $q_0 = -0.55$ and $j_0=1$ \citep{Riess2022a}.
We will refer to the \gls{scm} including the additional $a/e$ correction term as the extended \gls{scm}.

In the literature to date, the coefficients have been determined by a non-hierarchical likelihood function (see, e.g., \citealt{Jaeger2017,Jaeger2020a,Jaeger2020b}).
This approach faces several limitations, as observed \gls{sne} datasets are affected by, e.g., outliers, non-linear correlations, partially known uncertainties, selection effects and heterogeneity.
Although these issues can for the most part be addressed in separate steps (e.g. by manually culling the dataset for outliers), these steps usually do not account for the interconnected nature of these effects.
In case of \glspl{sneIa}, e.g. \cite{Kunz2007a} developed a Bayesian technique to account for outliers, but it does not consider selection effects.
For a more detailed discussion on the limitations of non-hierarchical models, see e.g. \cite{Rubin2015a}.
In this work, we introduce a hierarchical approach, closely following the UNITY framework of \citet{Rubin2015a}.
Moving to a hierarchical approach allows us, in principle, to address the above mentioned limitations in a single, unified framework.
In essence, this hierarchical approach assumes that the measured values ($v_{\mathrm{H}\beta}$, $col$, $a/e$) have underlying true values, represented by latent variables ($v^\mathrm{true}_{\mathrm{H}\beta}$, $col^\mathrm{true}$, $a/e^\mathrm{true}$).
These true values follow an unknown population distribution, which we parameterize using so-called hyperparameters ($R^v$, $v^*$, $R^c$, $c^*$, $R^{a/e}$, $a/e^*$, where $R^{param}$ and $param^*$ refer to the dispersion and of the respective latent variable; see Appendix~\ref{sec:app_model_expl}).
Both the latent variables and hyperparameters are fitted in our framework by optimizing a corresponding likelihood function, i.e. by minimizing the difference between observed quantities and latent variables.
Here, the sample-dependent unexplained dispersion, $\sigma_\mathrm{int}$, and systematic uncertainties, $\Delta\mathrm{sys}$, are taken into account.
Other effects, such as selection effects, are included in this likelihood function as well; see Appendix~\ref{sec:app_model_expl} for more details.
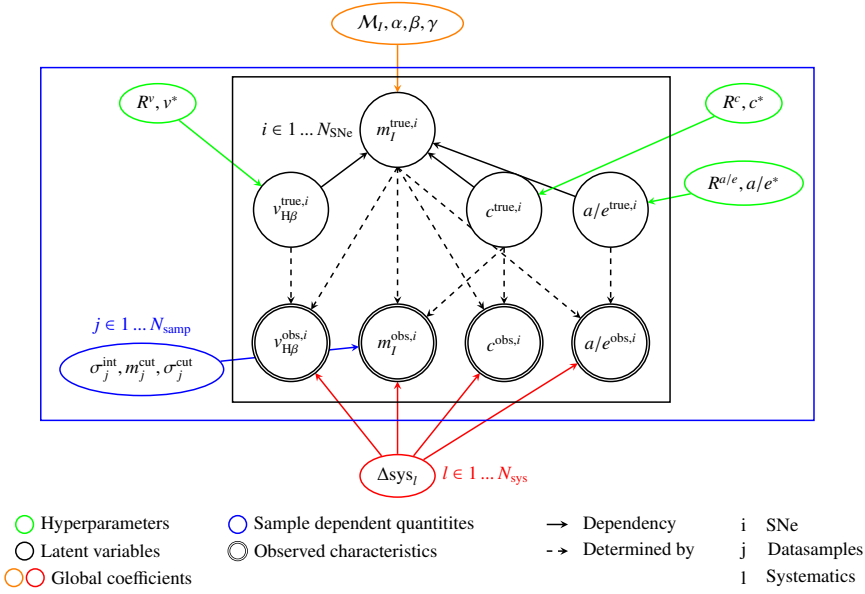
\begin{figure*}
    \begin{lrbox}{\modelbox}
    \begin{minipage}[c]{11.5cm}
    \centering
    \resizebox{\linewidth}{!}{%
    \begin{tikzpicture}[
        dot/.style = {circle,minimum size=#1,inner sep=0pt,outer sep=0pt,fill=white,thick},
        dot/.default = 40pt 
    ]
        \dimendef\prevdepth=0
        \node[ellipse,orange,draw,thick,text=black] (free) at (0, 2) {$\mathcal{M}_I, \alpha, \beta, \gamma$};

        \node[dot,draw,label=left:$i \in 1\,...\,N_\mathrm{SNe}$] (mtrue) at (0, 0) {$m_I^{\mathrm{true},i}$};
        \node[dot,draw] (vtrue) at (-2, -1.5) {$v_{\mathrm{H}\beta}^{\mathrm{true},i}$};
        \node[dot,draw] (ctrue) at (2, -1.5) {$c^{\mathrm{true},i}$};
        \node[dot,draw] (atrue) at (4, -1.5) {$a/e^{\mathrm{true},i}$};

        \node[dot,draw,accepting] (mobs) at (0, -4) {$m_I^{\mathrm{obs},i}$};
        \node[dot,draw,accepting] (vobs) at (-2, -4) {$v_{\mathrm{H}\beta}^{\mathrm{obs},i}$};
        \node[dot,draw,accepting] (cobs) at (2, -4) {$c^{\mathrm{obs},i}$};
        \node[dot,draw,accepting] (aobs) at (4, -4) {$a/e^{\mathrm{obs},i}$};

        \node[ellipse,green,draw,thick,text=black] (pv) at (-4.5, 0.5) {$R^v,v^*$};
        \node[ellipse,green,draw,thick,text=black] (pc) at (6.5, 0.5) {$R^c,c^*$};
        \node[ellipse,green,draw,thick,text=black] (pa) at (6.5, -1) {$R^{a/e},a/e^*$};

        \node[ellipse,blue,draw,thick,text=black,label={[text=blue]north:$j \in 1\,...\,N_\mathrm{samp}$}] (sint) at (-4.8, -4.5) {$\sigma_j^\mathrm{int}, m_j^\mathrm{cut},\sigma_j^\mathrm{cut}$};

        \node[ellipse,red,draw,thick,text=black,label={[text=red]east:$l \in 1\,...\,N_\mathrm{sys}$}] (sys) at (0, -6.5) {$\Delta\mathrm{sys}_l$};

        \draw[thick] ($(vtrue.north west)+(-0.6,2)$)  rectangle ($(aobs.south east)+(0.6,-0.6)$);
        \draw[thick,blue] ($(sint.south west)+(-0.8,-0.6)$)  rectangle ($(pc.north east)+(0.8,0.4)$);

        \node[circle,green,draw,thick,text=black,label=right:Hyperparameters] (hyp) at ($(sys.south)+(-7,-0.5)$) {};
        \node[circle,black,draw,thick,text=black,label=right:Latent variables] (lat) at ($(sys.south)+(-7,-1)$) {};
        \node[circle,orange,draw,thick,text=black] (glob1) at ($(sys.south)+(-7.2,-1.5)$) {};
        \node[circle,red,draw,thick,text=black,label=right:Global coefficients] (glob2) at ($(sys.south)+(-6.8,-1.5)$) {};
        \node[circle,blue,draw,thick,text=black,label=right:Sample dependent quantitites] (samp) at ($(sys.south)+(-3,-0.5)$) {};
        \node[circle,black,draw,accepting,text=black,label=right:Observed characteristics] (obs) at ($(sys.south)+(-3,-1.0)$) {};
        \draw[-stealth,thick] ($(sys.south)+(2.8,-0.5)$) to node[right]{\quad Dependency}($(sys.south)+(3.2,-0.5)$);
        \draw[-stealth,thick,dashed] ($(sys.south)+(2.8,-1.0)$) to node[right]{\quad Determined by}($(sys.south)+(3.2,-1.0)$);
        \node[circle,draw=none,text=black,label=right:SNe] (is) at ($(sys.south)+(6.5,-0.5)$) {i};
        \node[circle,draw=none,text=black,label=right:Datasamples] (is) at ($(sys.south)+(6.5,-1.0)$) {j};
        \node[circle,draw=none,text=black,label=right:Systematics] (is) at ($(sys.south)+(6.5,-1.5)$) {l};

        \begin{scope}[on background layer]
            \draw[stealth-,thick] (mtrue) to node[left]{}(vtrue);
            \draw[stealth-,thick] (mtrue) to node[left]{}(ctrue);
            \draw[stealth-,thick] (mtrue) to node[left]{}(atrue);

            \draw[-stealth,dashed,thick] (mtrue.south) to node[left]{}(vobs);
            \draw[-stealth,dashed,thick] (mtrue.south) to node[left]{}(mobs);
            \draw[-stealth,dashed,thick] (mtrue.south) to node[left]{}(cobs);
            \draw[-stealth,dashed,thick] (mtrue.south) to node[left]{}(aobs);

            \draw[-stealth,dashed,thick] (vtrue.south) to node[left]{}(vobs);
            \draw[-stealth,dashed,thick] (ctrue.south) to node[left]{}(cobs);
            \draw[-stealth,dashed,thick] (ctrue.south) to node[left]{}(mobs);
            \draw[-stealth,dashed,thick] (atrue.south) to node[left]{}(aobs);

            \draw[-stealth,orange,thick] (free.south) to node[left]{}(mtrue);

            \draw[-stealth,green,thick] (pv) to node[left]{}(vtrue);
            \draw[-stealth,green,thick] (pc) to node[left]{}(ctrue);
            \draw[-stealth,green,thick] (pa) to node[left]{}(atrue);

            \draw[-stealth,blue,thick] (sint) to node[left]{}(mobs);

            \draw[-stealth,red,thick] (sys) to node[left]{}(vobs);
            \draw[-stealth,red,thick] (sys) to node[left]{}(mobs);
            \draw[-stealth,red,thick] (sys) to node[left]{}(cobs);
            \draw[-stealth,red,thick] (sys) to node[left]{}(aobs);
        \end{scope}
    \end{tikzpicture}
    }
    \end{minipage}
    \end{lrbox}
    \sidecaption
    \usebox{\modelbox}
    \caption{Graphical representation of the framework used in our analysis.
    It should be noted that we only distinguish the sample into a Hubble-flow and a calibrator sample and not, e.g., by their observational campaign, due to the limited size of the overall data sample. For more details, see Appendix~\ref{sec:app_model_expl}.
    [Adapted from \cite{Rubin2015a}.]}
    \label{fig:model}
\end{figure*}
A graphical representation of this model can be seen in Figure~\ref{fig:model}, while a detailed explanation of the model, for example, the likelihood function used, can be found in Appendix~\ref{sec:app_model_expl}.
We fit this model to our data using the \textsc{Sccala} toolkit, which in turn relies on the \textsc{cmdstanpy}\footnote{\url{https://github.com/stan-dev/cmdstanpy}} wrapper of \textsc{Stan} \citep{Hoffman2014a,Betancourt2013a}.

Analogously to UNITY, we model a survey selection function as the CDF of a Gaussian having mean $m^\mathrm{cut}$ and dispersion $\sigma^\mathrm{cut}$.
As a prior for the \gls{snfactory} sample we assume a nominal survey depth of $m^\mathrm{nominal} = 18.5\,\mathrm{mag}$ and standard deviation $\sigma^\mathrm{cut} = 0.5\,\mathrm{mag}$. (We experimented with a broader prior with standard deviation $1$\,mag; it left the mean depth unchanged and broadened the roll-off by only around $0.01$~mag.)
This is about a magnitude brighter than the nominal limiting magnitude of the \gls{snfactory} SNIFS instrument, so this is really an empirical limit beyond which objects are less likely to be selected from the various discovery channels, which in turn have different limiting magnitudes independently. Since a Gaussian distribution of many such $m^\mathrm{cut}$ values results in another CDF of a Gaussian, and since
we marginalize over the survey depth in our modeling framework (see Appendix~\ref{sec:app_model_expl}), we find this a sufficient approximation of flux selection in the \gls{snfactory} dataset (see Appendix~\ref{sec:app_a} for details).

Similar to \citet{Rubin2015a}, we also test our framework on simulated data, comparing the fits of both the hierarchical and non-hierarchical models (see Appendix~\ref{sec:app_model_test} for details on the testing setup). The results are summarized in Table~\ref{tab:testdata}.
Here, it can be seen that the hierarchical model recovers the input coefficients much more accurately than the non-hierarchical model. A particular improvement can be seen with the color correction $\beta$.
However, since there is no way of knowing what the true model coefficients are outside of such artificial test scenarios, it is necessary to introduce a measure of quality to gauge the goodness of a certain fit versus another.
While other works \citep{Poznanski2009,Poznanski2010,Jaeger2017,Jaeger2020a,Jaeger2020b} use the fitted unexplained dispersion, $\sigma_\mathrm{int}$, as a measure of quality of a fit, we instead use the mean squared error (MSE)
\begin{equation}
    MSE = \frac{1}{N_\mathrm{SNe}}\sum_i^{N_\mathrm{SNe}} (\mu^\text{obs}(\mathcal{M}_I,\alpha,\beta,\gamma) - \mu(z_i, \text{cosmo}))^2
\end{equation}
as a measure of quality, since we find that the dispersion explained by the measurement uncertainties dominates over $\sigma_\mathrm{int}$ in our $H_0$-free SCM analysis.

\subsection{$H_0$ \gls{scm}}
While the $H_0$-free approach outlined above allows for a relative standardization of SNe II-P magnitudes with only a Hubble-flow sample, the extraction of a value for $H_0$ requires a calibrator sample with known distances.
In a nutshell, the calibrator sample fixes the value of $M_I$ in Equation \ref{eq:hubble-free-abs-mag}, i.e. providing an anchor for the Hubble diagram, while the Hubble-flow sample fixes the value for the slope, $H_0$.

In case of the Hubble-flow sample, Equation \ref{eq:h0free_mag} now reads:
\begin{equation}
\begin{split}
    m_I^{\mathrm{true}} & = M_I - \alpha \cdot \log_{10}\left(\frac{v_{\mathrm{H}\beta}}{\langle v_{\mathrm{H}\beta}\rangle}\right)\\
    & + \beta \cdot (col - \langle col \rangle) + \gamma \cdot \left(\frac{a}{e} - \langle \frac{a}{e}\rangle\right)\\
    & + 5\log_{10} \left(d_L(z_\mathrm{CMB}, H_0)\right) + 25.
\end{split}
\label{eq:h0scm_hf}
\end{equation}
Similarly, for the calibrator sample, the true magnitudes can be calculated as
\begin{equation}
\begin{split}
    m_{I,\mathrm{calib}}^{\mathrm{true}} & = M_I - \alpha \cdot \log_{10}\left(\frac{v_{\mathrm{H}\beta}}{\langle v_{\mathrm{H}\beta}\rangle}\right)\\
    & + \beta \cdot (col - \langle col \rangle) + \gamma \cdot \left(\frac{a}{e} - \langle \frac{a}{e}\rangle\right)\\
    & + \mu_\mathrm{calib},
\end{split}
\label{eq:h0scm_calib}
\end{equation}
utilizing the measured distance modulus $\mu_\mathrm{calib}$. Note that in case of the $H_0$ \gls{scm}, the free parameter $\mathcal{M_I}$ has been replaced by $M_I$.
Furthermore, the normalization averages $\langle \cdot \rangle$ in the Eqs.~\ref{eq:h0scm_hf} and \ref{eq:h0scm_calib} are based on the combined calibrator and Hubble-flow populations, i.e., they are identical.
For details on the full hierarchical model implementation, see Appendix~\ref{sec:app_model_expl}.

We also define a total uncertainty for each \gls{sne} as
\begin{equation}
\begin{split}
    \sigma_\mathrm{tot}^2 & = \sigma_{m_I}^2 + \left(\frac{\alpha}{\ln(10)}\frac{\sigma_{v_{\mathrm{H}\beta}}}{v_{\mathrm{H}\beta}}\right)^2\\
    & + \left(\beta~\sigma_{col}\right)^2 + \left(\gamma~\sigma_{a/e}\right)^2 + \sigma_z^2
\end{split}
\label{eq:sigma_tot_hf}
\end{equation}
in case of the Hubble-flow sample and as
\begin{equation}
\begin{split}
    \sigma_\mathrm{tot}^2 & = \sigma_{m_I}^2 + \left(\frac{\alpha}{\ln(10)}\frac{\sigma_{v_{\mathrm{H}\beta}}}{v_{\mathrm{H}\beta}}\right)^2\\
    & + \left(\beta~\sigma_{col}\right)^2 + \left(\gamma~\sigma_{a/e}\right)^2 + \sigma_{\mu_\mathrm{cal}}^2
\end{split}
\label{eq:sigma_tot_calib}
\end{equation}
for the calibrator sample.
Here, $\sigma_z$ refers to all external, redshift-dependent uncertainties that affect magnitudes (i.e. the first element of the diagonal of the covariance matrix component $\boldsymbol{\sigma}^\mathrm{ext}(z_i)$ defined in Appendix~\ref{sec:app_model_expl}).

\section{$H_0$-free \gls{scm} results}\label{sec:results}
We utilize the $H_0$-free \gls{scm} to investigate various aspects of the \gls{scm} and dataset while separating the analysis from the actual $H_0$ value, so as not to unconsciously bias it towards any particular value of $H_0$. 
A summary of the data used for the following analysis can be found in Table~\ref{tab:scmdata} in Appendix~\ref{app:data}.
During these investigations we address two main questions:
\begin{enumerate}
    \item How does the additional $a/e$ correction term improve the standardization?
    \item What is the optimal phase for standardization?
\end{enumerate}
\begin{figure}
    \centering
    \includegraphics[width=\linewidth]{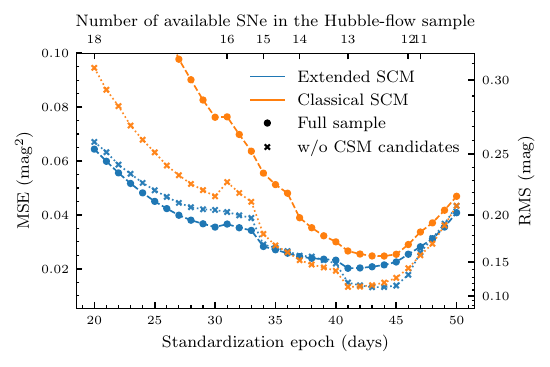}
    \caption{Illustration of the distance modulus MSE for various phases and modes utilizing the $H_0$-free \gls{scm}. Here, the full sample refers to the sample consisting of up to $18$ SNe as described in Section~\ref{sec:data}, while the sample ``w/o \acrshort{csm} candidates'' does not include PTF10xlr and LSQ12hnj. The MSE values of the classical \gls{scm} prior to $20$ days have also been computed, but have been cut off here for better visibility. After $50$ days, the number of available \glspl{sne} drops below $10$ and we no longer trust the resultant MSE values. The right hand $y$-axis shows the RMS value, defined as the square root of the respective MSE value. The upper $x$-axis lists the number of \glspl{sne} available at each standardization phase. Here only phases where the number of \glspl{sne} changes are counted; reading from left to right the number of \glspl{sne} stays constant, e.g. from $20$ to $30$ days, $18$ \glspl{sne} are available.}
    \label{fig:epoch_evolution}
\end{figure}

The introduction of an additional correction has been motivated mainly by the results of \citet{Jaeger2020a} that state that the correction strength of the color term is rather weak and that the main contribution to the standardization comes from the velocity term (see Figures 7 and 8 of \citealt{Jaeger2020a}).
While our sample is smaller than the one used by \citet{Jaeger2020a}, we can verify this behavior for our sample as well, at least up to a level that the sample size allows.
As a consequence of this, we introduce the additional $a/e$ correction term to increase the standardization strength of the \gls{scm}.

The addition of the $a/e$ correction term has been inspired by the correlation between the $a/e$ ratio and the magnitudes identified by \citet{Patat1994,Gutierrez2014,Gutierrez2017}\footnote{Technically, the $a/e$ ratio defined in their studies refers to the ratio of equivalent widths of the respective features.} (see Figure~7 therein), but it can also be physically motivated somewhat:
A substantial fraction of SNe~II are affected to varying degrees by \gls{csm} interaction.
\gls{csm} interaction can manifest itself in suppressed H$_\alpha$ emission lines and shallow absorption profiles, long phases of blue and featureless spectra, narrow emission lines during the early phases, and a boost in luminosity with a delayed onset of the recombination phase \citep{Hillier2019}.
The duration and strength of these effects are connected to the mass and spatial distribution of the \gls{csm}.
These effects can, in general, have a very complex influence on the observed appearance of a \gls{sneIIp} and ultimately affect the \gls{scm} if not corrected for.
\citet{Hillier2019} have investigated four of the supernovae in the calibrator sample (\sem, \et, \aw, \ej) and found that, at least for \ej{}, \gls{csm} interaction is necessary to explain the early \gls{sne} evolution. For the other \glspl{sne}, the evidence for \gls{csm} is less strong but small amounts of \gls{csm} still improve the agreement of the models with the data at very early times.

As far as our Hubble-flow sample is concerned, there are at least two objects showing signs of \gls{csm} interaction: \hnj{} and \xlr, which is indicated by their strongly suppressed H$_\alpha$ emission in the first few weeks after explosion following a blue featureless phase (see Figures \ref{fig:specxlr} and \ref{fig:spechnj}).
Excluding these \gls{csm} candidates does not improve the quality of the extended \gls{scm} itself.
In contrast, this exclusion brings the classical \gls{scm} to a level comparable to that of the extended \gls{scm} of the full sample (also see the ``No CSM cand.'' lines in Table~\ref{tab:h0vals}).
This can be seen as evidence that the $a/e$ term corrects for the negative effects of the \gls{csm} interaction on the \gls{scm} (e.g. \citealt{Hillier2019} argue that the \gls{csm} interaction influences the overall luminosity, i.e. the core quantity of the \gls{scm}).

An additional argument for this hypothesis is the fact that the exclusion of \gls{csm} candidates brings the most notable improvements during the early phases, as the \gls{csm} interaction is a phenomenon most dominant during early phases, at least for the currently favored scenario of a confined \gls{csm} \citep{Yaron2017a,Morozova2017a,Dessart2017a,Moriya2017a,Foerster2018a}.
Lastly, the fact that the $a/e$ correction has a negative sign (i.e. $\gamma$ is negative) agrees well with the expectation that \gls{csm} candidate \glspl{sne} (which should have a higher $a/e$) are brighter.
It should nonetheless be noted that the connection between the $a/e$ correction and \gls{csm} interaction (or in general any mechanism affecting the H$_\alpha$ feature) is provisional at this point, and requires more rigorous investigation.

The difference in the quality of the standardization between the classical and extended \gls{scm} is illustrated in Figure~\ref{fig:epoch_evolution} for various phases (see the  ``Classical SCM'' lines Table~\ref{tab:h0vals} as well). It becomes immediately apparent that the extended \gls{scm} achieves a lower MSE than the classical \gls{scm}, particularly at earlier phases.
We have also explored other combinations of correction terms (e.g. replacing the velocity correction by the $a/e$ correction), yet none proves as effective as the \gls{scm} containing a velocity, color, and $a/e$ correction.

It should be noted that although the introduction of an additional fit parameter will inevitably decrease the MSE, we observe that the improvement of the MSE is comparable to a model where we replace the color term by the $a/e$ correction.
In general, we find the color correction subdominant for our data sample improving the overall standardization only by a small margin for the phase ranges of interest.
Although \citet{Jaeger2020b} suggest that the color variety of \glspl{sneIIp} is mostly caused by intrinsic dispersion rather than host galaxy extinction, we cannot exclude that the color correction accounts for at least some amount of extinction.
A more extensive data sample is needed for further investigations into this issue.

The second goal of studying the $H_0$-free \gls{scm} is the determination of the optimal phase for standardization.
The works of other authors find the optimal phase for standardization to be around $43$ days \citep{Jaeger2020a} or $50$ days \citep{Poznanski2009}, which we also find in the case of the classical \gls{scm}.
However, the introduction of the extended \gls{scm} drastically widens this rather narrow band of optimal phases: We find that between the phases from $\sim 30$ to $\sim 50$ days, reasonably good MSE values can be achieved.
In our case, an additional factor is the limited number of \glspl{sne} available at later phases as we do not extrapolate the data far beyond the last observation.
As such, although the minimal MSE can be found at $41$ days, we choose
$30$ days as the optimal phase for standardization, where we can use the full
$18$ \glspl{sne} contained in our Hubble-flow sample (which is a $\sim 38\%$ increase in
the usable sample size, compared to a standardization at $41$ days with only $13$ \glspl{sne}).
It should be pointed out that for earlier phases the rather uncertain \glspl{toe} affect the \gls{scm} more than at later phases. As the H$_\beta$ velocities and $a/e$ values (on average) exhibit a larger slope at earlier phases, a large \gls{toe} uncertainty introduces a larger uncertainty in the interpolated value than at later, ``flatter'' phases, given an otherwise constant interpolation quality.
Hence we expect the standardization quality to improve even further at earlier phases for a sample with better known \glspl{toe}.
Not only is this shift in optimal phase (or rather the now much broader range of acceptable phases) convenient for the present data set, but it will also increase the number of usable \glspl{sneIIp} in general.

\section{$H_0$ \gls{scm} results}\label{sec:res_h0}
Following the results of the previous Section, we apply the \gls{scm} to the combined calibrator and Hubble-flow sample, i.e. a total of $30$ \glspl{sne}, at $30$ days after the explosion.
In this Section, \gls{scm} refers to the extended \gls{scm} unless stated otherwise.
We summarize all $H_0$ values discussed in this Section in Table~\ref{tab:h0vals}.

\subsection{The calibrator sample}\label{sec:calib_res}
\begin{figure}
    \centering
    \includegraphics[]{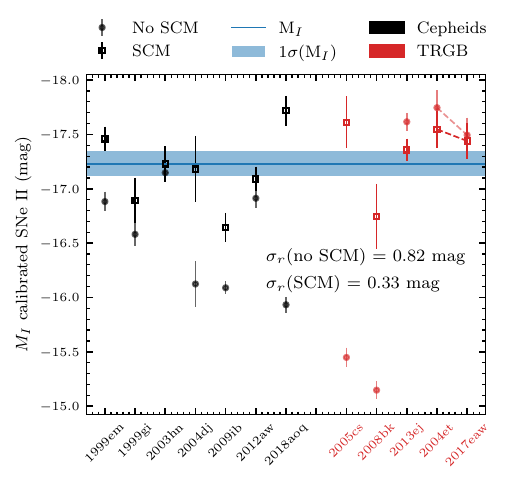}
    \caption{Standardized absolute magnitudes of the calibrator sample at $30$ days utilizing the $H_0$ \gls{scm}. The color coding in the label indicates the origin of the distance calibration, i.e. through Cepheid or \gls{trgb} distances. This is only for ease of identification; the \gls{scm} is unaware of this information. The largest difference between the \gls{scm} and no-\gls{scm} value is observed for \cs{} and \bk{}, due to them being low luminosity \gls{sneII} \citep{Pastorello2006}.
    \et{} and \eaw{} are connected by a dashed line to indicate their shared host galaxy.
    }
    \label{fig:calib_standard}
\end{figure}

Figure~\ref{fig:calib_standard} shows the standardized absolute magnitudes ($m^\mathrm{true}_{I,\mathrm{calib}} - \mu_\mathrm{calib}$) of the calibrator sample obtained from the $H_0$ \gls{scm} fit of the combined calibrator and Hubble-flow sample.
Whereas the unstandardized magnitudes show a scatter of $0.82$\,mag, the
application of the \gls{scm} reduces this value to $0.33$\,mag, demonstrating the effectiveness of the standardization.
Overall, we find good agreement with the results of \citet{Jaeger2022a}, particularly in the relative standardized magnitudes\footnote{We note that their Figure~1 is in a different photometric system that should be accounted for when comparing to our Figure~\ref{fig:calib_standard}.}. 
Looking at the sibling \glspl{sne} in NGC~6946 (i.e. \et{} and \eaw{}) we find that the difference between their standardized magnitudes is $\sim 0.5$\,mag in case of \citet{Jaeger2022a} and $\sim 0.1$\,mag in our case.
While it is hard to draw any deeper insights from this due to the different methodologies and the spectral flux corrections applied for \et{}, it is nonetheless a satisfying consistency check to see that \glspl{sne} in the same host galaxy yield roughly similar distances.

Lastly, we find that \ib{} shows the largest residual --- $0.59$\,mag.
Although this is consistent within the given uncertainties and the resultant unexplained dispersion $\sigma_\mathrm{int,Calib}$ (see Section~\ref{sec:res_main}), it nonetheless stands out.
Particularly since \citet{Jaeger2022a} find a smaller residual of around $0.4$\,mag (see their Figure~1).
We further explore this issue further in Section~\ref{sec:res_main}.

\subsection{The Hubble-Lemaître constant}\label{sec:res_main}
Applying the $H_0$ \gls{scm} to our combined analysis and calibrator sample yields a Hubble-Lemaître constant of
\begin{equation*}
    H_0 = 70.6^{+4.5}_{-4.3}\,\mathrm{km}\,\mathrm{s}^{-1}\,\mathrm{Mpc}^{-1},
\end{equation*}
where the uncertainties are purely statistical. These uncertainties correspond to a precision of $6.2$\%.
The MCMC results satisfy all the tests provided in the \textsc{cmdstan} diagnose tool, including the convergence criterion $\hat{R} < 1.05$ \citep{vehtari2021a}\footnote{The maximum $\hat{R}$ value of all chains is $1.00019$.}.

\begin{figure*}
    \sidecaption
    \includegraphics[width=12cm]{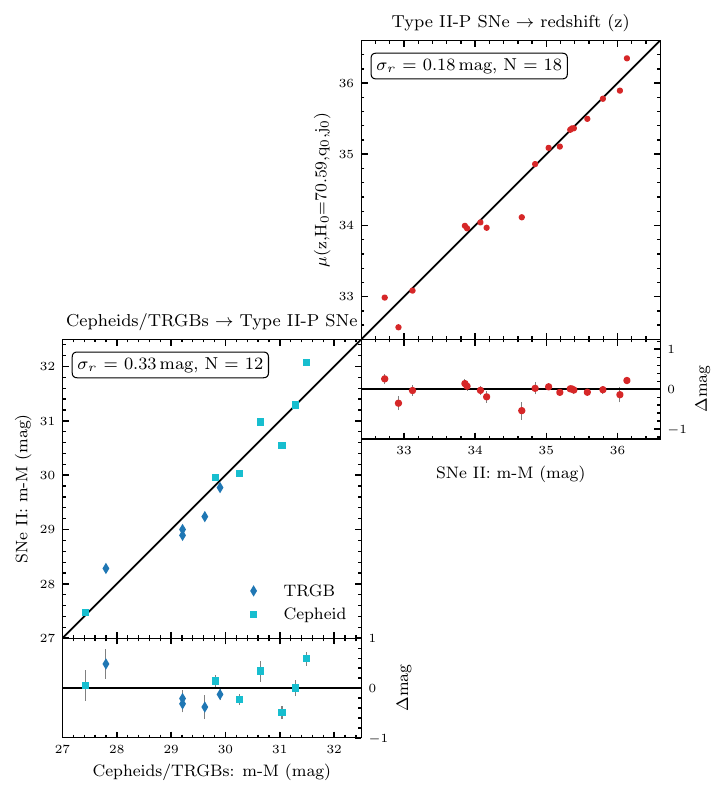}
    \caption{Two-rung Hubble diagram presented in the form used in \citet{Jaeger2022a} or \citet{Riess2022a}, illustrating the calibrator \glspl{sne} (blue, cyan) and the Hubble-flow sample (red). Here the different labels for the calibrator \glspl{sne} are only for convenience in identifying the distance calibration source; both subsets constitute the calibrator sample.}
    \label{fig:hubble_diag}
\end{figure*}

\begin{figure*}
    \centering
    \includegraphics[width=\linewidth]{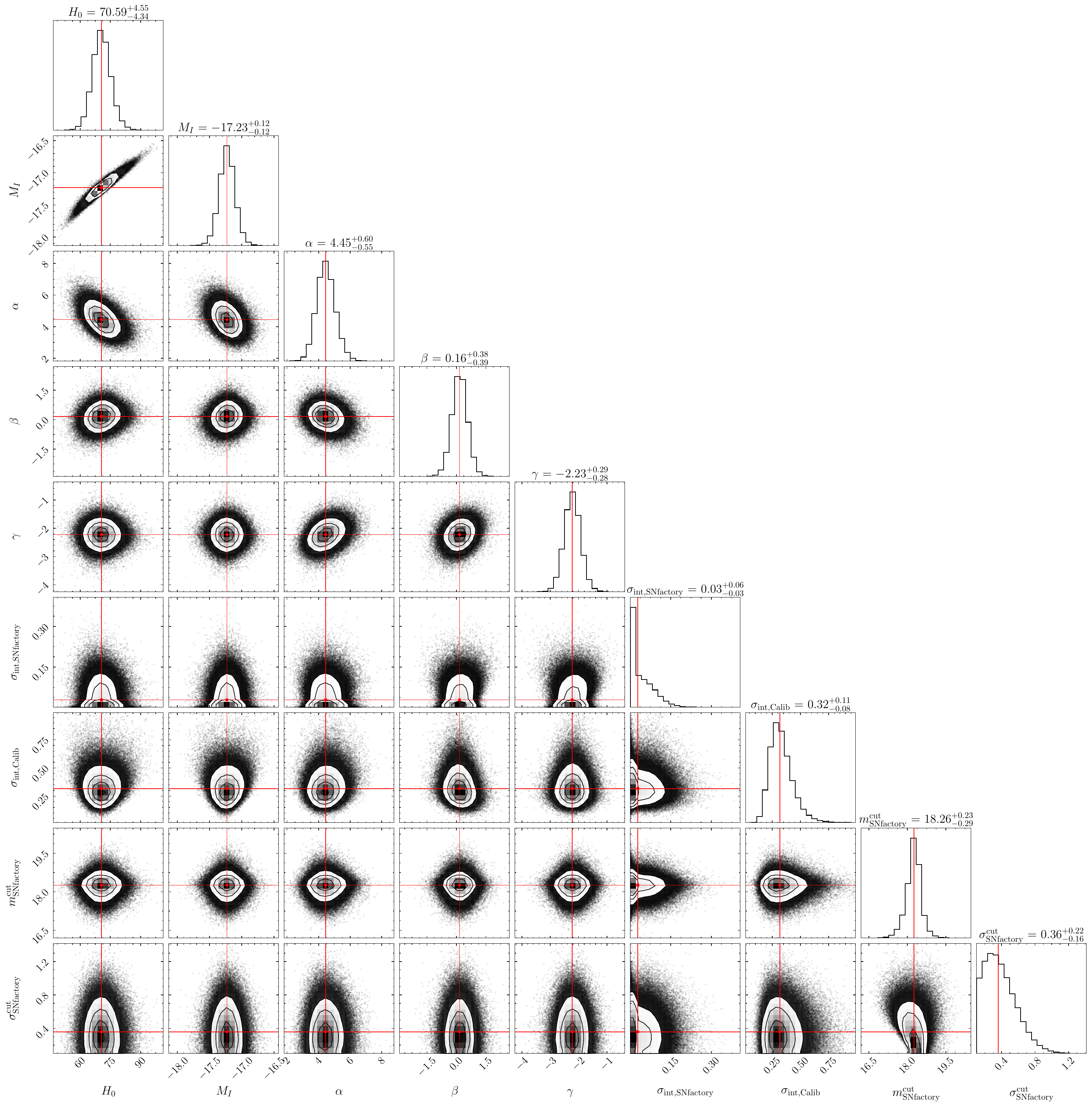}
    \caption{Corner plot of the posterior sample distributions of the main model coefficients obtained from the \gls{scm} at $30$ days corresponding to the Hubble diagram in Figure~\ref{fig:hubble_diag}. Here, the values in the titles of the individual plots refer to the $16$th, $50$th and $84$th percentile. The red lines indicate the $50$th percentile value.}
    \label{fig:corner}
\end{figure*}

The respective Hubble diagram and posterior distributions are illustrated in Figure~\ref{fig:hubble_diag} and \ref{fig:corner}, respectively.
In the Hubble diagram, both the distances for the calibrator and \gls{snfactory} \glspl{sne} can be seen, as well as the corresponding residuals. 
Looking at the \gls{snfactory} objects in the upper rung, we find that these standardize fairly well, with considerably smaller residuals compared to, e.g., \citet[Figure~3]{Jaeger2022a}. While they find an overall scatter of $\sigma = 0.28$\,mag, our residual scatter only amounts to $\sigma = 0.18$\,mag.

For the posterior distributions shown in Figure~\ref{fig:corner}, several things can be pointed out. Importantly, the posterior shapes of the model coefficients follow the expected shapes, as, e.g., found by \citet[Figure~2]{Jaeger2022a} as well.
Here, we find the value of $\beta=0.2\pm 0.4$ to be consistent with the statement made in the previous Section~\ref{sec:h0freescm}, i.e. that the color correction is a subdominant correction for our data sample.

The shape of the two $\sigma_\mathrm{int}$ posteriors stands out somewhat.
Here, the $\sigma_\mathrm{int,SNfactory}$ value of the Hubble-flow sample is indicative of the comparably good standardization.
We remind the reader that due to our choice of a log-uniform prior, the posterior will collapse towards zero once the model can explain the data using the given measurement uncertainties. Thus for the \gls{snfactory} sample only $0.03$\,mag of additional dispersion is required, leading to an overall dispersion of $0.18$\,mag.
In contrast, $\sigma_\mathrm{int,Calib}$ requires a large added dispersion relative to the uncertainties, leading to a  $\sigma_\mathrm{int,Calibrator} = 0.32^{+0.11}_{-0.08}$, essentially as large as the total dispersion of $\sigma_\mathrm{r}=0.33\,$mag (see Figures~\ref{fig:calib_standard}, \ref{fig:hubble_diag} and \ref{fig:corner}).

The survey depth posterior value of $m^\mathrm{cut}_\mathrm{SNfactory}=18.3^{+0.2}_{-0.3}$\,mag, when compared to the $m_I$ values in Table~\ref{sec:results}, suggests that there is no strong brightness selection effect contributing to the overall analysis.
We find the same value for $m^\mathrm{cut}_\mathrm{SNfactory}$ irrespective of the input value of $m^\mathrm{nominal}$; while we assume $m^\mathrm{nominal}_\mathrm{SNfactory}=18.5$\,mag, we obtain a nearly identical posterior values from input values of $16.5$ or $20.5$~mag.
The posterior distributions of the hyperparameters can be found in Figure~\ref{fig:corner_snfactory} in Appendix~\ref{sec:app_posteriors}.


Focusing next on the poorly-standardizing \ib{} (see Section~\ref{sec:calib_res} and Figure~\ref{fig:calib_standard}), we find that removing it yields $H_0 = 68.3\pm3.3$\,km\,s$^{-1}$\,Mpc$^{-1}$.
This is consistent with our fiducial result, but with an improved precision of $4.8\%$.
Of course this reduces the residual scatter of the calibrators in the Hubble diagram, to $\sigma = 0.29$\,mag, more in line with the total dispersion found by \citet{Jaeger2022a}.
While \ib{} formally has sufficient data available to constrain the quantities needed for the \gls{scm}, with only one spectrum at approximately $45$~days bridging the range from around $20$ to $80$~days after the explosion, the interpolation of $v_{\mathrm{H}\beta}$ and $a/e$ is highly dependent on this single spectrum (see Figure~\ref{fig:09ib_interp})\footnote{Manual inspection of its spectra and photometry did not reveal any peculiarities. Furthermore, experiments with an outlier model (see our comments on outlier modeling in Appendix~\ref{sec:app_model_expl}) did not reveal \ib{} to be an outlier within our framework.}.
We also note that \ib{} has been extensively studied due to its particularly long plateau phase \citep{Takats2015}, raising the question of whether it should be considered in the normal \gls{sneIIp} population.
Thus, while \ib\ may have some issue interfering with its standardization, we could not firmly identify any issue that demonstrates that it does not belong to the normal \gls{sneIIp} population.
Once one starts to dig deeper to find peculiarities, this question can be also raised for several other \glspl{sne} in the calibrator sample such as \cs{} or \bk{}; e.g., \citet{Pastorello2006} attribute to \cs{} some degree of peculiarity.
Only better data for more calibrator \glspl{sneIIp} can resolve the questions such as velocity interpolation accuracy requirements, plateau duration, etc.

\begin{table*}
	\centering
	\caption{Overview of various fit coefficients of the results discussed in Sections~\ref{sec:res_h0} and \ref{sec:djcomp_head}.}
	\label{tab:h0vals}
	\begingroup
	\setlength{\tabcolsep}{4pt}
	\renewcommand{\arraystretch}{1.5}
	\begin{tabular}{l|c|c|c|c|c|c|c|c}
		\hline\hline
	Sample name & $N_\mathrm{calib}$ & $N_\mathrm{HF}$ & $H_0$ (km\,s$^{-1}$\,Mpc$^{-1}$) & $\alpha$ & $\beta$ & $\gamma$ & $M_I$ (mag) & $\Delta H_0^\mathrm{fiducial}$\tablefootmark{a} \\    
        \hline
Fiducial & $12$ & $18$ & $70.6^{+4.5}_{-4.3}$ & $4.4^{+0.6}_{-0.6}$ & $0.2^{+0.4}_{-0.4}$ & $-2.2^{+0.3}_{-0.3}$ & $-17.23^{+0.12}_{-0.12}$ & $\ldots$ \\
Fiducial, non-hier. & $12$ & $18$ & $71.3^{+4.5}_{-4.2}$ & $4.3^{+0.6}_{-0.5}$ & $0.2^{+0.4}_{-0.4}$ & $-2.1^{+0.3}_{-0.3}$ & $-17.22^{+0.12}_{-0.11}$ & $0.2\sigma$ \\
No color correction & $12$ & $18$ & $71.0^{+4.6}_{-4.3}$ & $4.3^{+0.6}_{-0.5}$ & $\ldots$ & $-2.4^{+0.3}_{-0.3}$ & $-17.22^{+0.12}_{-0.12}$ & $0.1\sigma$ \\
TRGB calibration & $5$ & $18$ & $63.7^{+3.3}_{-3.2}$ & $5.1^{+0.6}_{-0.5}$ & $-0.2^{+0.4}_{-0.4}$ & $-2.2^{+0.3}_{-0.3}$ & $-17.60^{+0.10}_{-0.10}$ & $-1.6\sigma$ \\
Cepheid calibration & $7$ & $18$ & $74.9^{+6.0}_{-5.9}$ & $4.2^{+0.8}_{-0.7}$ & $0.2^{+0.4}_{-0.4}$ & $-2.2^{+0.3}_{-0.3}$ & $-17.20^{+0.16}_{-0.16}$ & $1.0\sigma$ \\
\hline
No CSM cand., ext. SCM, hier. & $12$ & $16$ & $71.0^{+4.7}_{-4.4}$ & $4.2^{+0.6}_{-0.6}$ & $0.1^{+0.4}_{-0.4}$ & $-2.6^{+0.4}_{-0.4}$ & $-17.16^{+0.12}_{-0.12}$ & $0.1\sigma$ \\
No CSM cand., class. SCM, hier. & $12$ & $16$ & $70.4^{+5.8}_{-5.5}$ & $5.4^{+0.9}_{-0.8}$ & $1.0^{+0.7}_{-0.7}$ & $\ldots$ & $-17.16^{+0.14}_{-0.14}$ & $-0.05\sigma$ \\
\hline
Classical SCM, hier. & $12$ & $18$ & $71.7^{+6.1}_{-6.0}$ & $5.5^{+0.9}_{-0.8}$ & $1.1^{+0.8}_{-0.8}$ & $\ldots$ & $-17.19^{+0.14}_{-0.15}$ & $0.3\sigma$ \\
Classical SCM, non-hier. (30 days)\tablefootmark{b} & $12$ & $18$ & $74.7^{+5.6}_{-5.2}$ & $5.0^{+0.7}_{-0.7}$ & $0.9^{+0.7}_{-0.7}$ & $\ldots$ & $-17.17^{+0.12}_{-0.12}$ & $0.9\sigma$ \\
Classical SCM, non-hier. (43 days)\tablefootmark{b,c} & $12$ & $13$ & $73.0^{+6.9}_{-6.1}$ & $3.8^{+0.8}_{-0.8}$ & $-0.5^{+0.5}_{-0.5}$ & $\ldots$ & $-17.10^{+0.13}_{-0.13}$ & $0.5\sigma$ \\
\hline
No low-$v_{\mathrm{H}\beta}$ Calibs. & $9$ & $18$ & $70.5^{+4.5}_{-4.1}$ & $5.3^{+1.0}_{-0.9}$ & $0.3^{+0.5}_{-0.4}$ & $-2.1^{+0.3}_{-0.3}$ & $-17.36^{+0.12}_{-0.11}$ & $-0.03\sigma$ \\
$v_{\mathrm{H}\beta}$ overlap only & $6$ & $9$ & $70.8^{+3.3}_{-3.4}$ & $6.0^{+2.7}_{-2.2}$ & $0.8^{+0.7}_{-0.7}$ & $-1.9^{+0.5}_{-0.5}$ & $-17.29^{+0.08}_{-0.09}$ & $0.04\sigma$ \\
$a/e$ overlap only & $9$ & $15$ & $72.1^{+4.5}_{-4.2}$ & $4.0^{+0.7}_{-0.7}$ & $0.4^{+0.4}_{-0.4}$ & $-2.6^{+0.4}_{-0.4}$ & $-17.25^{+0.12}_{-0.12}$ & $0.4\sigma$ \\
\hline
\multirow{2}{*}{Split populations\tablefootmark{d}} & \multirow{2}{*}{12} & \multirow{2}{*}{18} & \multirow{2}{*}{$69.5^{+4.6}_{-4.4}$} & $4.6^{+0.8}_{-0.7}$ & $0.2^{+0.4}_{-0.4}$ & $-2.2^{+0.3}_{-0.3}$ & \multirow{2}{*}{$-17.25^{+0.13}_{-0.13}$} & \multirow{2}{*}{$-0.2\sigma$} \\
 & & & & $-0.4^{+0.8}_{-0.8}$ & $0.4^{+0.8}_{-0.8}$ & $-0.2^{+0.9}_{-0.8}$ & &  \\
\multirow{2}{*}{No CSM cand., split populations\tablefootmark{d}} & \multirow{2}{*}{12} & \multirow{2}{*}{16} & \multirow{2}{*}{$70.1^{+4.7}_{-4.5}$} & $4.3^{+0.8}_{-0.8}$ & $0.1^{+0.4}_{-0.4}$ & $-2.5^{+0.5}_{-0.4}$ & \multirow{2}{*}{$-17.17^{+0.12}_{-0.12}$} & \multirow{2}{*}{$-0.1\sigma$} \\
 & & & & $-0.3^{+0.8}_{-0.8}$ & $0.4^{+0.8}_{-0.8}$ & $-0.2^{+0.9}_{-0.9}$ & &  \\
        \hline
	\end{tabular}
	\endgroup
\tablefoot{
\tablefoottext{a}{This value refers to the statistical tension between our fiducial $H_0$ value and the respective variation. Since variants are correlated, we compare to the appropriate positive or negative uncertainty on our fiducial value.
}
\tablefoottext{b}{Here, only one global $\sigma_\mathrm{int}$ value was used (as in the model of \citealt{Jaeger2022a}) instead of a separate value for calibrator and Hubble-flow sample. This introduces only a small shift subdominant to our analysis. Note that the value of $M_I$ changes whenever the sample changes, since it is defined relative to the means of the sample independent variables.}
\tablefoottext{c}{For this value, only $13$ \glspl{sne} are used in the Hubble-flow sample, as illustrated in Figure~\ref{fig:epoch_evolution}}.
\tablefoottext{d}{Here, the second row of the $\alpha$, $\beta$, and $\gamma$ parameter refers to the value of $\Delta\alpha_\mathrm{Calib}$, $\Delta\beta_\mathrm{Calib}$, and $\Delta\gamma_\mathrm{Calib}$, i.e., the split population parameters as outlined in Appendix~\ref{sec:app_h0scm}.}
}
\end{table*}

\subsection{Bootstrap resampling}\label{sec:bootstrap}
\begin{figure}
    \centering
    \includegraphics[]{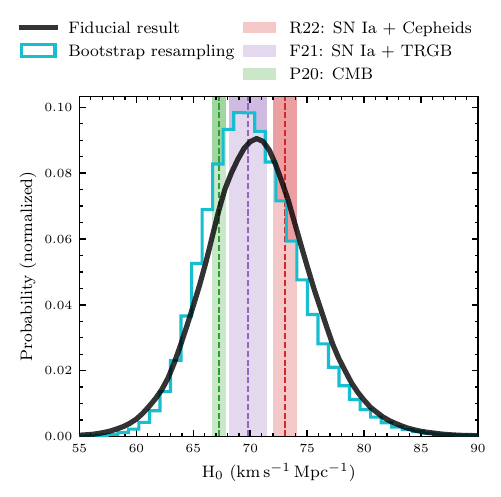}
    \caption{Posterior distributions of our fiducial $H_0$ measurement (black), and bootstrap resampling of the calibrator sample (cyan). For comparison, the $H_0$ value of \citet{Riess2022a} (R22), \citet{Freedman2021a} (F21) and \citet{Collaboration2018} (P20) are plotted as well. The range of shown $H_0$ values has been adjusted for better visibility, only some very infrequent values can be found outside this range.}
    \label{fig:bootstrap}
\end{figure}
Given the limited calibrator sample size and the fact that several other calibrator SNe (such as, e.g., \fourdj{}) besides those mentioned already may face similar questions, 
we performed a bootstrap resampling with replacement analysis. For our combined calibrator sample of $12$ \glspl{sne}, this results in $1\,352\,078$ possible combinations. The resulting $H_0$ values are illustrated in Figure~\ref{fig:bootstrap}, as well as the posterior distribution of our fiducial $H_0$ value from Section~\ref{sec:res_h0}. 

Overall the spread in possible $H_0$ values through variations of the calibrators, $\sigma = 4.4$\,km\,s$^{-1}$\,Mpc$^{-1}$, is captured rather well by the uncertainties given in the previous section (see Figure~\ref{fig:bootstrap}). 
This reinforces the evidence that our analysis is robust despite the outlying \ib{} or modest inconsistencies between the calibrator and Hubble-flow samples (see Section~\ref{sec:djcomp_head}).

\subsection{Phase dependence}
\begin{figure}
    \centering
    \includegraphics[]{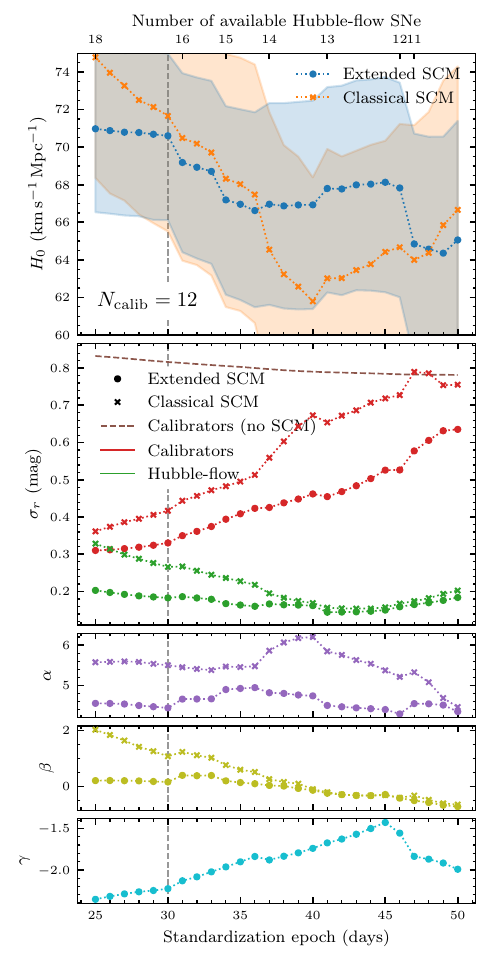}
    \caption{Illustration of the impact of the standardization on the overall $H_0$ value, the residual scatter in the Hubble diagram $\sigma_\mathrm{r}$, and the standardization coefficients $\alpha$, $\beta$, and $\gamma$. Here we also show the results using the classical \gls{scm}, indicated by the x-markers. It should be noted that after $45$ days, Hubble-flow sample size effects most likely become a driving factor behind some of the changes and thus all values after this point should be taken with care. The size of the calibrator sample is constant ($N_\mathrm{calib} = 12$ for all phases shown here).}
    \label{fig:h0_epoch_evo}
\end{figure}
As discussed in Section~\ref{sec:h0freescm}, we standardize at $30$ days after the explosion, in contrast to other authors applying the \gls{scm} at around $43$ to $50$ days.
Here, we examine the impact of this choice on the $H_0$ \gls{scm}, as illustrated in Figure~\ref{fig:h0_epoch_evo}.

Most importantly, we find that the resultant $H_0$ values are all consistent. Only considering phases where we retain at least $2/3$ of our fiducial sample size, i.e. up to and including $46$ days\footnote{After that point, we no longer fully trust the resultant $H_0$ values due to issues in the sampling procedure.}, we find a peak-to-peak variation that is $0.5\times$ the size of our reported uncertainty. The dispersion on $H_0$ across these phases is $1.5$\,km\,s$^{-1}$\,Mpc$^{-1}$ --- less than half of our reported statistical uncertainty.
When only considering phases with the full \gls{snfactory} sample. i.e. up to $30$~days, the dispersion in $H_0$ with phase is only $0.1$\,km\,s$^{-1}$\,Mpc$^{-1}$.
Naturally, we find that the statistical precision rapidly deteriorates with a reduction in the number of Hubble-flow \glspl{sne}.

Looking at the results of the classical \gls{scm}, we find, as in Section~\ref{sec:h0freescm}, a rather substantial phase dependence.
In particular at early phases --- where the size of the Hubble-flow sample is constant --- the extended \gls{scm} is almost constant in $H_0$, whereas the classical \gls{scm} changes rapidly with a change in standardization phase.
In general, the results of the classical \gls{scm} have a peak-to-peak variation for the same phases as above that is $1.1\times$ the width of the uncertainty on our classical \gls{scm} result, with a dispersion on $H_0$ of $4.2$\,km\,s$^{-1}$\,Mpc$^{-1}$ (for the phases with the full \gls{snfactory} sample the dispersion on $H_0$ is $1.1$\,km\,s$^{-1}$\,Mpc$^{-1}$), whereby this nominal small tension is most likely driven by a comparatively lower precision in $H_0$.

Moving on to the residual scatter in the Hubble diagram, $\sigma_\mathrm{r}$, we find that in the case of the Hubble-flow sample, the evolution of this value resembles the one expected from the MSE evolution in the $H_0$-free \gls{scm} (see Figure~\ref{fig:epoch_evolution}), with a relatively low dependence on the chosen phase and a nominal minimum at around $41$ days.
In contrast, the classical \gls{scm} shows a much more pronounced variation with phase, particularly at earlier phases, and an overall slightly higher scatter.
Surprisingly, the residual scatter $\sigma_\mathrm{r}$ of the calibrators exhibits a significantly different structure, with its minimum at $25$ days and an almost monotonic increase in scatter with later phases.
At later phases it becomes only marginally better than the unstandardized \glspl{sne}, especially in case of the classical \gls{scm}.
This difference could potentially be caused by an underlying time-dependent difference between our Hubble-flow and calibrator sample and should be re-examined in the future with larger Hubble-flow and calibrator data sets having quality comparable to our Hubble-flow data set\footnote{We note that this is indeed a property of the calibrator SNe, and not the hierarchical approach. Applying the classical, non-hierarchical \gls{scm}, i.e. the same model as \citet{Jaeger2022a} shows a similar trend.}.
Nonetheless, the extended \gls{scm} shows an average improvement of around $0.1$\,mag compared to its classical counterpart.

Lastly, examining the standardization coefficients $\alpha$, $\beta$, and $\gamma$, we find that they are strongly phase dependent, particularly the latter two.
For the case of $\beta$ we see that, at early phases, it has a far more substantial contribution in the classical \gls{scm} than in the extended \gls{scm}.
At later phases, this difference vanishes.
Here, the correction also switches signs, most likely related to the general reddening of \glspl{sne} at these times.
Connected to this, it can be seen that the $a/e$ correction becomes less important at later phases, presumably due to the diminishing contribution of \gls{csm} interaction.
After $45$ days this trend in $\gamma$ reversed. 
However, the decreasing Hubble-flow sample size (the calibrator sample size stays constant) becomes a potentially significant factor driving this change.

\subsection{Further sample and model variants}\label{sec:variants}

Having explored the influence of specific \glspl{sne} and phase dependence, this section explores the broader question of model and sample variants that is motivated in part by the observed population distribution functions for the standardization characteristics as discussed in Section~\ref{sec:consistency}.
All results of these additional variants are also given in Table~\ref{tab:h0vals}.

As already mentioned in Section~\ref{sec:res_main} we find the color correction to be rather weak.
Indeed, leaving out the color correction yields an $H_0$ value of $71.0^{+4.6}_{-4.3}$\,km\,s$^{-1}$\,Mpc$^{-1}$ which is consistent with our fiducial result above.
Similarly, removing the two \gls{csm} candidates, \xlr{} and \hnj{}, from the Hubble-flow sample does not significantly change the result giving $H_0 = 71.0^{+4.7}_{-4.4}$\,km\,s$^{-1}$\,Mpc$^{-1}$.
If we additionally leave out the $a/e$ correction in this case, i.e. apply the classical \gls{scm}, we obtain $H_0 = 70.4^{+5.8}_{-5.5}$\,km\,s$^{-1}$\,Mpc$^{-1}$, which agrees well with our fiducial result.
However, the reduced precision of $8.0\,\%$ over the model including the $a/e$ correction suggests that $a/e$ correction is still useful even in the absence of clear \gls{csm}, possibly due to a more subtle \gls{csm} contamination of the remaining objects.

Next, we investigate the impact of the low-$v_{\mathrm{H}\beta}$ calibrators (i.e. $v_{\mathrm{H}\beta}$ significantly below\footnote{We still include \fourdj{} as it is not significantly below this threshold considering its \vhb{} uncertainty. Whether we include or exclude this \gls{sne} does not significantly impact the resulting $H_0$.} $5\,000$\,km\,s$^{-1}$), underrepresented in our and the \cite{Jaeger2022a} Hubble-flow samples (see Section~\ref{sec:djcomp}).
Without these three \glspl{sne} in our calibrator sample we find $H_0 = 70.5^{+4.5}_{-4.1}$\,km\,s$^{-1}$\,Mpc$^{-1}$.
Similarly, when removing the 15 \glspl{sne} outside the mutual overlap in \vhb\, $H_0$ increases by only $0.2$\,km\,s$^{-1}$\,Mpc$^{-1}$.
For this variant the uncertainty in $H_0$ decreases even though the sample size is reduced substantially, due to a reduction in the intrinsic scatter $\sigma_\mathrm{int}$ for the calibrators.
In should be noted that this variant excludes several peculiar calibrators (e.g., \ib{}) further reducing the uncertainty in $H_0$, see the discussion at the end of Section~\ref{sec:res_main}.
These results suggest that our results are not dominated by \glspl{sne} on either end of the velocity range, despite the differing \vhb\ distributions, as the results are consistent with the ones obtained from the full calibrator sample.

In a similar vein, we investigate restricting the sample to \glspl{sne} that overlap in $a/e$, finding this variant to be slightly more discrepant.
Here six \glspl{sne} are removed and $H_0$ shifts by $+1.5$\,km\,s$^{-1}$\,Mpc$^{-1}$ compared to our fiducial result.
Notably, the strength of the $\gamma$ correction is now comparable to the case where we removed the \gls{csm} candidates.
Even so, this is only a $0.4\,\sigma$ shift.

Moreover, we consider different approaches to standardizing based on the standardization coefficients.
An extreme would be to assume that differences in the standardization characteristic population distributions identified in Section~\ref{sec:consistency} imply that the standardization coefficients for the calibrator and Hubble-flow sample are also different.
To model this ``split populations'' alternative, we assign individual standardization coefficients ($\alpha$, $\beta$, and $\gamma$) latent parameters ($v^*$, $R^v$, etc.) to the calibrator versus Hubble-flow \glspl{sne}.
(See Appendix~\ref{sec:app_h0scm} for implementation details).
The resulting $H_0$ differs from our fiducial result by only $0.2\,\sigma$ (see Table~\ref{tab:h0vals}).
When removing the \gls{csm} candidates as a further alternative, the difference is reduced to $0.1\,\sigma$. 
Consequently, the mismatch between the standardization characteristics of the two data sets does not negatively affect the resulting $H_0$, within our statistical uncertainties.

Since the latent population standardization characteristic distribution functions impact Malmquist bias correction\footnote{While our hierarchical Bayesian model never makes such an explicit correction, it is implicitly made in any survey-selection correction procedure.}, we examine the effect of turning it off.
One key thing to note there is that Malmquist bias for standardized candles goes as $\sigma_M \sim \sigma_r^2/\sigma_{\text{raw}}$ \citep{malmquist1922, rubin2023a}, where $\sigma_r$ is the Hubble diagram scatter after standardization and $\sigma_{\text{raw}}$ is the scatter before standardization.
Our Hubble-flow sample has $\sigma_r= 0.18$~mag, and taking the value $\sigma_{\text{raw}}=0.82$~mag, as for our the calibrator sample under the assumption it represents the true unstandardized dispersion, yields a value of $\sigma_M =0.04$~mag.
That is, another result of our better standardization is that our Hubble-flow sample should require a smaller Malmquist bias correction.
Indeed, without selection included gives an $H_0$ value only $1$\,km\,s$^{-1}$\,Mpc$^{-1}$ higher.

Following, e.g., \citet{Jaeger2022a}, we also calculate $H_0$ values using only calibrators with a Cepheid or \gls{trgb} distance
\footnote{As mentioned in Section~\ref{sec:data_calib}, several calibrator \gls{sne} host galaxies have multiple distance measurements using various methods, but we simply adopt the distances (and therefore calibration source) of \citet{Jaeger2020b}}.
Using only the \gls{trgb} calibration, we find $H_0 = 63.7^{+3.3}_{-3.2}$\,km\,s$^{-1}$\,Mpc$^{-1}$.
In contrast, using the Cepheid calibration, we obtain $H_0 = 74.9^{+6.0}_{-5.9}$\,km\,s$^{-1}$\,Mpc$^{-1}$.
Our results in this respect are similar to those of \citet{Jaeger2022a}, who also find that the \gls{trgb} result is shifted towards lower, and the Cepheid result towards higher values of $H_0$ compared to the combined result.
Whether this difference is dominated by the actual distance measurements, higher sensitivity to peculiar calibrator \glspl{sne} due to the resulting reduction in sample sizes, or the difference in sample characteristics, is unclear; an answer awaits larger and more homogenous \glspl{sne} samples.

In some published $H_0$ analyses all of the presented variants are used to estimate something called a modeling or systematic error.
However, this approach has two statistical pitfalls.
First, when the sample members change, simply reporting the RMS of the differences is statistically unsound because they are highly correlated.
Second, when model parameterizations or prior parameter settings are changed in a manner that is not nested, the spread of their results cannot be interpreted as a formal statistical uncertainty.
If we were to adopt this approach, considering all of the Table~\ref{tab:h0vals} values for hierarchical models we would find a value of $\Delta H_0 = 1.4\,\mathrm{km}\,\mathrm{s}^{-1}\,\mathrm{Mpc}^{-1}$. Adding in the the non-hierarchical results gives $\Delta H_0 = 2.5\,\mathrm{km}\,\mathrm{s}^{-1}\,\mathrm{Mpc}^{-1}$.
However, we argue that the hierarchical approach is {\it a priori} the more realistic model and therefore the non-hierarchical approach does not constitute an alternative model.
Regardless, these scatters are well below the statistical variance of our fiducial result.
As they likely partially double-count statistical uncertainties, instead, we recommend consulting Table~\ref{tab:h0vals} directly for questions about variants.

\section{Literature comparison}\label{sec:djcomp_head}

Here, we compare our result with previous SCM findings, and then with other prominent late-time measurements of $H_0$.

\subsection{Comparison to \citet{Jaeger2022a}}\label{sec:djcomp}

In this Section, we take the study of \citet{Jaeger2022a} as the exemplar of the cumulative studies of the SCM to date.
After having replacing the Hubble-flow sample, extended the SCM, standardizing at a different phase, and using a hierarchical Bayesian model in place of a frequentist model, our $H_0$value of $70.6^{+4.5}_{-4.3}$\,km\,s$^{-1}$\,Mpc$^{-1}$ differs from the $75.4^{+3.8}_{-3.7}$\,km\,s$^{-1}$\,Mpc$^{-1}$ found by \citet{Jaeger2022a} by only $0.8\,\sigma$. 

It might seem surprising that our uncertainties on $H_0$ are only $17\%$ larger given that our Hubble-flow sample has only $18$ \glspl{sneIIp} compared to the $89$ in \cite{Jaeger2022a}.
The total error in column $\sigma_{\mathrm{tot}}$ of our Table~\ref{tab:scmdata} implies an error of $0.021$\,mag on the mean due to measurement uncertainties alone, or $0.042$\,mag with the $0.18$\,mag of Hubble diagram dispersion we find. By comparison, the measurement error in Table~D1 of \cite{Jaeger2020a} indicates an error on the mean of $0.019$\,mag from measurement uncertainties alone, or $0.030$\,mag when using their $0.28$\,mag Hubble diagram dispersion. Thus, the higher precision of our spectroscopic and photometric characteristics for the Hubble-flow \glspl{sne} results in similar statistical weight for our much smaller sample, and this conclusion holds up even after standardization dispersion dilutes the statistical weight of the per-\gls{sne} measurements.

The modest $H_0$ difference with \citet{Jaeger2022a} could arise from the different Hubble-flow data, the different \gls{scm} models and methodologies, or both. If we were to apply the classical, non-hierarchical \gls{scm} model using only one global $\sigma_\mathrm{int}$\footnote{Using a separate $\sigma_\mathrm{int}$ for calibrator and Hubble-flow sample mostly impacts only the uncertainties, giving $H_0 = 74.7^{+6.4}_{-5.8}$\,km\,s$^{-1}$\,Mpc$^{-1}$} (i.e. the model used by \citealt{Jaeger2022a}) to our sample, we would obtain an $H_0$ of $74.7^{+5.6}_{-5.2}$\,km\,s$^{-1}$\,Mpc$^{-1}$.

We emphasize that this value applies the \gls{scm} at $30$ days after the explosion, a phase where the quality of the standardization is found to rapidly deteriorate for the classical variant (see Figure~\ref{fig:epoch_evolution}).
Using the same phase as \citet{Jaeger2022a}, $43$ days, we obtain $H_0 = 73.0^{+6.9}_{-6.1}$\,km\,s$^{-1}$\,Mpc$^{-1}$.
Here, the loss in precision is at least partially driven by the loss of a few SNe from the Hubble-flow sample that did not have coverage out to $+43$~days.\footnote{Unlike the de~Jaeger analyses, we do not extrapolate in phase for our main analysis, but only for the velocity distribution tests discussed below. The only exception is \fourdj, in which case we extrapolate by around $10$ to obtain measurements at $43$~days.} This result indicates that our data are reasonably consistent with the results of other works.
In contrast, the classical, hierarchical \gls{scm} (at $30$ days) yields $H_0 = 71.7^{+6.1}_{-6.0}$\,km\,s$^{-1}$\,Mpc$^{-1}$.

Of the standardization terms, \vhb\ has the largest impact on unstandardized magnitudes --- against which flux selection is operative.\footnote{The brightness range induced by \vhb\ is 1.4~mag larger for the calibrators, and that from $a/e$ is 0.7~mag smaller.} With \vhb\ values in hand for a phase of $43$~days\footnote{Extrapolation of several days was required in order to estimate \vhb\ at a phase of $43$~days for $5$ of the SNfactory Hubble-flow \glspl{sne}.}, we next perform a K-S test\footnote{In order to compare with the de Jaeger sample tables, here we switch to the frequentist K-S test} in order to see whether the $H_0$ difference might come from disagreement between the distribution of our \vhb\ values and those used by \citet{Jaeger2022a} for our different Hubble-flow samples. We find agreement --- a \mbox{K-S} probability of $0.53$ after allowing for a small shift of $2.4$\,\AA\ derived via comparison of the different velocity fitting methods (Section~\ref{sec:spec_fit}).
Thus, the \vhb\ distributions --- at least at $43$~days --- are consistent between the two Hubble-flow samples.

As noted in Section~\ref{sec:consistency}, SNfactory Hubble-flow sample \vhb\ values are statistically inconsistent with the calibrator \vhb\ distribution.
Interestingly, the \citet{Jaeger2022a} \vhb\ distributions are not very consistent either, with a K-S score of 0.018 between their calibrator and Hubble-flow sample.
This slightly better agreement is made possible primarily by 19 low-\vhb\ \glspl{sneIIp} in their sample. Scaling by the relative sizes of the Hubble-flow samples, we would expect 3.8 such \glspl{sne} in our sample. 
The Poisson probability of observing none is $2.3\%$. However, since this comparison was made post hoc, we cannot rule out the possibility of a Poisson fluctuation.

When removing low-\vhb\ \gls{sneIIp} in Section~\ref{sec:variants} we found that $H_0$ increased by less than $0.1$\,km\,s$^{-1}$\,Mpc$^{-1}$, in the direction of the \citet{Jaeger2022a} value. Thus, the discrepancy between the \vhb\ distributions between our Hubble-flow and calibrator samples, while real, does not appear to explain the difference in $H_0$ between this work and that of \citet{Jaeger2022a}.

One might also wonder whether the selection effect differences between the two samples are coming into play in some other way.
Following the discussion of Section~\ref{sec:variants}, we estimate the Malmquist bias correction for the \citet{Jaeger2022a} sample to be $\sigma_M =0.10$~mag compared to our 0.04~mag.
Thus, there is less room for bias on $H_0$ due to our Hubble-flow sample because it should have less Malmquist bias.

Further, one might wonder whether our smaller $\sigma_r$ is caused by selection bias, after all, the unstandardized dispersion of our Hubble-flow sample is 0.40~mag, about half that for the calibrator sample, and thus similar to the ratio for their standardized dispersions. The unstandardized dispersion for the \citet{Jaeger2022a} Hubble flow sample is 0.55~mag, so here too the standardized dispersion scales vary roughly in proportion to the unstandardized dispersion.
Functionally, selection like that considered in Malmquist bias does not necessarily lead to a smaller standardized dispersion. Rather, selection might be eliminating \gls{sneIIp} that the SCM does not standardize well.
In this context it is important to note that \gls{sneIIp} at the edges of the \vhb, $col$, $a/e$ distributions have larger error bars after standardization due to the lever-arm (Eq.~\ref{eq:h0free_mag}; Fig.~\ref{fig:calib_standard}) by which their measurement errors must be multiplied.
These larger statistical uncertainties may make it harder to detect breakdowns in the simple linear relations assumed for the empirical standard candle standardization method. We can perform a limited test of this idea by removing the low-\vhb\ calibrators; while reducing the unstandardized dispersion of the calibrator sample from 0.82~mag to 0.52~mag --- a factor of 1.6$\times$ reduction --- the standardized dispersion is reduced to 0.29~mag --- only a 1.1$\times$ reduction.
Thus, any selection against such low-\vhb\ \gls{sneIIp} that may affect our Hubble-flow sample does not alone explain our small dispersion of 0.18~mag.
Interestingly, removing the normal-\vhb\ but outlying \ib\ {\it in addition} to the low-\vhb\ \gls{sneIIp} does reduce the calibrator unstandardized dispersion to 0.40~mag and the standardized dispersion to 0.20~mag; these are similar to the same values we find for our Hubble-flow sample in this scenario.
However, Figure~\ref{fig:epoch_evolution} shows that for classical SCM our sample has a scatter much like that of \citet{Jaeger2022a}. Thus,  even if an artifact of selection is removing some \glspl{sneIIp} that the SCM standardizes poorly, the improved dispersion with the addition of the $a/e$ parameter is not explained by selection.

Thus, while selection may play a role, the comparisons in this Section suggest that the difference in the $H_0$ values using \gls{scm} is predominantly caused by the different models rather than the data, particularly the introduction of the $a/e$ correction and the hierarchical approach.
If we apply the $a/e$ correction in a non-hierarchical approach, we obtain $H_0 = 71.3^{+4.5}_{-4.2}$\,km\,s$^{-1}$\,Mpc$^{-1}$, indicating a small, but noticeable impact from the change to a hierarchical model. This change is only 0.1\,$\sigma$ and is not surprising given that the hierarchical and non-hierarchical approaches weight the data --- especially the comparatively noisier measurements for the calibrators --- slightly differently.

\subsection{Hubble tension}\label{sec:h0tension}
\begin{figure*}
    \sidecaption
    % \centering
    % \includegraphics[height=0.9\textheight]{figures/H0Whisker.pdf}
    \includegraphics[width=12cm,keepaspectratio]{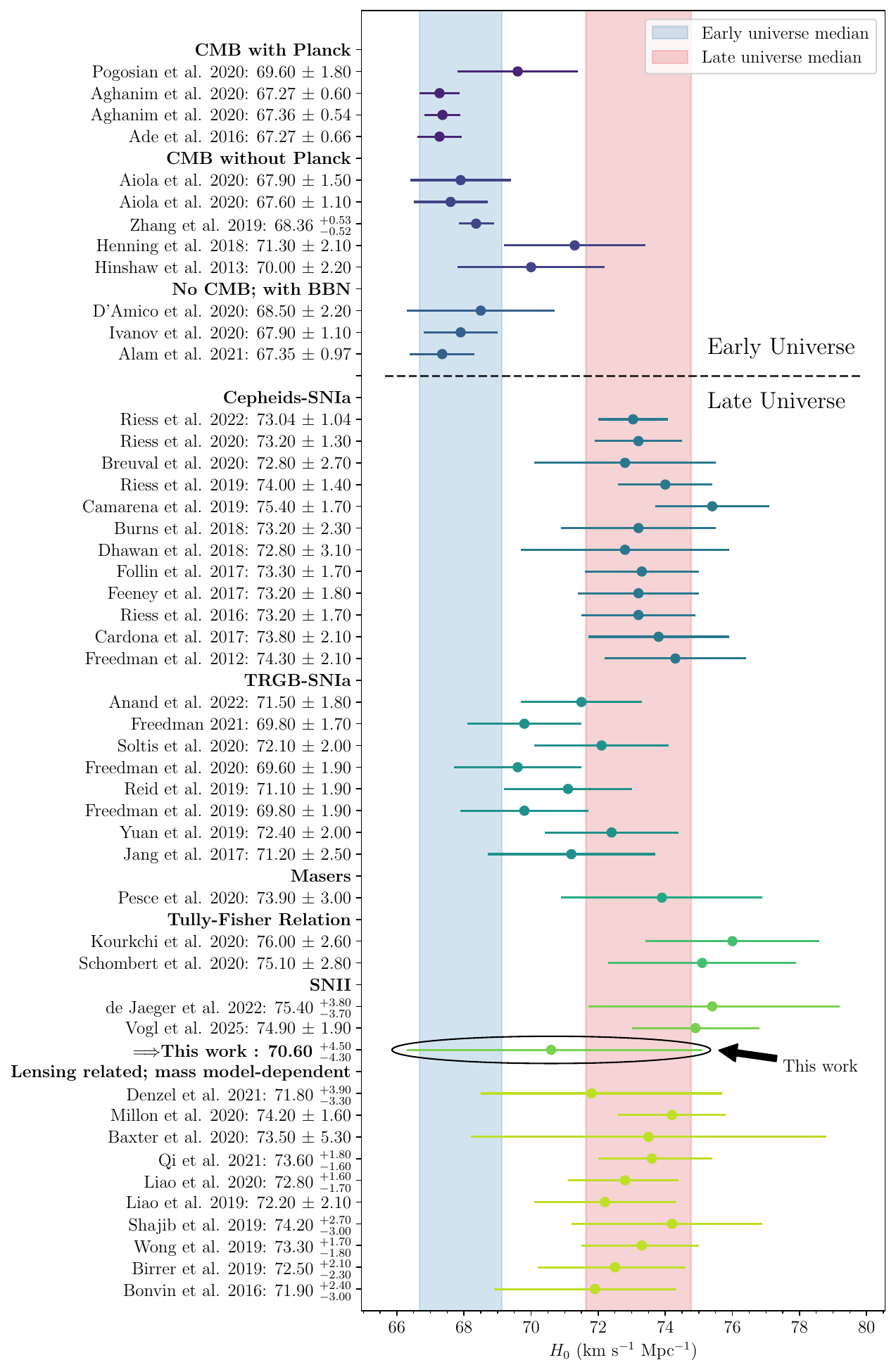}
    \caption{The $H_0$ obtained in this work in the context of several other $H_0$ values, both from early and late universe techniques. It should be noted that many of these studies are highly correlated with each other (e.g., by using similar underlying data). The shaded vertical regions represent the standard deviation around the median of the displayed early universe (blue) and late universe (red) methods. Here, the red shaded region takes the value obtained in this work into account as well, shifting the mean by $0.07$\,km\,s$^{-1}$\,Mpc$^{-1}$. This illustration was created with an adapted selection of the data compiled by \citet{diValentino2021}, taken from \url{https://github.com/lucavisinelli/H0TensionRealm}.}
    \label{fig:whisker}
\end{figure*}
\nocite{Pogosian2020,Collaboration2018,Collaboration2015,Aiola2020,Zhang2018,Henning2017,Hinshaw2012,DAmico2019,Ivanov2019,Alam2021a}
\nocite{Riess2022a,Riess2020,Breuval2020,Riess2019,Camarena2019,Burns2018,DhawanSuhail2018,Follin2017,Feeney2017,Riess2016,Cardona2016,Freedman2012}
\nocite{Anand2022a,Freedman2021a,Soltis2020,Freedman2020,Reid2019,Freedman2019,Yuan2019,Jang2017}
\nocite{Pesce2020,Kourkchi2020,Schombert2020,Denzel2020,Millon2019,Baxter2020,Qi2020,Liao2020,Liao2019,Shajib2019,Wong2019,Birrer2018,Bonvin2017}
In this section, we compare our resulting Hubble-Lemaître constant with the $H_0$ obtained by other studies; Figure ~\ref{fig:whisker} presents a graphical representation based on $H_0$ measurements compiled by \citet{diValentino2021}.
We find the best agreement with the \gls{trgb}-calibrated \gls{sneIa} result of \citet{Freedman2021a}, $H_0 = 69.8\pm1.7$\,km\,s$^{-1}$\,Mpc$^{-1}$, with only a $0.17\,\sigma$ difference.
Similarly, we only find a $0.19\,\sigma$ tension with the \gls{trgb} measurement of \citet[][$H_0 = 71.5\pm1.8$\,km\,s$^{-1}$\,Mpc$^{-1}$]{Anand2022a}.
The Planck \gls{cmb} result of \citet[][$H_0 = 67.27\pm0.60$\,km\,s$^{-1}$\,Mpc$^{-1}$]{Collaboration2018}, shows a similarly small difference of only $0.8\,\sigma$.
This is noteworthy insofar as a considerable fraction of recent late universe measurements show substantial tension with early universe measurements.
Our results, similar to \gls{trgb}-calibrated \gls{sneIa} distance measurements, do not exhibit such a discrepancy.
Looking at the (at the time of writing) most precise late universe measurement of \citet{Riess2022a}, $H_0=73.04\pm1.04$\,km\,s$^{-1}$\,Mpc$^{-1}$, we see a $0.5\,\sigma$ disagreement, which is not a statistically significant tension.
Lastly, considering other measurements using \glspl{sneII}\footnote{We do not consider results such as \citet{Schmidt1994} ($H_0=73\pm13$\,km\,s$^{-1}$\,Mpc$^{-1}$, \gls{epm}), \citet{OlivaresE.2010} ($H_0=69\pm16$\,km\,s$^{-1}$\,Mpc$^{-1}$; \gls{scm}) or \citet{Rodriguez2019} ($H_0\approx71\pm8$\,km\,s$^{-1}$\,Mpc$^{-1}$; \gls{pmm}) as their lack of, precision limits their value here.} we arrive at a $0.9\,\sigma$ tension with the result of \citet{vogl2024a}, which uses the tailored \gls{epm} --- independent of any \gls{trgb} or Cepheid distance calibration. 

While this good agreement is in part due to our larger statistical uncertainty compared to, e.g., SNe~Ia, the framework we have introduced (both the extended \gls{scm} and the
hierarchical Bayesian approach) shows such a substantially reduced scatter --- $\sigma_r = 0.18$\,mag --- for the Hubble-flow sample that future samples measured as well could be quite competitive.

\section{Conclusion}\label{sec:conclusion}
In this work, we present a unique \gls{sneIIp} data sample consisting of $21$ \glspl{sne} observed by the \gls{snfactory} in the form of flux-calibrated spectra, allowing for synthetic photometry, spectral velocity and line shape measurements all at the same phases. Its high quality make these data especially suited for use in a cosmological application such as the \gls{scm}.
Further improving on existing methods, we develop a novel \gls{scm} framework based on the UNITY framework for \glspl{sneIa}, which is encapsulated in the \textsc{Sccala} \textsc{Python} package.
Closely following the UNITY approach, we use a hierarchical Bayesian approach to fit the model.
The classical \gls{scm} is extended by an additional correction based on the H$_\alpha$ absorption to emission ratio $a/e$, which possibly is able to correct for \gls{csm} interaction impacting the observed magnitudes. Regardless of the physical interpretation, we find that it substantially improves the standardization quality, reducing the scatter of our Hubble-flow sample using the $H_0$ \gls{scm} from a value like the $0.28$\,mag found by \cite{Jaeger2022a} down to $0.18$\,mag in our work.
We find that, using the $H_0$-free \gls{scm}, the standardization can be applied at $30$ days instead of $43$ with only a small loss in standardization quality but a larger overall \gls{sne} sample.
At that phase we are able to utilize a total of $30$ \glspl{sne} for the determination of $H_0$. With this, we arrive at $H_0 = 70.6^{+4.5}_{-4.3}$\,km\,s$^{-1}$\,Mpc$^{-1}$ for the combined sample.

While our results appear to be robust against individual calibrators (see Section~\ref{sec:bootstrap}), the currently available \glspl{sneIIp} calibrators are the limiting factor here, both in terms of available \glspl{sne} as well as the quality of their data.
Consequently, any $H_0$ measurement will be sensitive to potential outliers unless the calibrator sample can be improved upon.
Therefore, we emphasize that the next step in investigating the Hubble tension using \glspl{sneII} and the SCM should enlarge the sample of well-observed events with new phase-conscious spectrophotometry, and then obtain TRGB or Cepheid distances to their host galaxies.
The existing sample of well-measured Hubble-flow \glspl{sneIIp} also is still rather limited, both in its sample size and its physical diversity (e.g. it does not include the lowest-luminosity, lowest-velocity \glspl{sneIIp}), and should be improved upon with a volume-limited \gls{sne} search reaching much fainter, e.g., than needed for \glspl{sneIa}. Using our hierarchical Bayesian framework, we can easily account for e.g. selection effects or outliers once the sample grows sufficiently large, which will help us to further constrain the source of the presently found $H_0$ tensions.

\begin{acknowledgements}
    The authors thank Assaf Sternberg for his help during the data acquisition phase.
    AH acknowledges support by the Klaus Tschira Foundation. AH is a Fellow of
the International Max Planck Research School for Astronomy and Cosmic Physics at
the University of Heidelberg (IMPRS-HD).
    CV and WH were supported
for part of this work by the Excellence Cluster ORIGINS, which is funded by the
Deutsche Forschungsgemeinschaft (DFG, German Research Foundation) under
Germany’s Excellence Strategy-EXC-2094-39078331.
ST acknowledges support by the Transregional Collaborative Research Centre TRR33 "The Dark Universe" of the German Research Foundation (DFG) and by the European Research Council (ERC) under the European Union’s Horizon 2020 research and innovation program (LENSNOVA: grant agreement No 771776).
    This research has made use of the NASA/IPAC Extragalactic Database (NED), which is operated by the Jet Propulsion Laboratory, California Institute of Technology, under contract with the National Aeronautics and Space Administration.
    We thank the technical staff of the University of
Hawaii 2.2-m telescope for observing assistance. We recognize the significant cultural role of Mauna Kea within the indigenous Hawaiian community, and we appreciate the opportunity to conduct observations from this revered site. This work was supported in part by the Director, Office of Science, Office of High Energy Physics of the U.S. Department of Energy under Contract No. DE-AC025CH11231. Additional support was provided by NASA under the Astrophysics Data Analysis Program grant 15-ADAP15-0256 (PI: Aldering). Support in France was provided by CNRS/IN2P3, CNRS/INSU, and PNC; LPNHE acknowledges support from LABEX ILP, supported by French state funds managed by the ANR within the Investissements d’Avenir programme under reference ANR-11-IDEX-0004-02. Support in Germany was provided by DFG through TRR33 “The Dark Universe” and by DLR through grants FKZ 50OR1503 and FKZ 50OR1602. In China support was provided by Tsinghua University 985 grant and NSFC grant No. 11173017. We thank the Gordon and Betty Moore Foundation for their support. This project has received funding from the European Research Council (ERC) under the European Union’s Horizon 2020 research and innovation programme (grant agreement No. 759194 – USNAC).

Corner plots have been created using the \textsc{Python} package \textsc{corner} \cite{ForemanMackey2016}.
\end{acknowledgements}

\section*{Data availability}
The spectra of the \gls{snfactory} data sample presented in this work can be found on the \gls{snfactory} website\footnote{\url{https://snfactory.lbl.gov/snf/data/index.html}}. Interpolated data and spectra are available on Zenodo\footnote{\url{https://doi.org/10.5281/zenodo.23038309}}.

\bibliographystyle{aa}
\bibliography{References}{}

\begin{appendix}

\section{A hierarchical SCM framework}\label{sec:app_a}
\subsection{The framework in detail}\label{sec:app_model_expl}
\subsubsection{$H_0$-free SCM}\label{sec:app_h0freescm}
In case of the $H_0$-free \gls{scm}, our model is constructed analogously to the one described by \citet{Rubin2015a}.
There are some notable differences and simplifications to the UNITY framework worth mentioning:
\begin{itemize}
    \item The unexplained dispersion is only added to the magnitudes, as the fractional partitioning of the dispersion into $v_{\mathrm{H}\beta}$, $col$, and $a/e$ contributions (see Equation 6 of \citealt{Rubin2015a}) could not be constrained.
    \item Here, the standardization coefficients $\alpha$, $\beta$, and $\gamma$ do not follow a redshift-dependent broken-linear relation (or any redshift dependence) as is the case in the UNITY model. In addition to the limited redshift range of the present dataset, there exists no evidence for such a relation in the case of \glspl{sneIIp}.
     \item  Since our sample comes from several surveys contributing at most 6 SNe each, we do not implement a fixed per-survey depth and roll-off as in \citet{Rubin2015a} but rather, we marginalize over the depth as in \citet{rubin2023a}. Due to the various survey depths, we allow broad priors on the depth and roll-over.
    \item An outlier model analogous to that of \citet{Rubin2015a} was initially implemented, but it found no outliers (outlier fraction $<1\%$ or $<0.5$ \glspl{sne}) and left the results unchanged. This may be due in part to the fact that we first manually inspected for outliers (see e.g. the exclusion of \jc).
    Since including it in the model made sampling take longer with no impact on the parameters of interest we do not include it here. 
\end{itemize}
The likelihood function of our model is given by
\begin{gather}
    \mathcal{L} = \prod_\mathrm{SNe} \frac{P\left(\mathrm{obs}\mid\mathrm{params}\right)\times P\left(\mathrm{detect} \mid \mathrm{obs}\right)}{\epsilon + P\left(\mathrm{detect}\;\middle|\;z_i\right)}, \\
    P\left(\mathrm{obs}\mid\mathrm{params}\right) = \mathcal{N} \left( \begin{bmatrix} m_I^\mathrm{obs} \\ v_{\mathrm{H}\beta}^\mathrm{obs} \\ col^\mathrm{obs} \\ a/e^\mathrm{obs}\end{bmatrix} \;\middle|\; \begin{bmatrix} m_I^\mathrm{true} + \Delta m_I \\ v_{\mathrm{H}\beta}^\mathrm{true} + \Delta v_{\mathrm{H}\beta} \\ col^\mathrm{true} + \Delta col \\ a/e^\mathrm{true} +\Delta a/e\end{bmatrix}, \boldsymbol{\Sigma} \right),\\
    P\left(\mathrm{detect}\mid \mathrm{obs}\right) = \Phi\left(\frac{m^\mathrm{cut}_I - m^\mathrm{obs}_I}{\sigma^\mathrm{cut}}\right),\\
    P\left(\mathrm{detect}\mid z_i\right) = \Phi\left(\frac{m^\mathrm{cut}_I - mean}{\sqrt{V^{m_I}}}\right),
\end{gather}
where $\mathcal{N}$ represents a multivariate normal distribution and $\Phi$ the Gaussian CDF. $\Delta\left\{m_I, v_{\mathrm{H}\beta}, col, a/e\right\}$ are connected to the systematic uncertainty terms through, e.g.,
\begin{equation}
    \Delta m_I \equiv \sum_l \frac{\partial m_I^\mathrm{obs}}{\partial \Delta \mathrm{sys}_l} \Delta \mathrm{sys}_l.
\end{equation}
In this work, we only consider a $150$\,km\,s$^{-1}$ systematic H$_\beta$ velocity uncertainty accounting for the unknown galaxy rotation velocity.
$\boldsymbol{\Sigma}$ is the covariance matrix, which is constructed analogously to the one by \citet{Rubin2015a}, i.e. $\boldsymbol{\Sigma} = \boldsymbol{\sigma}^\mathrm{ext}(z_i) + \boldsymbol{\Sigma}^\mathrm{obs}_i + \boldsymbol{\Sigma}^\mathrm{samp}_i$. Here $\boldsymbol{\sigma}^\mathrm{ext}(z_i)$ contains all external, redshift dependent uncertainties such as a redshift uncertainty, a peculiar velocity error of $150$\,km\,s$^{-1}$ \citep{peterson2022a} and a gravitational lensing uncertainty $\sigma_\mathrm{lensing}(z_i) = 0.055z_i$ \citep{Joensson2010a}. $\boldsymbol{\Sigma}^\mathrm{obs}_i$ contain all observational uncertainties, or more precisely the uncertainties of the interpolated observational quantities. $\boldsymbol{\Sigma}^\mathrm{samp}_i$ captures the sample-dependent unexplained dispersion of the \gls{sne}-distribution, which in our case is only $\sigma_\mathrm{int}$.

The terms $P(\mathrm{detect}\mid\mathrm{obs})$ and $P(\mathrm{detect}\mid z_i)$ model selection effects and are derived analogously to \citet{Rubin2015a}.
Specifically, we obtain
\begin{equation}
\begin{split}
    mean = \mathcal{M}_I &- \alpha\cdot\log_{10}\left(\frac{v^*}{\langle v_{H\beta}\rangle}\right)+\beta\cdot\left(col^*-\langle col\rangle\right)\\ &+ \gamma\cdot\left(\frac{a}{e}^* - \langle\frac{a}{e}\rangle\right) + 5\log_{10}\left(\mathcal{D}_L(z_i,H_0)\right)
\end{split}
\end{equation}
and
\begin{equation}
    V^{m_I}=C^{m_I} + (\sigma^{cut})^2 + \left(\alpha \frac{R^v}{\ln(10) \cdot v^*}\right)^2 + (\beta R^{col})^2 + (\gamma R^{a/e})^2.
\end{equation}
Here, $C^{m_I}$ is the full covariance matrix entry with index $(1,1)$.
Our derived expressions already include the approximations of \citet{Rubin2015a}, i.e. $b^\mathrm{cut}_j=a^\mathrm{cut}_j=0$.
Additionally, since our fit coefficients $\alpha$, $\beta$, and $\gamma$ do not follow a broken linear relation with redshift, their derivatives vanish, further simplifying the expressions.
Similar to \citet{Rubin2015a}, we use a finite choice for the numerical factor $\epsilon$, in our case $\epsilon=0.0001$.
We find no noticeable differences in the resulting fits between our choice and the choice of $\epsilon=0.01$ of \citet{Rubin2015a}.

We use the following priors:
\begin{align*}
    \mathcal{M}_I & \sim \mathcal{U}(-30,0) \tag*{Absolute Hubble-free magnitude}\\
    \alpha & \sim \mathcal{U}(-20,20) \tag*{Velocity correction coefficient}\\
    \beta & \sim \mathcal{U}(-20,20) \tag*{Color correction coefficient}\\
    \gamma & \sim \mathcal{U}(-20,20) \tag*{\text{$a/e$ correction coefficient}}\\
    \log_{10}\sigma^\mathrm{int}_j & \sim \mathcal{U}(-3,0) \tag*{Sample unexplained dispersion}\\
    v^* & \sim \text{Cauchy}(7.5, 1.5) \tag*{\text{Mean of Latent $v$ (in $10^3$\,km\,s$^{-1}$)}}\\
    col^* & \sim \text{Cauchy}(0, 0.5) \tag*{\text{Mean of Latent $col$}}\\
    a/e^* & \sim \text{Cauchy}(0.5, 0.5) \tag*{\text{Mean of Latent $a/e$}}\\
    R^v & \sim \mathcal{N}(0, 1.5) \tag*{\text{Dispersion of Latent $v$ (in $10^3$\,km\,s$^{-1}$)}}\\
    R^{col} & \sim \mathcal{N}(0, 0.5) \tag*{\text{Dispersion of Latent $col$}}\\
    R^{a/e} & \sim \mathcal{N}(0, 0.5) \tag*{\text{Dispersion of Latent $a/e$}}\\
    v_{\mathrm{H}\beta}^{\mathrm{true},i} & \sim \mathcal{N}(v_j^*, R^v_j) \tag*{\text{Modeled Latent $v$}}\\
    col^{\mathrm{true},i} & \sim \mathcal{N}(col_j^*, R^{col}_j) \tag*{\text{Modeled Latent $col$}}\\
    a/e^{\mathrm{true},i} & \sim \mathcal{N}(a/e_j^*, R^{a/e}_j) \tag*{\text{Modeled Latent $a/e$}}\\
    m^\mathrm{cut}_j & \sim \mathcal{N}(m^\mathrm{nominal}_j, 0.5)\mathcal{U}(14,30) \tag*{\text{Median Survey Depth}}\\
    \sigma^\mathrm{cut}_j & \sim \mathcal{N}(\sigma^\mathrm{depth}_j, 0.25)\mathcal{U}(0.1,3) \tag*{\text{Survey Depth $1\,\sigma$ Around Median}}
\end{align*}
Here, the prior choices are motivated by the priors used by the UNITY model, with the prior ranges selected empirically.
We validated through testing that our results are insensitive to the exact prior choices, e.g., choosing a wider prior for the dispersion of the latent $col$ does not change the results outside of sampling fluctuations.
The only noteworthy difference here is the prior choice for $\sigma_\mathrm{int}$, i.e. the choice of a log-uniform prior (``Jefferey's prior'') instead of a uniform prior. This choice was made to force the model to prefer smaller $\sigma_\mathrm{int}$ values and only make use of this unexplained scatter when the other parameters cannot explain the remaining scatter in the fit within the given uncertainties. When using a uniform prior, the model tends to use a higher $\sigma_\mathrm{int}$ than necessary to explain the fit.
\begin{figure}[!hbt]
    \centering
    \includegraphics[]{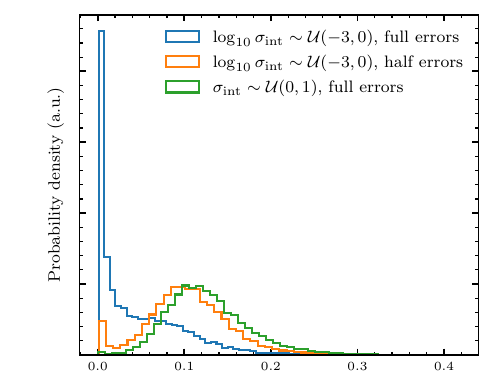}
    \caption{Illustration of the effect of different $\sigma_\mathrm{int}$ prior choices on the posterior. Here, the result using the $H_0$-free SCM with the \gls{snfactory} data at a phase of $30$ days is shown. The results marked with full errors used full uncertainties, whereas the half errors result was obtained with the uncertainties artificially reduced by one half.}
    \label{fig:sigmaintprior}
\end{figure}
To illustrate this effect, we use the $H_0$-free \gls{scm} described in Section~\ref{sec:h0freescm} with the \gls{snfactory} sample data at a phase of $30$ days with both a log-uniform prior and a uniform prior. As can be seen in Figure~\ref{fig:sigmaintprior}, the uniform prior yields a much higher $\sigma_\mathrm{int}$ value. However, when artificially reducing the uncertainties of the data by one half, the log-uniform prior yields a similar $\sigma_\mathrm{int}$ value. I.e. with the log-uniform prior the model will only use the unexplained scatter when it cannot otherwise create a good fit within the uncertainties, whereas the uniform prior will be used even when the fit could be well explained with the given uncertainties. We leave a more detailed study introduced by this prior choice for further studies. It should be noted that the log-uniform prior choice is also very similar to the $1/\sigma_\mathrm{int}$ prior frequently used in a \gls{scm} context (see e.g. \citealt{Jaeger2020b,Jaeger2022a}).

\subsubsection{$H_0$ \gls{scm}}\label{sec:app_h0scm}
The framework for the $H_0$ \gls{scm} is similar to the previously described $H_0$-free \gls{scm} framework, but here the value of the Hubble-Lemaître constant is a free parameter fixed by the calibrator sample. As such the basic likelihood function now reads
\begin{equation}
\begin{split}
    \mathcal{L} &= \prod_\mathrm{SNe} \mathcal{N}  \left( \begin{bmatrix} m_I^\mathrm{obs} \\ v_{\mathrm{H}\beta}^\mathrm{obs} \\ col^\mathrm{obs} \\ a/e^\mathrm{obs}\end{bmatrix} \;\middle|\; \begin{bmatrix} m_I^\mathrm{true} + \Delta m_I \\ v_{\mathrm{H}\beta}^\mathrm{true} + \Delta v_{\mathrm{H}\beta} \\ col^\mathrm{true} + \Delta col \\ a/e^\mathrm{true} +\Delta a/e\end{bmatrix}, \boldsymbol{\Sigma} \right)\\
    &\cdot \prod_\mathrm{Calib-SNe} \mathcal{N} \left( \begin{bmatrix} m_I^\mathrm{obs} \\ v_{\mathrm{H}\beta}^\mathrm{obs} \\ col^\mathrm{obs} \\ a/e^\mathrm{obs}\end{bmatrix} \;\middle|\; \begin{bmatrix} m_{I,\mathrm{calib}}^\mathrm{true} + \Delta m_I \\ v_{\mathrm{H}\beta}^\mathrm{true} + \Delta v_{\mathrm{H}\beta} \\ col^\mathrm{true} + \Delta col \\ a/e^\mathrm{true} +\Delta a/e\end{bmatrix}, \boldsymbol{\Sigma} \right).
\end{split}
\end{equation}
Here, the selection effect model is added analogously to the $H_0$-free model; the expression for $mean$ is adapted to include $H_0$.
We note that here, we do not model selection effects for the calibrator sample, as it is a volume-limited sample.
During the testing of our model, we found no difference whether or not we model magnitude-limiting selection for our calibrator sample, based on simulated data mimicking our actual data sample.
The priors remain the same with the additional priors
\begin{align*}
    M_I & \sim \mathcal{U}(-30,0) \tag*{Absolute magnitude}\\
    H_0 & \sim \mathcal{U}(0,200) \tag*{Hubble-Lemaître constant}
\end{align*}
for the coefficients introduced in the $H_0$ \gls{scm} case.

\subsubsection{Split Populations}\label{sec:split_pops}

In Section~\ref{sec:djcomp} we investigated whether there is evidence that the calibrator \glspl{sne}  originate from a different underlying population that follows a different standardization relation.
To capture this behavior, in addition to assuming different latent parameters (i.e., $v^*$, $col^*$, and $a/e^*$) for calibrator and Hubble-flow \glspl{sne}, we introduce the standardization coefficient offsets $\Delta\alpha_\mathrm{calib}$, $\Delta\beta_\mathrm{calib}$, and $\Delta\gamma_\mathrm{calib}$ that connect the standardization coefficients of the Hubble flow sample, $\alpha$, $\beta$, and $\gamma$, to those of the calibrator sample:
\begin{align}
    \alpha_\mathrm{calib} &= \alpha + \Delta\alpha_\mathrm{calib},\\
    \beta_\mathrm{calib} &= \beta + \Delta\beta_\mathrm{calib},\\
    \gamma_\mathrm{calib} &= \gamma + \Delta\gamma_\mathrm{calib}.\\
\end{align}
The calibrator \glspl{sne} are then assumed to standardize according to the modified Eq.~\ref{eq:h0scm_calib}, that is
\begin{equation}
\begin{split}
    m_{I,\mathrm{calib}}^{\mathrm{true}}  = M_I &- \alpha_\mathrm{calib} \cdot \log_{10}\left(\frac{v_{\mathrm{H}\beta}}{\langle v_{\mathrm{H}\beta}\rangle}\right) + \beta_\mathrm{calib} \cdot (col - \langle col \rangle)\\ &+ \gamma_\mathrm{calib} \cdot \left(\frac{a}{e} - \langle \frac{a}{e}\rangle\right) + \mu_\mathrm{calib}.
\end{split}
\end{equation}
This requires the following additional priors
\begin{align*}
    \Delta\alpha_\mathrm{calib} & \sim \mathcal{N}(0,1) \tag*{Velocity correction coefficient offset}\\
    \Delta\beta_\mathrm{calib} & \sim \mathcal{N}(0,1) \tag*{Color correction coefficient offset}\\
    \Delta\gamma_\mathrm{calib} & \sim \mathcal{N}(0,1) \tag*{\text{$a/e$ correction coefficient offset}}\\
\end{align*}
for this standardization mode.
It should be noted that this is a rather simplified model that assumes that both population follow a standardization that is largely unconnected.
For the present discussion we deem this sufficient, but this model can easily be generalized, accounting for correlations between the two (or more) populations.

\subsection{Testing procedure}\label{sec:app_model_test}
\begin{table}[!hbt]
    \centering
    \caption{Comparison of the global standardization coefficients between the hierarchical and non-hierarchical approach.}\label{tab:testdata}
	\begingroup
	\setlength{\tabcolsep}{6pt}
	\renewcommand{\arraystretch}{1.25}
    \begin{tabular}{l|c|c|c}
        \hline\hline
        & Input & Non-hierarchical & Hierarchical\\
        \hline
        $\mathcal{M}_I$ & $-1.6$ & $-1.61\pm0.06$ & $-1.60\pm0.05$ \\
        $\alpha$ & $3.4$ & $2.85\pm0.37$ & $3.36\pm0.47$ \\
        $\beta$ & $1.8$ & $1.13\pm0.53$ & $1.84\pm0.44$ \\
        $\gamma$ & $-1.5$ & $-1.45\pm0.15$ & $-1.49\pm0.14$\\
        \hline
    \end{tabular}
    \endgroup
    \tablefoot{
    The results illustrated here are averaged from $150$ synthetic datasets.
    }
\end{table}
Our testing procedure, much like the hierarchical framework, closely follows the work described by \cite{Rubin2015a}. In order to test the framework, we generate simulated data based on the following inputs:
\begin{itemize}
    \item First, we generate random redshifts in the ranges $0.01-0.04$, $0.03-0.2$ and $0.1-0.4$ with $250$ \glspl{sne} in each redshift set. These redshifts are drawn from a triangular distribution as in \cite{Rubin2015a}. The ranges have been chosen to represent the redshift range of the \gls{snfactory} data sample, a sample well above the $z > 0.023$ cut suggested by \cite{Riess2022a} and a sample well within the Hubble flow, respectively.
    \item Second, we generate velocity, color and $a/e$ populations from normal distributions centered around the mean values of the \gls{snfactory} data sample and with its dispersion.
    \item Third, we calculate the resulting model magnitudes based on Eq.~\eqref{eq:h0free_mag} with the free parameters values listed in Table~\ref{tab:testdata}.
    \item Each of these model magnitudes is then passed through a magnitude selection function, chosen for $50\,\%$ completeness (i.e. $m^\mathrm{cut}$) at $18.5$, $22$, and $24$\,mag, respectively. If a simulated object is rejected, another one is generated until the target $250$ \glspl{sne} are obtained. 
    \item For the uncertainties we assume a color uncertainty of $0.05$, an $a/e$ uncertainty of $0.013$ and a redshift uncertainty of $0.0001$. The H$_\beta$ velocity uncertainty is assumed to be $200$\,km\,s$^{-1}$ to which we also add the host galaxy rotation uncertainty of $150$\,km\,s$^{-1}$. The magnitude uncertainty is calculated analogously to the one described in Appendix~\ref{sec:app_model_expl}, i.e. including a peculiar velocity error and a gravitational lensing uncertainty. Here we assume a base magnitude uncertainty of $0.05\,\mathrm{mag}$.
    \item Lastly, we randomly scatter the magnitude, velocity, color, and $a/e$ populations by their respective uncertainty. In case of the magnitudes we add $\sigma_\mathrm{int} = 0.25\,\mathrm{mag}$ as additional scatter.
\end{itemize}
The main difference between the simulated date here and the one described by \cite{Rubin2015a} is the inclusion of outliers, which are not accounted for by our hierarchical framework and as such do not need to be tested. Consequently our sample data might lack some sophistication, but is still sufficient for testing our framework and comparing the hierarchical and non-hierarchical approach.\\
For the total of $150\times250$ simulated \glspl{sne} we fit both the hierarchical and non-hierarchical model and record the fitted coefficients. A summary averaged over all three redshift subsets (i.e. $150$ synthetic datasets) can be found in Table~\ref{tab:testdata}.
While the non-hierarchical model manages to capture the overall trend, it systematically underestimates the dominant velocity correction by a substantial amount.
Consequentially, this also leads to a small offset value for $\mathcal{M}_I$ which is directly correlated to the resultant $H_0$ value.
In contrast, the hierarchical model recovers this value rather well.
This means that, on average, the hierarchical model will fit the correct $\mathcal{M}_I$ values and therefore also the correct $H_0$, whereas the non-hierarchical model will introduce a systematic offset.

\FloatBarrier

\onecolumn
\section{The SNfactory sample}
\subsection{Sample properties}\label{app:sample_prop}
This section lists miscellanious sample properties in Table~\ref{tab:analysissample}.
\begin{sidewaystable*}
	\caption{Overview of the \gls{snfactory} \gls{sneII} Hubble-flow sample used in this work.}
	\label{tab:analysissample}
	\begingroup
	\setlength{\tabcolsep}{6pt}
	\renewcommand{\arraystretch}{1.25}
	\begin{tabular}{l|c|c|c|c|c|c|c|c}
		\hline\hline
		\gls{sne} name & RA & DEC & Host galaxy & Discovery\tablefootmark{a} & \gls{toe} (MJD) & \gls{toe} Method\tablefootmark{b} & $z_\text{CMB}$\tablefootmark{c} & $N_\text{phases}$\tablefootmark{d,e} \\
		\hline
		\tpa & 23:04:40.95 & -09:38:27.7 & \href{https://ned.ipac.caltech.edu/byname?objname=WISEA\%20J230440.55-093830.3\&hconst=67.8\&omegam=0.308\&omegav=0.692\&wmap=4\&corr\_z=1}{WISEA J230440.55-093830.3} & 2010-08-23 (1) & $55408.0^{+5.7}_{-5.5}$ & SNID & $0.03204\pm 0.00024$ & $6$ \\
		\hb & 03:07:01.66 & 46:37:20.20 &                  \href{https://ned.ipac.caltech.edu/byname?objname=UGC\%2002537\&hconst=67.8\&omegam=0.308\&omegav=0.692\&wmap=4\&corr\_z=1}{UGC 02537} & 2010-08-24 (2) & $55414.2^{+4.0}_{-3.8}$ & SNID & $0.01612\pm 0.00005$ & $8^*$ \\
		\wmf & 21:56:15.69 & 02:10:13.80 &               \href{https://ned.ipac.caltech.edu/byname?objname=CGCG\%20377-004\&hconst=67.8\&omegam=0.308\&omegav=0.692\&wmap=4\&corr\_z=1}{CGCG 377-004} & 2010-09-24 (1) & $55459.3^{+2.0}_{-2.0}$ & LC & $0.02672\pm 0.00014$ & $7$ \\
		\xlr & 23:09:22.76 & 01:00:03.60 & \href{https://ned.ipac.caltech.edu/byname?objname=SDSS+J230922.92\%2B010002.0&hconst=67.8&omegam=0.308&omegav=0.692&wmap=4&corr_z=1}{SDSS J230922.92+010002.0} & 2010-10-03 (1) & $55468.0^{+2.0}_{-2.0}$  & LC & $0.0141\pm 0.00009$ & $7^*$ \\
		\jc & 02:40:13.97 & -08:46:25.7 & \href{https://ned.ipac.caltech.edu/byname?objname=NGC\%201033\&hconst=67.8\&omegam=0.308\&omegav=0.692\&wmap=4\&corr\_z=1}{NGC 1033} & 2010-10-29 (3,4) & $55489.0^{+1.5}_{-1.6}$ & SNID & $0.02358\pm 0.00007$ & $6^*$ \\
		\icthree & 12:53:53.43 & 36:05:19.30 & \href{https://ned.ipac.caltech.edu/byname?objname=IC\%203862\&hconst=67.8\&omegam=0.308\&omegav=0.692\&wmap=4\&corr\_z=1}{IC 3862} & 2011-03-18 (5) & $55637.0^{+1.7}_{-1.8}$ & SNID & $0.01481\pm 0.00006$ & $7$ \\
		\ngctwo & 06:46:39.85 & 60:21:02.80 & \href{https://ned.ipac.caltech.edu/byname?objname=NGC\%202273B\&hconst=67.8\&omegam=0.308\&omegav=0.692\&wmap=4\&corr\_z=1}{NGC 2273B} & 2011-08-20 (6) & $55772.0^{+2.0}_{-2.2}$ & SNID & $0.007122\pm 0.000017$ & $10^*$ \\
		\ngcfour & 01:18:07.80 & 17:33:29.80 & \href{https://ned.ipac.caltech.edu/byname?objname=NGC\%20459\&hconst=67.8\&omegam=0.308\&omegav=0.692\&wmap=4\&corr\_z=1}{NGC 459} & 2011-08-26 (7) & $55793.6^{+2.0}_{-2.1}$ & SNID & $0.04134\pm 0.00008$ & $6$ \\
		\pgc & 18:39:53.93 & 40:01:43.7 & \href{https://ned.ipac.caltech.edu/byname?objname=IC\%204772\&hconst=67.8\&omegam=0.308\&omegav=0.692\&wmap=4\&corr\_z=1}{IC 4772} & 2011-09-17 (8) & $55814.1^{+0.5}_{-0.5}$ & SNID & $0.02326\pm 0.0001$ & $8^*$ \\
		\cer & 13:17:08.45 & -20:24:41.8 & \href{https://ned.ipac.caltech.edu/byname?objname=WISEA\%20J131708.58-202434.9\&hconst=67.8\&omegam=0.308\&omegav=0.692\&wmap=4\&corr\_z=1}{WISEA J131708.58-202434.9} & 2012-04-24 (9) & $56039.2^{+0.5}_{-0.5}$ & SNID & $0.02464\pm 0.00017$ & $7^*$ \\
		\css & 15:06:02.54 & 41:25:32.70 & \href{https://ned.ipac.caltech.edu/byname?objname=WISEA\%20J150602.64\%2B412535.3\&hconst=67.8\&omegam=0.308\&omegav=0.692\&wmap=4\&corr\_z=1}{CSS 120517} & 2012-05-17 (10) & $56060.6^{+0.4}_{-0.4}$ & SNID & $0.00902\pm 0.00021$ & $12$ \\
		\icone & 01:27:31.45 & 14:49:05.80 & \href{https://ned.ipac.caltech.edu/byname?objname=IC\%201706\&hconst=67.8\&omegam=0.308\&omegav=0.692\&wmap=4\&corr\_z=1}{IC 1706} & 2012-08-20 (11) & $56147.2^{+3.4}_{-3.3}$ & SNID & $0.02065\pm 0.00009$ & $6$ \\
		\icthreefive & 05:58:51.93 & -23:20:24.22 & \href{https://ned.ipac.caltech.edu/byname?objname=IC+35&hconst=67.8&omegam=0.308&omegav=0.692&wmap=4&corr_z=1}{IC 35} & 2012-10-07 (12) & $56202.6^{+1.0}_{-1.1}$ & SNID & $0.01415\pm 0.00008$ & $5$ \\
		\fvq &  01:16:36.17 & -31:27:07.40 & \href{https://ned.ipac.caltech.edu/byname?objname=WISEA+J011636.34-312714.2&hconst=67.8&omegam=0.308&omegav=0.692&wmap=4&corr_z=1}{WISEA J011636.34-312714.2} & 2012-11-01 (9) & $56228.7^{+4.6}_{-4.5}$ & SNID & $0.0342\pm 0.0004$ & $6^*$ \\
		\ljg & 08:12:48.97 & 46:17:00.10 & \href{https://ned.ipac.caltech.edu/byname?objname=SDSS\%20J081248.90\%2B461701.9\&hconst=67.8\&omegam=0.308\&omegav=0.692\&wmap=4\&corr\_z=1}{SDSS J081248.90+461701.9} & 2012-11-22 (1) & $56250.1^{+1.8}_{-1.8}$ & SNID & $0.03219\pm 0.00004$& $8$ \\
		\hi & 08:39:42.02 & 60:58:16.00 & \href{https://ned.ipac.caltech.edu/byname?objname=UGC\%204512\&hconst=67.8\&omegam=0.308\&omegav=0.692\&wmap=4\&corr\_z=1}{UGC 4512} & 2012-12-02 (13) & $56256.0^{+1.0}_{-1.0}$ & LC & $0.02673\pm 0.00005$ & $9$ \\
		\zw & 22:40:17.02 & -02:25:34.10 & \href{https://ned.ipac.caltech.edu/byname?objname=WISEA+J224015.14-022526.7&hconst=67.8&omegam=0.308&omegav=0.692&wmap=4&corr_z=1}{Zw 183} & 2012-12-06 (14,15) & $56264.2^{+0.5}_{-0.5}$ & SNID & $0.00870\pm 0.00009$ & $6^*$ \\
		\hnj & 05:12:24.82 & -25:46:57.4 & \href{https://ned.ipac.caltech.edu/byname?objname=WISEA\%20J051224.61-254658.0\&hconst=67.8\&omegam=0.308\&omegav=0.692\&wmap=4\&corr\_z=1}{WISEA J051224.61-254658.0} & 2012-12-12 (9) & $56271.0^{+1.0}_{-1.0}$ & LC & $0.01479\pm 0.00015$ & $6$ \\
		\ugc & 10:48:26.57 & 38:24:07.90 & \href{https://ned.ipac.caltech.edu/byname?objname=UGC\%2005910\&hconst=67.8\&omegam=0.308\&omegav=0.692\&wmap=4\&corr\_z=1}{UGC 05910} & 2013-04-16 (16,17) & $56396.9^{+0.5}_{-0.5}$ & SNID & $0.02642\pm 0.00017$ & $6^*$ \\
		\bjx & 14:14:51.95 & 36:47:28.90 & \href{https://ned.ipac.caltech.edu/byname?objname=KUG\%201412\%2B370\&hconst=67.8\&omegam=0.308\&omegav=0.692\&wmap=4\&corr\_z=1}{KUG 1412+370} & 2013-05-29 (18) & $56442.20^{+0.05}_{-0.05}$ & \cite{Vogl2020} & $0.02846\pm 0.00012$ & $6$ \\
		\ds & 16:11:29.58 & 57:22:51.71 & \href{https://ned.ipac.caltech.edu/byname?objname=WISEA+J161131.46\%2B572257.1&hconst=67.8&omegam=0.308&omegav=0.692&wmap=4&corr_z=1}{WISEA J161131.46+572257.1} & 2013-07-01 (19) & $56474.0^{+6.1}_{-6.3}$ & SNID & $0.025114\pm 0.000005$ & $3^*$ \\
		\hline
	\end{tabular}
    \tablefoot{
    \tablefoottext{a}{The number in parenthesis refers to the relevant discovery reference listed below this table.}
    \tablefoottext{b}{Method with which the \gls{toe} was determined. SNID refers to the method of fitting the phases of SNID template matches, whereas LC indicates that photometry was used.}
    \tablefoottext{c}{The \gls{cmb} redshifts were all taken from the respective host galaxies (as listed on the NED) as named in the host galaxy column.}
    \tablefoottext{d}{This number refers to the number of observed spectra and not necessarily to the number of phases usable in the SCM.}
    \tablefootmark{e}{In cases where the phase number has been marked with $^*$, a host galaxy subtraction has not been applied to any spectrum of the respective \glspl{sne} as described in Section~\ref{sec:data_snfactory}.}
    }
	\endgroup 
    \tablebib{
    (1)~\citet{Law2009};
    (2)~\citet{2010CBET.2424....1B};
    (3)~\citet{2010CBET.2521....1H};
    (4)~\citet{2010CBET.2522....1M};
    (5)~\citet{2011CBET.2680....1B};
    (6)~\citet{2011CBET.2791....1K};
    (7)~\citet{2011CBET.2828....1C};
    (8)~\citet{2011CBET.2831....1A};
    (9)~\citet{Baltay2013};
    (10)~\citet{2012CBET.3118....1D};
    (11)~\citet{2012CBET.3208....1C};
    (12)~\citet{2012CBET.3278....1D};
    (13)~\citet{2012CBET.3332....1B};
    (14)~\citet{2012CBET.3338....1I};
    (15)~\citet{2012CBET.3339....1H};
    (16)~\citet{2013ATel.4989....1T};
    (17)~\citet{2013CBET.3487....1C};
    (18)~\citet{Kulkarni2013};
    (19)~\citet{2013CBET.3577....1C}
    }
\end{sidewaystable*}

\subsection{Spectra and photometry}
\begin{figure}[!htb]
	\centering
	\parbox{.35\linewidth}{\includegraphics[width=\linewidth,keepaspectratio]{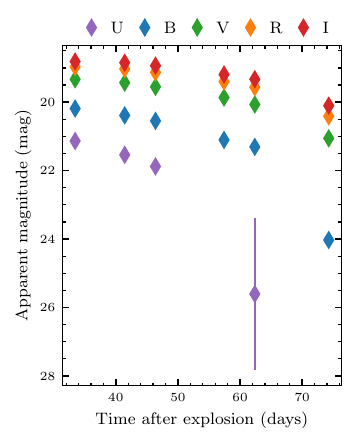}}
	\begin{minipage}{0.6\linewidth}%
		\includegraphics[width=\linewidth,keepaspectratio]{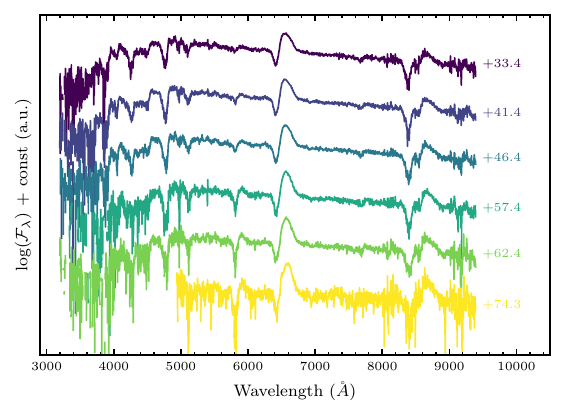}
	\end{minipage}
	\caption{Synthetic \gls{bessell12} photometry (left) and spectral time series (right) of \tpa. The numbers next to the spectra indicate the time after the explosion in days.}\label{fig:app_sample1}
\end{figure}

This section (Figures~\ref{fig:app_sample1} to \ref{fig:app_sample2}) contains an overview of the spectral time series and synthetic photometry of the Hubble-flow sample \glspl{sne} as used in our analysis. All spectra are corrected for Milky Way extinction and transformed into the \gls{sne} restframe.

\begin{figure}[!htb]
	\centering
	\parbox{.35\linewidth}{\includegraphics[width=\linewidth,keepaspectratio]{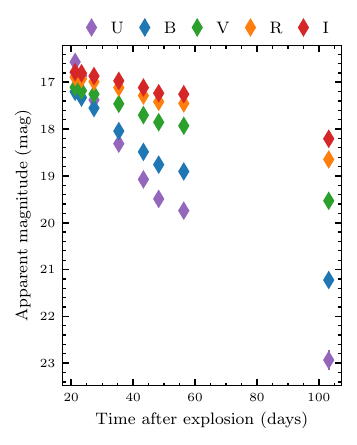}}
	\qquad
	\begin{minipage}{0.6\linewidth}%
		\includegraphics[width=\linewidth,keepaspectratio]{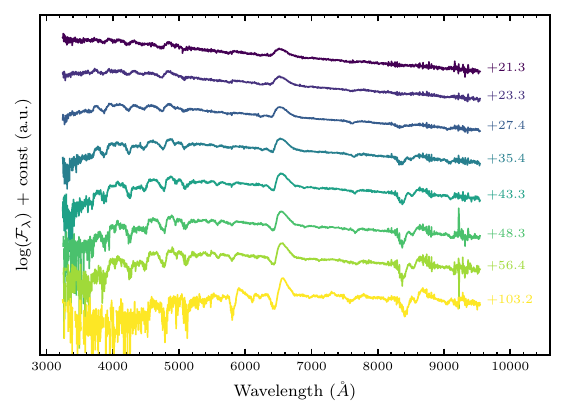}
	\end{minipage}
	\caption{Synthetic \gls{bessell12} photometry (left) and spectral time series (right) of \hb. The numbers next to the spectra indicate the time after the explosion in days. For this \gls{sne}, no host galaxy template subtraction was performed.}
\end{figure}

\begin{figure}[!htb]
	\centering
	\parbox{.35\linewidth}{\includegraphics[width=\linewidth,keepaspectratio]{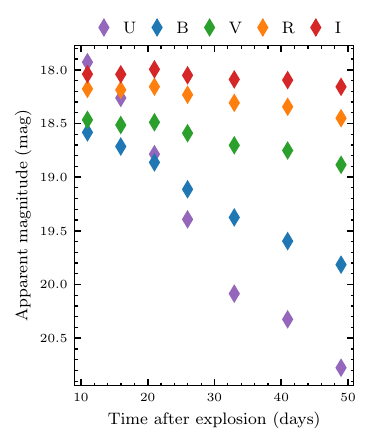}}
	\qquad
	\begin{minipage}{0.6\linewidth}%
		\includegraphics[width=\linewidth,keepaspectratio]{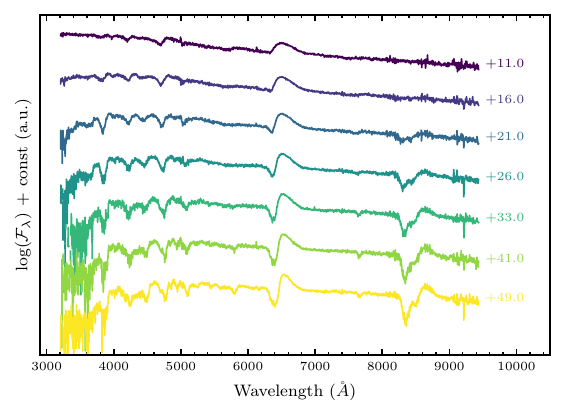}
	\end{minipage}
	\caption{Synthetic \gls{bessell12} photometry (left) and spectral time series (right) of \wmf. The numbers next to the spectra indicate the time after the explosion in days.}
\end{figure}

\begin{figure}[!htb]
	\centering
	\parbox{.35\linewidth}{\includegraphics[width=\linewidth,keepaspectratio]{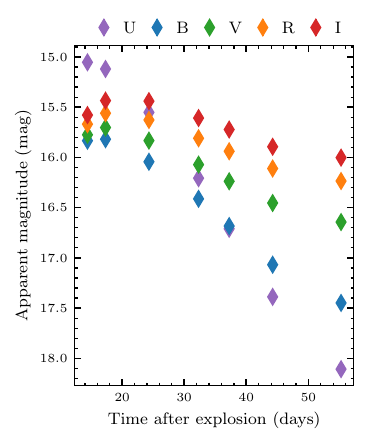}}
	\qquad
	\begin{minipage}{0.6\linewidth}%
		\includegraphics[width=\linewidth,keepaspectratio]{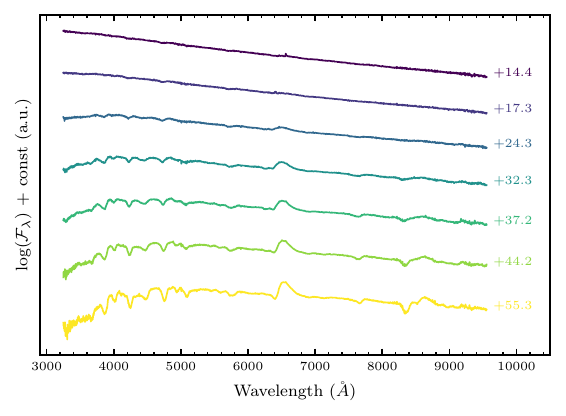}
	\end{minipage}
	\caption{Synthetic \gls{bessell12} photometry (left) and spectral time series (right) of \xlr. The numbers next to the spectra indicate the time after the explosion in days. For this \gls{sne}, no host galaxy template subtraction was performed.}\label{fig:specxlr}
\end{figure}

\begin{figure}[!htb]
	\centering
	\parbox{.35\linewidth}{\includegraphics[width=\linewidth,keepaspectratio]{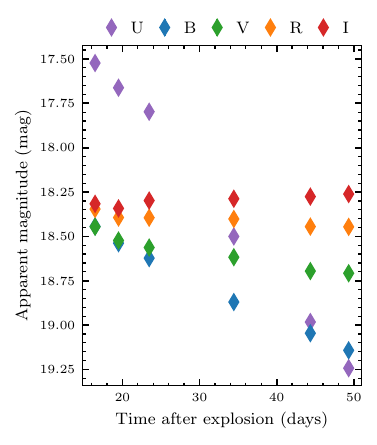}}
	\qquad
	\begin{minipage}{0.6\linewidth}%
		\includegraphics[width=\linewidth,keepaspectratio]{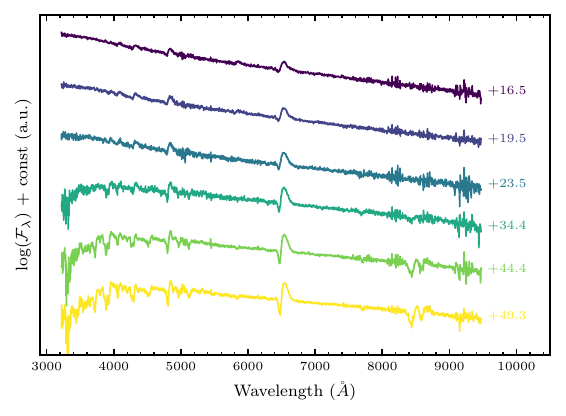}
	\end{minipage}
	\caption{Synthetic \gls{bessell12} photometry (left) and spectral time series (right) of \jc. The numbers next to the spectra indicate the time after the explosion in days. For this \gls{sne}, no host galaxy template subtraction was performed.}
\end{figure}

\begin{figure}[!htb]
	\centering
	\parbox{.35\linewidth}{\includegraphics[width=\linewidth,keepaspectratio]{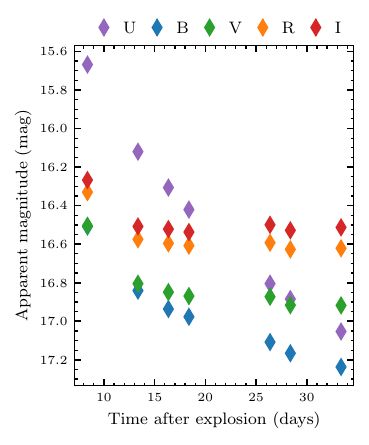}}
	\qquad
	\begin{minipage}{0.6\linewidth}%
		\includegraphics[width=\linewidth,keepaspectratio]{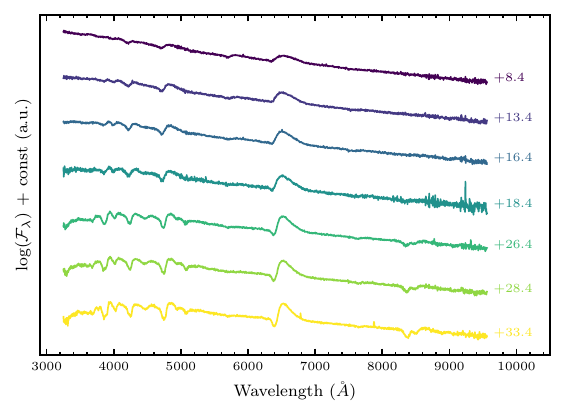}
	\end{minipage}
	\caption{Synthetic \gls{bessell12} photometry (left) and spectral time series (right) of \icthree. The numbers next to the spectra indicate the time after the explosion in days.}
\end{figure}

\begin{figure}[!htb]
	\centering
	\parbox{.35\linewidth}{\includegraphics[width=\linewidth,keepaspectratio]{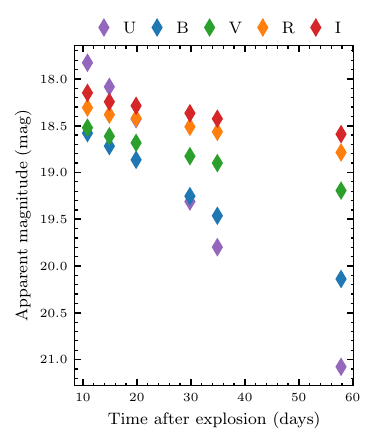}}
	\qquad
	\begin{minipage}{0.6\linewidth}%
		\includegraphics[width=\linewidth,keepaspectratio]{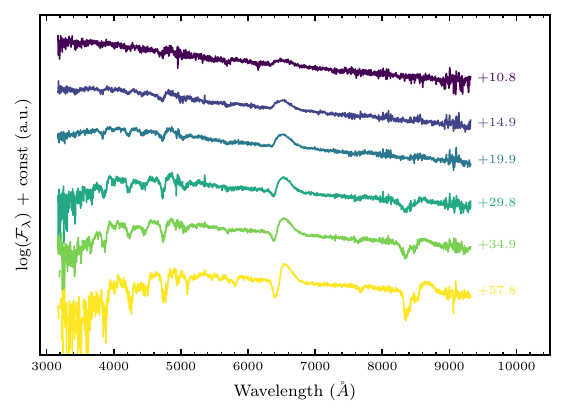}
	\end{minipage}
	\caption{Synthetic \gls{bessell12} photometry (left) and spectral time series (right) of \ngcfour. The numbers next to the spectra indicate the time after the explosion in days.}
\end{figure}

\begin{figure}[!htb]
	\centering
	\parbox{.35\linewidth}{\includegraphics[width=\linewidth,keepaspectratio]{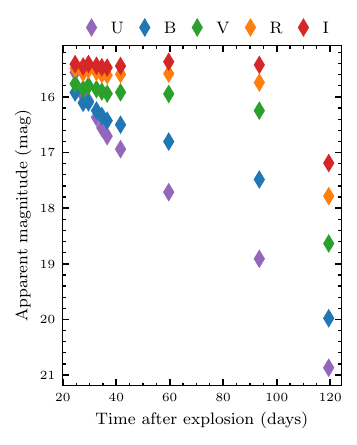}}
	\qquad
	\begin{minipage}{0.6\linewidth}%
		\includegraphics[width=\linewidth,keepaspectratio]{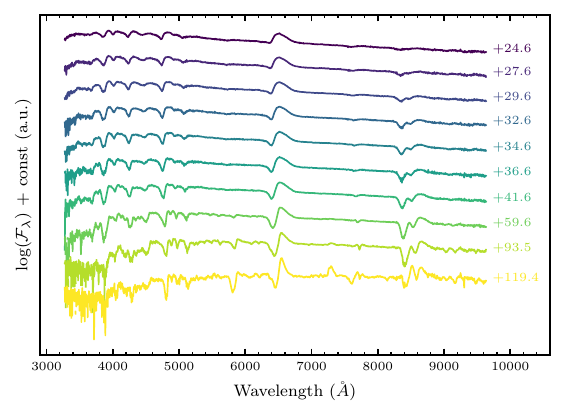}
	\end{minipage}
	\caption{Synthetic \gls{bessell12} photometry (left) and spectral time series (right) of \ngctwo. The numbers next to the spectra indicate the time after the explosion in days. For this \gls{sne}, no host galaxy template subtraction was performed.}
\end{figure}

\begin{figure}[!htb]
	\centering
	\parbox{.35\linewidth}{\includegraphics[width=\linewidth,keepaspectratio]{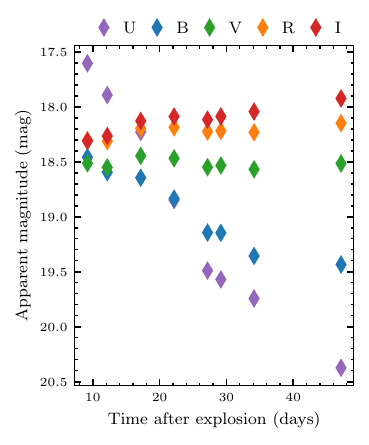}}
	\qquad
	\begin{minipage}{0.6\linewidth}%
		\includegraphics[width=\linewidth,keepaspectratio]{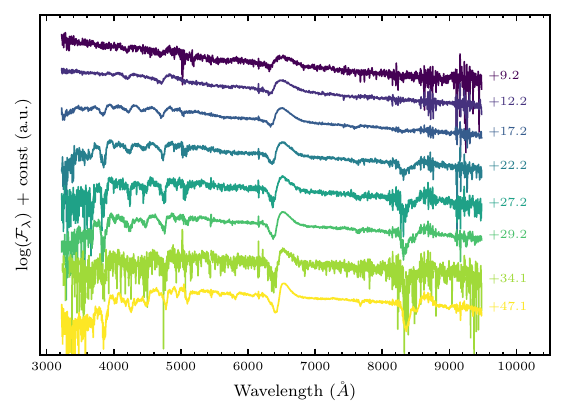}
	\end{minipage}
	\caption{Synthetic \gls{bessell12} photometry (left) and spectral time series (right) of \pgc. The numbers next to the spectra indicate the time after the explosion in days. For this \gls{sne}, no host galaxy template subtraction was performed.}
\end{figure}

\begin{figure}[!htb]
	\centering
	\parbox{.35\linewidth}{\includegraphics[width=\linewidth,keepaspectratio]{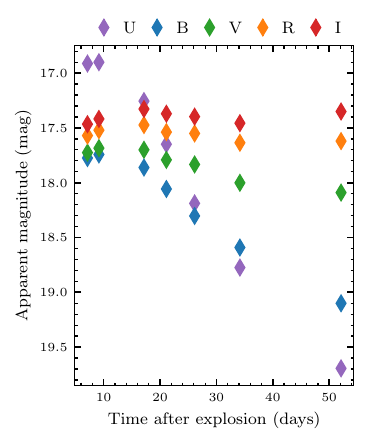}}
	\qquad
	\begin{minipage}{0.6\linewidth}%
		\includegraphics[width=\linewidth,keepaspectratio]{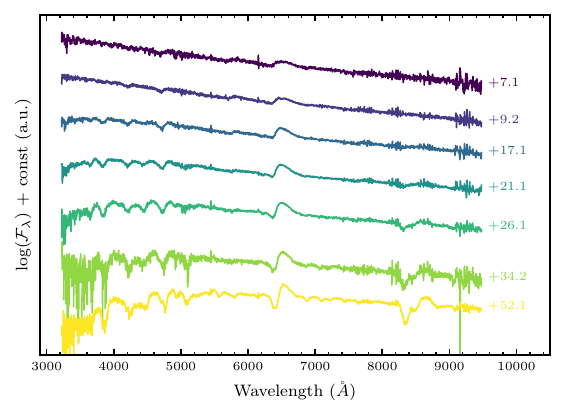}
	\end{minipage}
	\caption{Synthetic \gls{bessell12} photometry (left) and spectral time series (right) of \cer. The numbers next to the spectra indicate the time after the explosion in days. For this \gls{sne}, no host galaxy template subtraction was performed.}
\end{figure}

\begin{figure}[!htb]
	\centering
	\parbox{.35\linewidth}{\includegraphics[width=\linewidth,keepaspectratio]{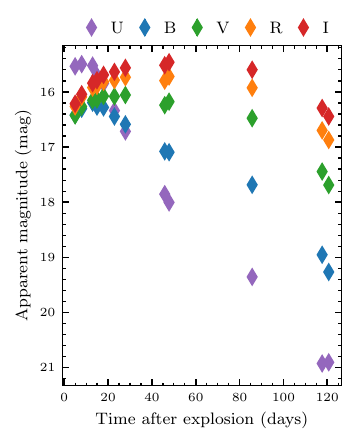}}
	\qquad
	\begin{minipage}{0.6\linewidth}%
		\includegraphics[width=\linewidth,keepaspectratio]{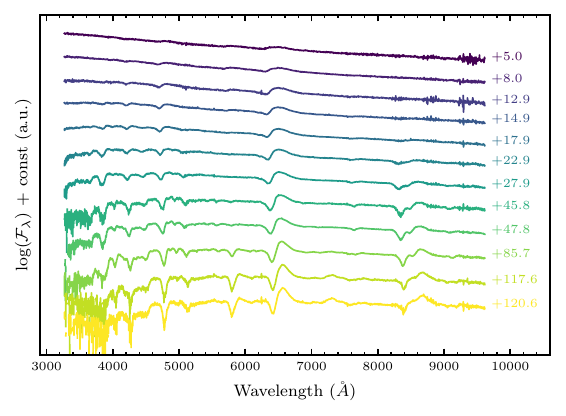}
	\end{minipage}
	\caption{Synthetic \gls{bessell12} photometry (left) and spectral time series (right) of \css. The numbers next to the spectra indicate the time after the explosion in days.}
\end{figure}

\begin{figure}[!htb]
	\centering
	\parbox{.35\linewidth}{\includegraphics[width=\linewidth,keepaspectratio]{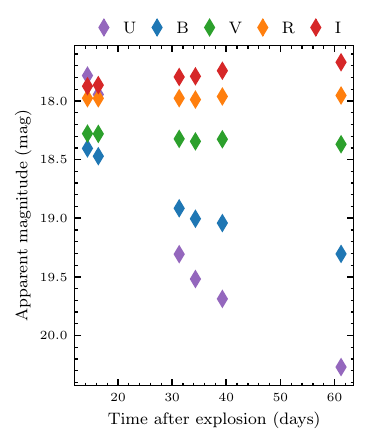}}
	\qquad
	\begin{minipage}{0.6\linewidth}%
		\includegraphics[width=\linewidth,keepaspectratio]{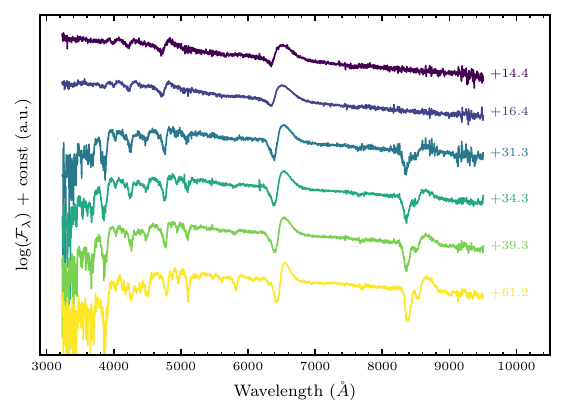}
	\end{minipage}
	\caption{Synthetic \gls{bessell12} photometry (left) and spectral time series (right) of \icone. The numbers next to the spectra indicate the time after the explosion in days.}
\end{figure}

\begin{figure}[!htb]
	\centering
	\parbox{.35\linewidth}{\includegraphics[width=\linewidth,keepaspectratio]{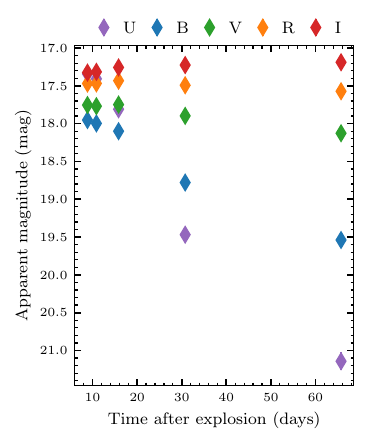}}
	\qquad
	\begin{minipage}{0.6\linewidth}%
		\includegraphics[width=\linewidth,keepaspectratio]{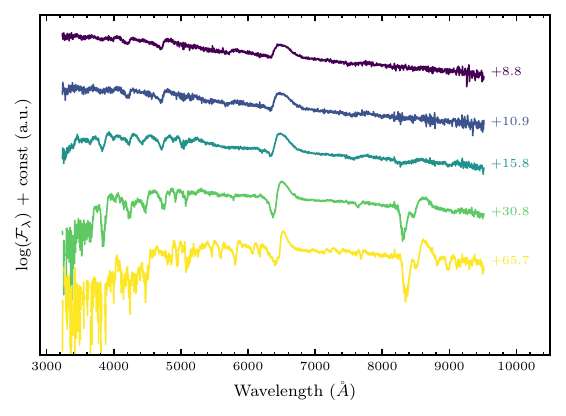}
	\end{minipage}
	\caption{Synthetic \gls{bessell12} photometry (left) and spectral time series (right) of \icthreefive. The numbers next to the spectra indicate the time after the explosion in days.}
\end{figure}

\begin{figure}[!htb]
	\centering
	\parbox{.35\linewidth}{\includegraphics[width=\linewidth,keepaspectratio]{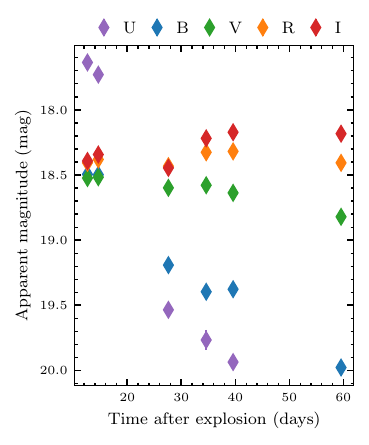}}
	\qquad
	\begin{minipage}{0.6\linewidth}%
		\includegraphics[width=\linewidth,keepaspectratio]{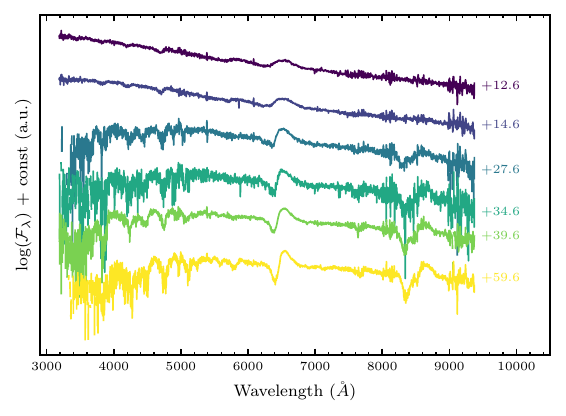}
	\end{minipage}
	\caption{Synthetic \gls{bessell12} photometry (left) and spectral time series (right) of \fvq. The numbers next to the spectra indicate the time after the explosion in days. For this \gls{sne}, no host galaxy template subtraction was performed.}
\end{figure}

\begin{figure}[!htb]
	\centering
	\parbox{.35\linewidth}{\includegraphics[width=\linewidth,keepaspectratio]{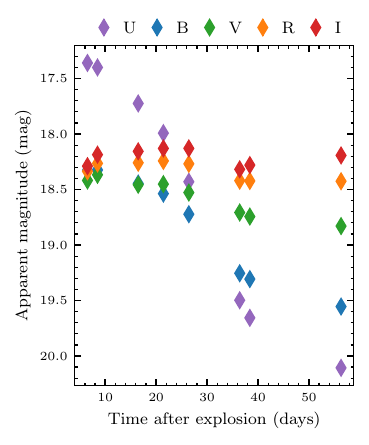}}
	\qquad
	\begin{minipage}{0.6\linewidth}%
		\includegraphics[width=\linewidth,keepaspectratio]{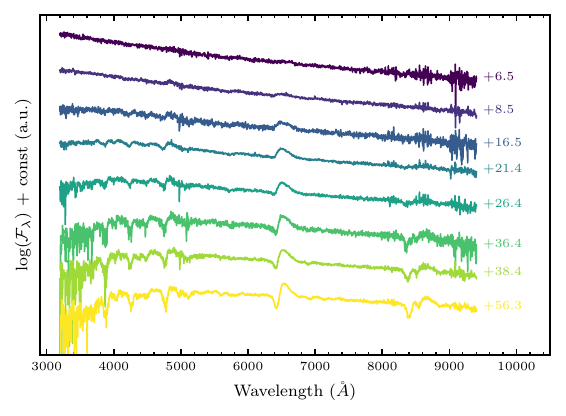}
	\end{minipage}
	\caption{Synthetic \gls{bessell12} photometry (left) and spectral time series (right) of \ljg. The numbers next to the spectra indicate the time after the explosion in days.}
\end{figure}

\begin{figure}[!htb]
	\centering
	\parbox{.35\linewidth}{\includegraphics[width=\linewidth,keepaspectratio]{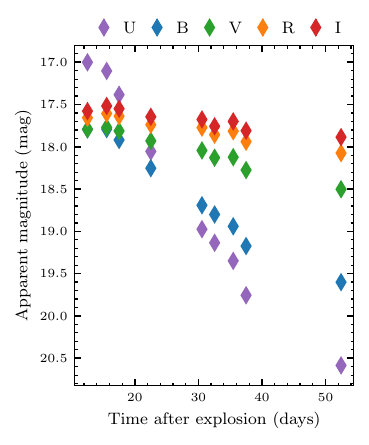}}
	\qquad
	\begin{minipage}{0.6\linewidth}%
		\includegraphics[width=\linewidth,keepaspectratio]{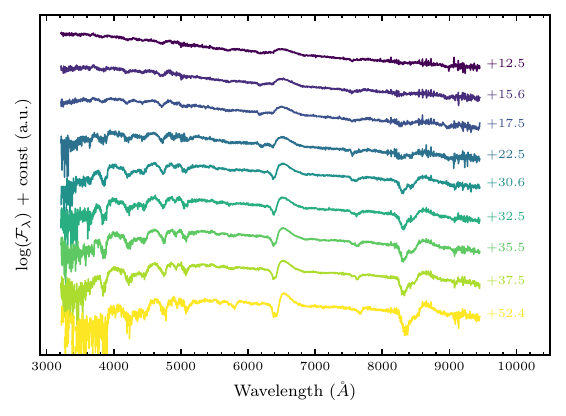}
	\end{minipage}
	\caption{Synthetic \gls{bessell12} photometry (left) and spectral time series (right) of \hi. The numbers next to the spectra indicate the time after the explosion in days.}
\end{figure}

\begin{figure}[!htb]
	\centering
	\parbox{.35\linewidth}{\includegraphics[width=\linewidth,keepaspectratio]{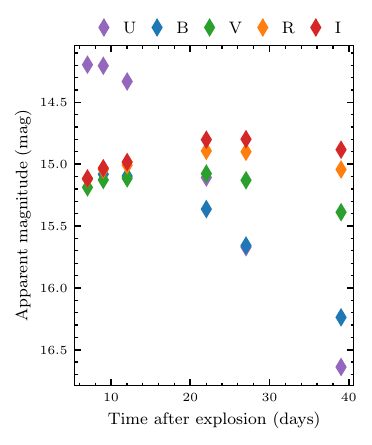}}
	\qquad
	\begin{minipage}{0.6\linewidth}%
		\includegraphics[width=\linewidth,keepaspectratio]{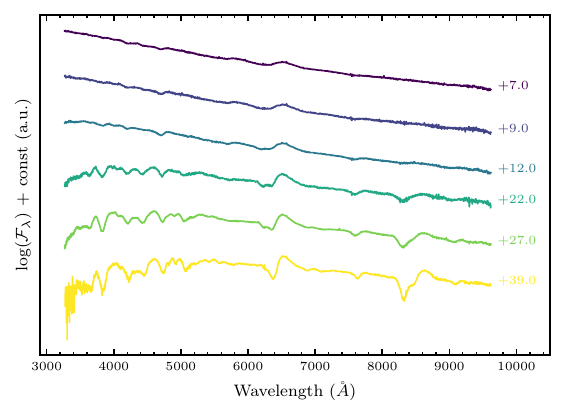}
	\end{minipage}
	\caption{Synthetic \gls{bessell12} photometry (left) and spectral time series (right) of \zw. The numbers next to the spectra indicate the time after the explosion in days. For this \gls{sne}, no host galaxy template subtraction was performed.}
\end{figure}

\begin{figure}[!htb]
	\centering
	\parbox{.35\linewidth}{\includegraphics[width=\linewidth,keepaspectratio]{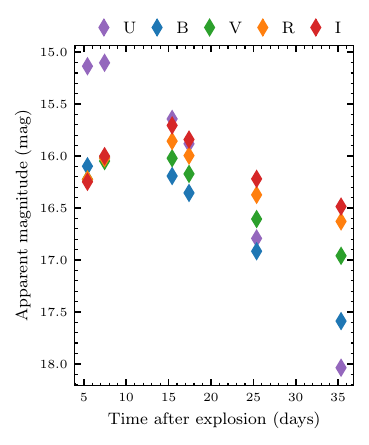}}
	\qquad
	\begin{minipage}{0.6\linewidth}%
		\includegraphics[width=\linewidth,keepaspectratio]{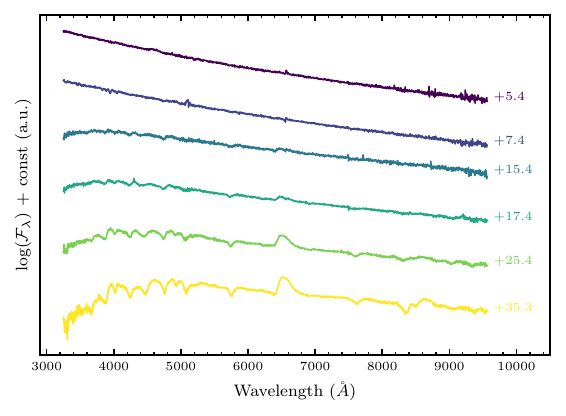}
	\end{minipage}
	\caption{Synthetic \gls{bessell12} photometry (left) and spectral time series (right) of \hnj. The numbers next to the spectra indicate the time after the explosion in days.}\label{fig:spechnj}
\end{figure}

\begin{figure}[!htb]
	\centering
	\parbox{.35\linewidth}{\includegraphics[width=\linewidth,keepaspectratio]{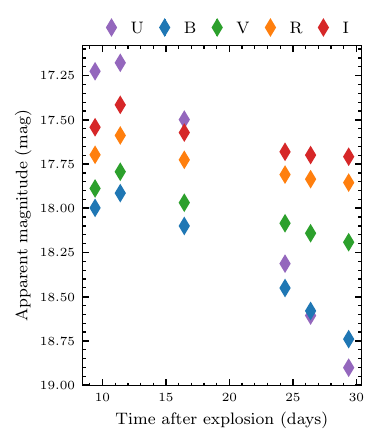}}
	\qquad
	\begin{minipage}{0.6\linewidth}%
		\includegraphics[width=\linewidth,keepaspectratio]{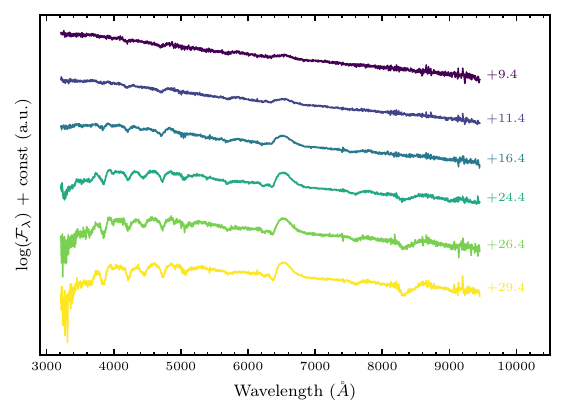}
	\end{minipage}
	\caption{Synthetic \gls{bessell12} photometry (left) and spectral time series (right) of \ugc. The numbers next to the spectra indicate the time after the explosion in days. For this \gls{sne}, no host galaxy template subtraction was performed.}
\end{figure}

\begin{figure}[!htb]
	\centering
	\parbox{.35\linewidth}{\includegraphics[width=\linewidth,keepaspectratio]{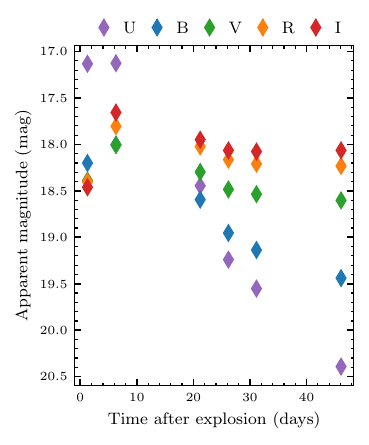}}
	\qquad
	\begin{minipage}{0.6\linewidth}%
		\includegraphics[width=\linewidth,keepaspectratio]{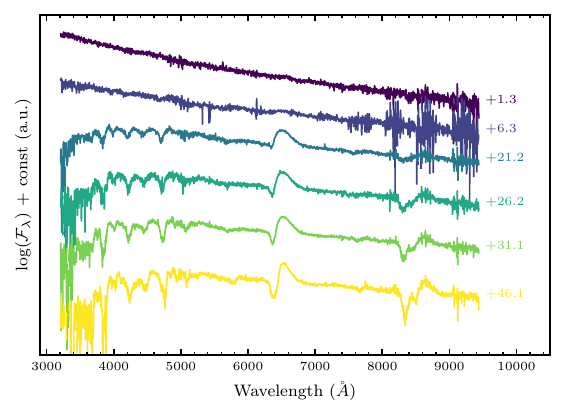}
	\end{minipage}
	\caption{Synthetic \gls{bessell12} photometry (left) and spectral time series (right) of \bjx. The numbers next to the spectra indicate the time after the explosion in days.}
\end{figure}

\begin{figure}[!htb]
	\centering
	\parbox{.35\linewidth}{\includegraphics[width=\linewidth,keepaspectratio]{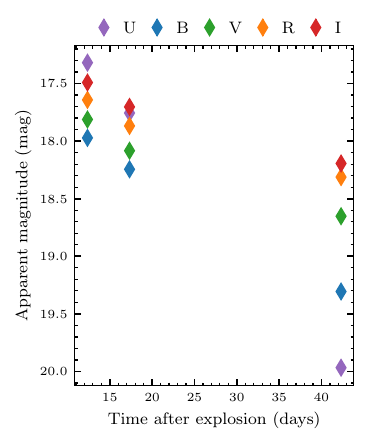}}
	\qquad
	\begin{minipage}{0.6\linewidth}%
		\includegraphics[width=\linewidth,keepaspectratio]{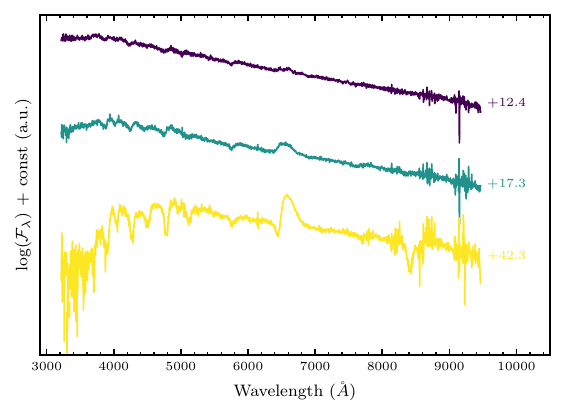}
	\end{minipage}
	\caption{Synthetic \gls{bessell12} photometry (left) and spectral time series (right) of \ds. The numbers next to the spectra indicate the time after the explosion in days. For this \gls{sne}, no host galaxy template subtraction was performed.}\label{fig:app_sample2}
\end{figure}

\FloatBarrier

\subsection{Interpolated data}\label{sec:app_interp}
\begin{figure}[!htb]
	\centering
    \includegraphics{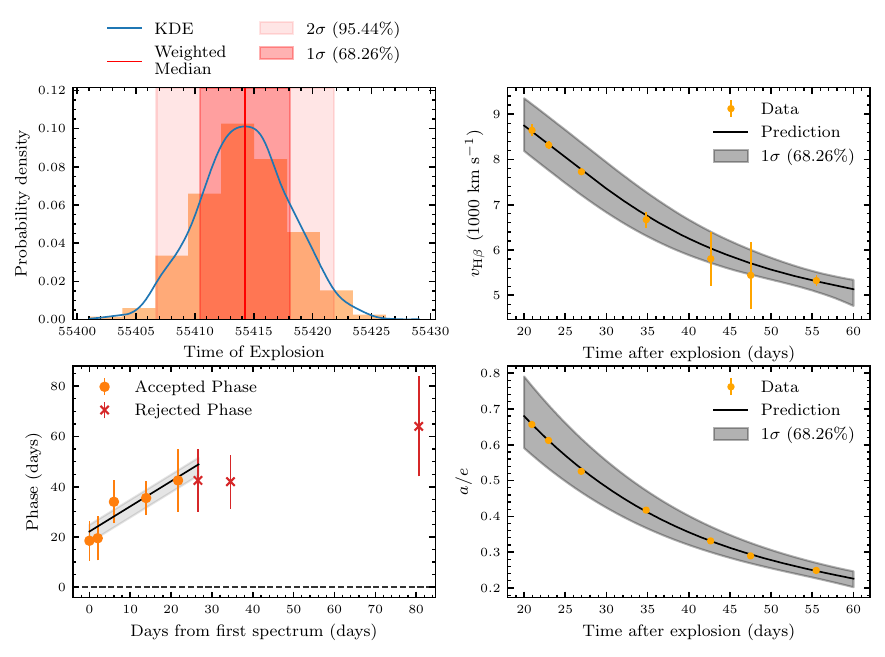}
	\caption{\gls{toe} fit and resultant KDE (left) as well as the interpolated $v_{\mathrm{H}\beta}$ and $a/e$ values (right) of \hb.}\label{fig:app_sample3}
\end{figure}

This section (Figures~\ref{fig:app_sample3} to \ref{fig:app_sample4}) contains illustrations of the \gls{toe} fits and the resultant KDEs, as well as the interpolation of the $v_{\mathrm{H}\beta}$ and $a/e$ values of the \gls{snfactory} \glspl{sne} used in our fiducial Hubble-flow sample.

\begin{figure}[!htb]
	\centering
    \includegraphics{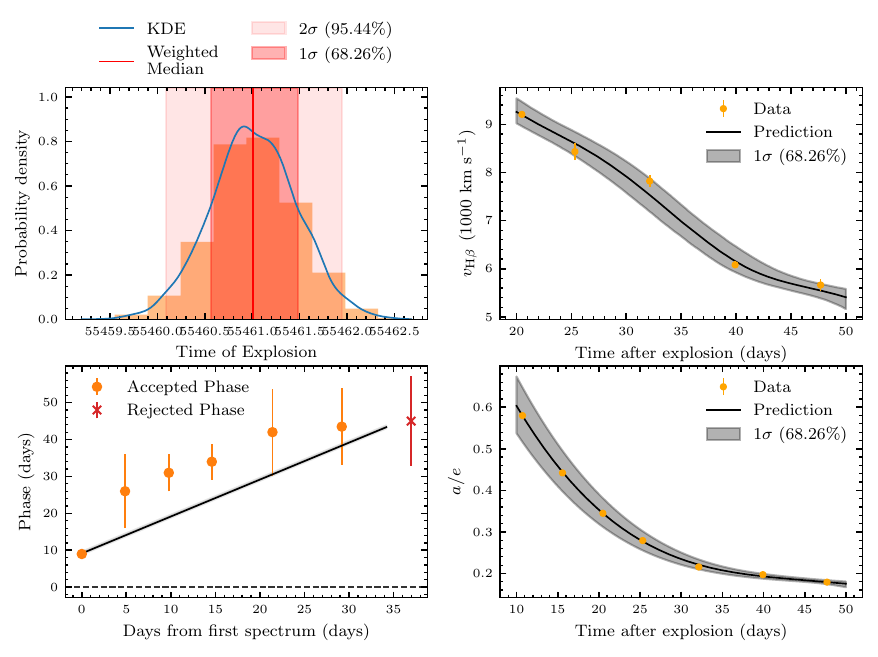}
	\caption{\gls{toe} fit and resultant KDE (left) as well as the interpolated $v_{\mathrm{H}\beta}$ and $a/e$ values (right) of \wmf.}
\end{figure}

\begin{figure}[!htb]
	\centering
    \includegraphics{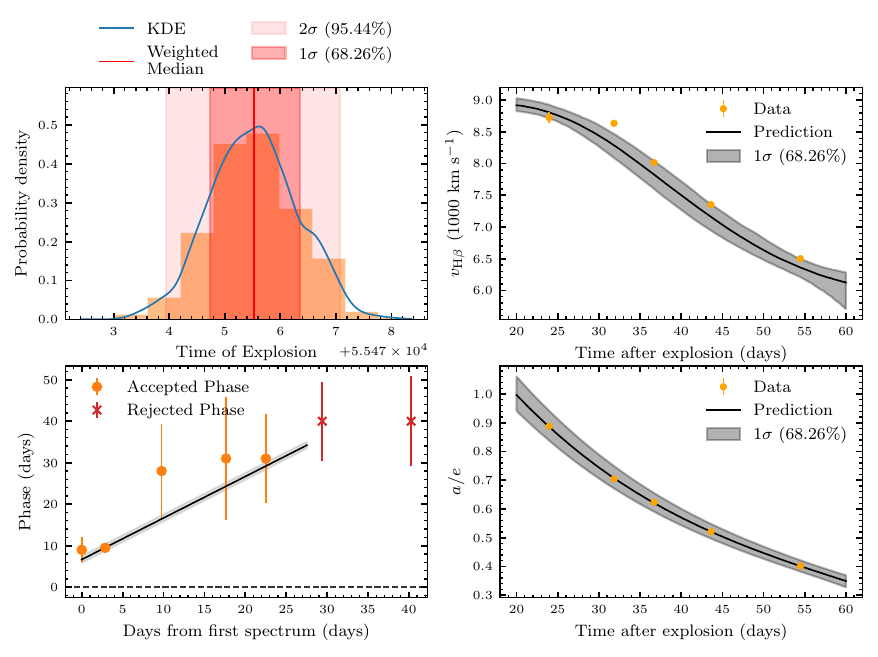}
	\caption{\gls{toe} fit and resultant KDE (left) as well as the interpolated $v_{\mathrm{H}\beta}$ and $a/e$ values (right) of \xlr.}
\end{figure}

\begin{figure}[!htb]
	\centering
    \includegraphics{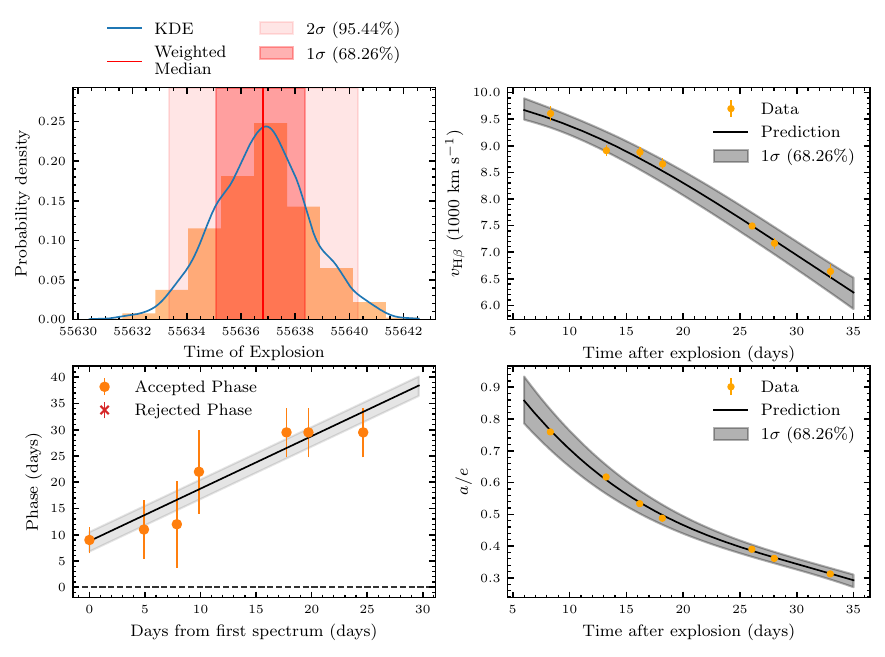}
	\caption{\gls{toe} fit and resultant KDE (left) as well as the interpolated $v_{\mathrm{H}\beta}$ and $a/e$ values (right) of \icthree.}
\end{figure}

\begin{figure}[!htb]
	\centering
    \includegraphics{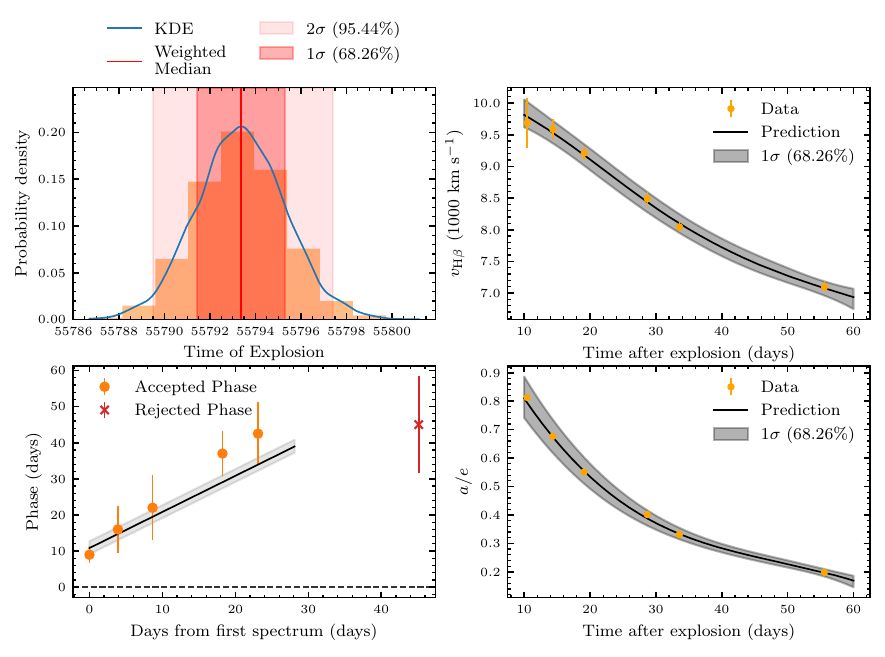}
	\caption{\gls{toe} fit and resultant KDE (left) as well as the interpolated $v_{\mathrm{H}\beta}$ and $a/e$ values (right) of \ngcfour.}
\end{figure}

\begin{figure}[!htb]
	\centering
    \includegraphics{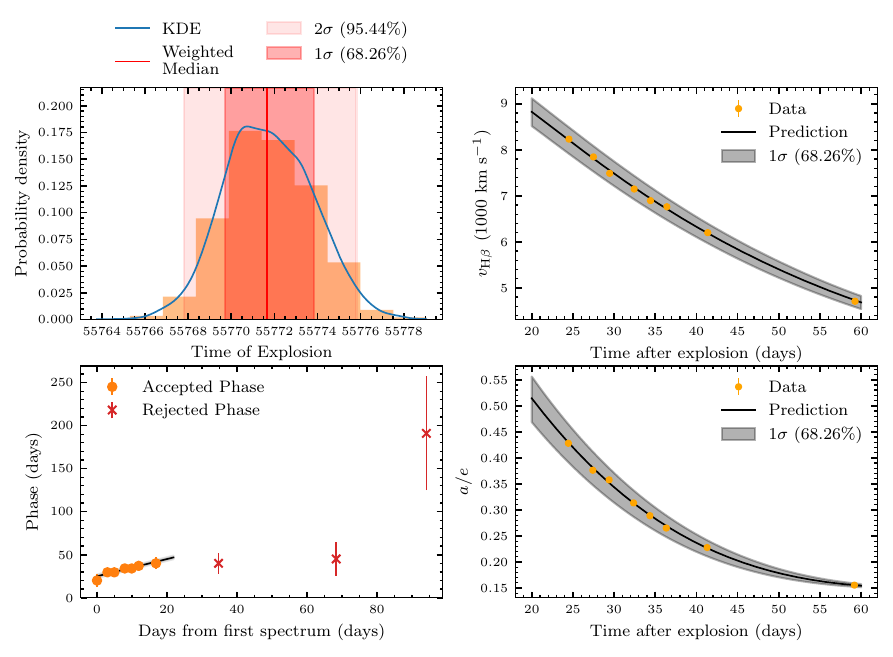}
	\caption{\gls{toe} fit and resultant KDE (left) as well as the interpolated $v_{\mathrm{H}\beta}$ and $a/e$ values (right) of \ngctwo.}
\end{figure}

\begin{figure}[!htb]
	\centering
    \includegraphics{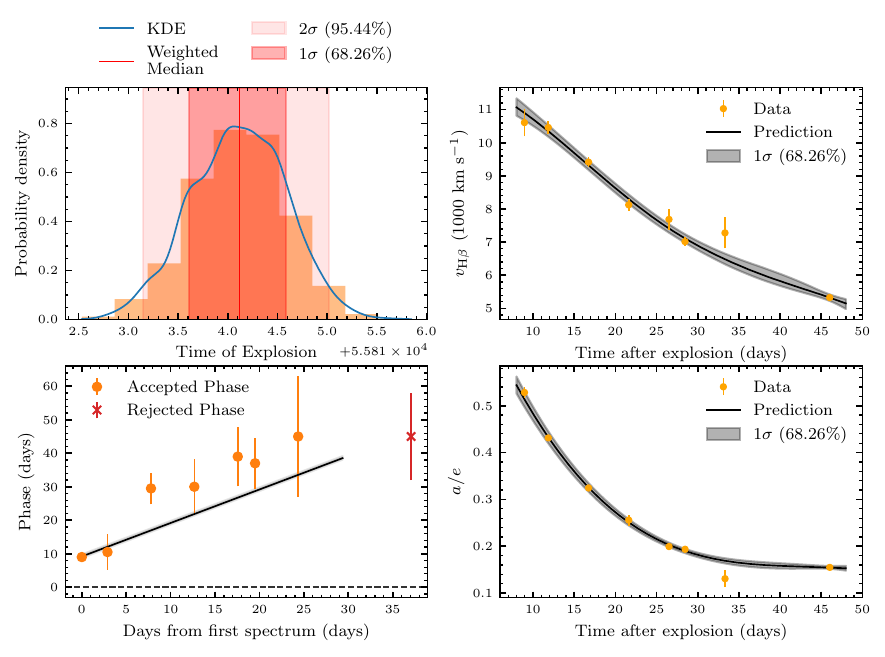}
	\caption{\gls{toe} fit and resultant KDE (left) as well as the interpolated $v_{\mathrm{H}\beta}$ and $a/e$ values (right) of \pgc.}
\end{figure}

\begin{figure}[!htb]
	\centering
    \includegraphics{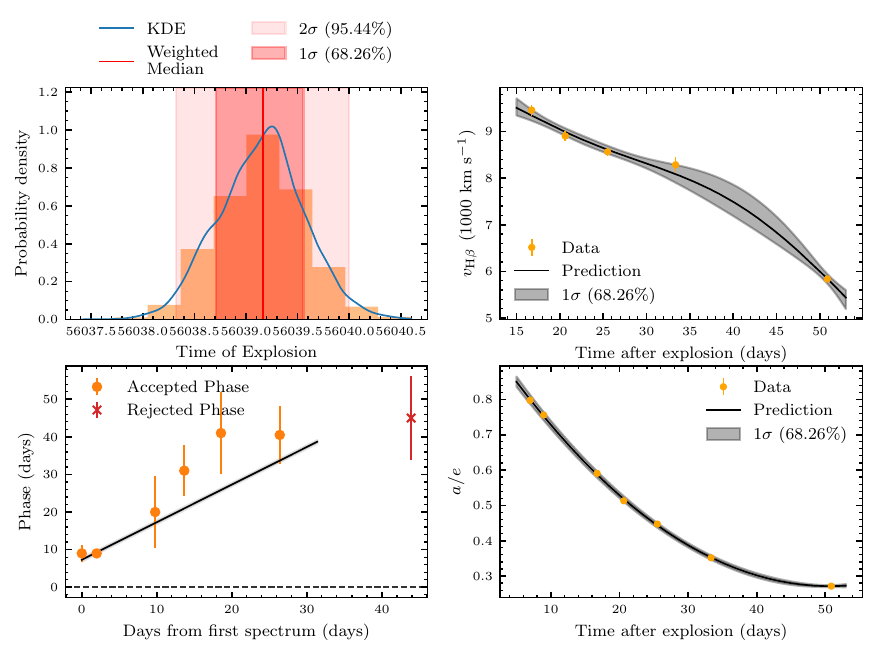}
	\caption{\gls{toe} fit and resultant KDE (left) as well as the interpolated $v_{\mathrm{H}\beta}$ and $a/e$ values (right) of \cer.}
\end{figure}

\begin{figure}[!htb]
	\centering
    \includegraphics{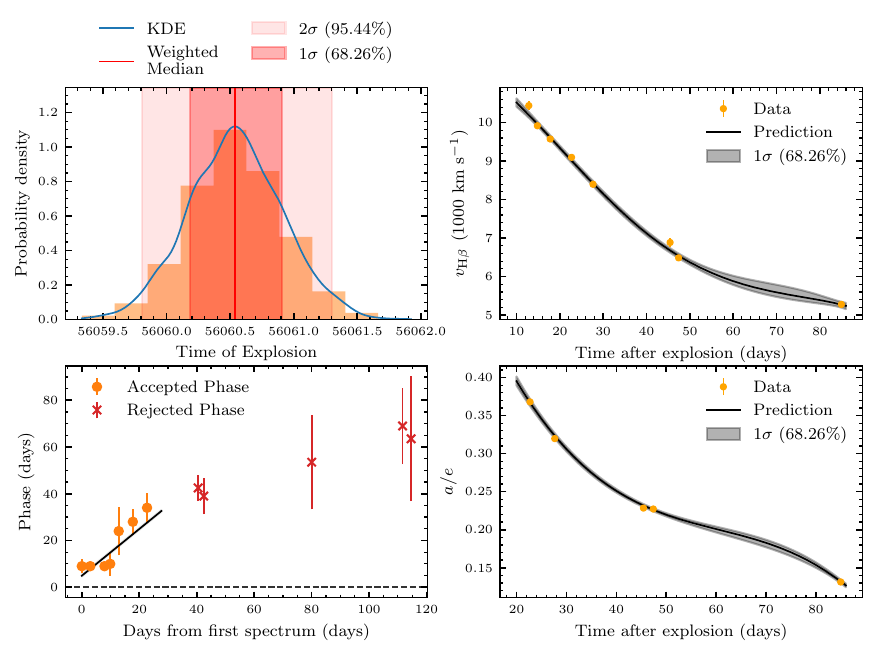}
	\caption{\gls{toe} fit and resultant KDE (left) as well as the interpolated $v_{\mathrm{H}\beta}$ and $a/e$ values (right) of \css.}
\end{figure}

\begin{figure}[!htb]
	\centering
    \includegraphics{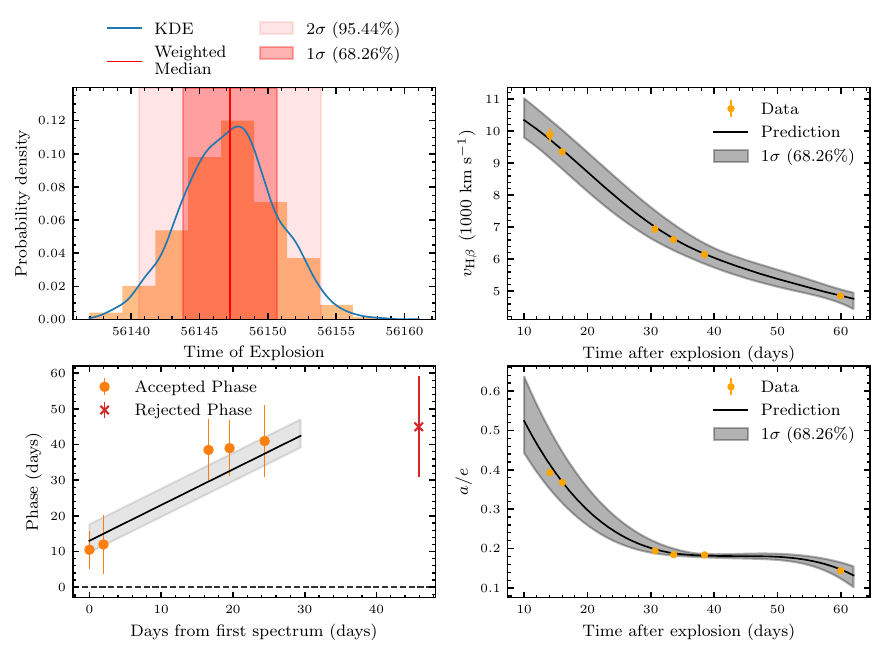}
	\caption{\gls{toe} fit and resultant KDE (left) as well as the interpolated $v_{\mathrm{H}\beta}$ and $a/e$ values (right) of \icone.}
\end{figure}

\begin{figure}[!htb]
	\centering
    \includegraphics{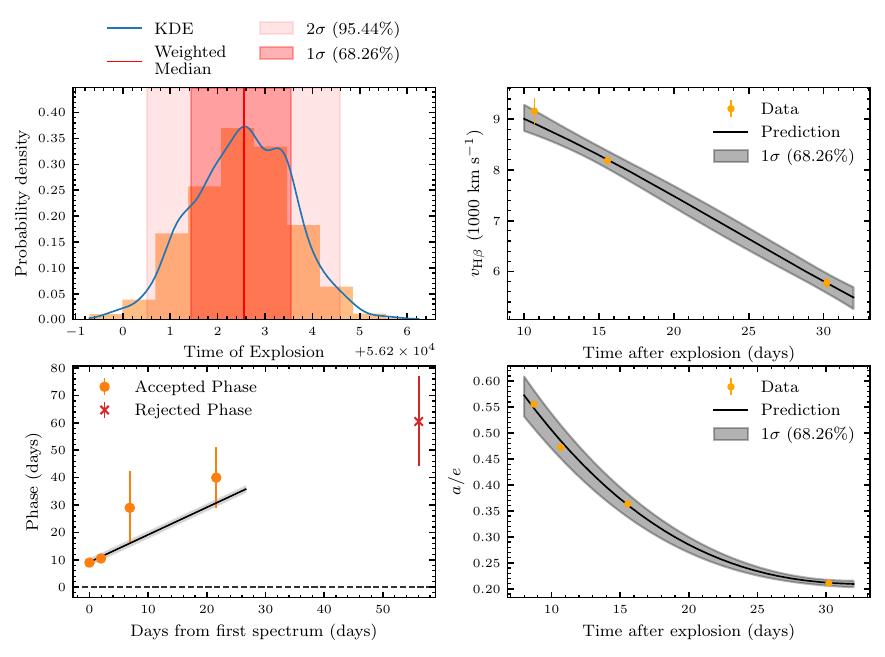}
	\caption{\gls{toe} fit and resultant KDE (left) as well as the interpolated $v_{\mathrm{H}\beta}$ and $a/e$ values (right) of \icthreefive.}
\end{figure}

\begin{figure}[!htb]
	\centering
    \includegraphics{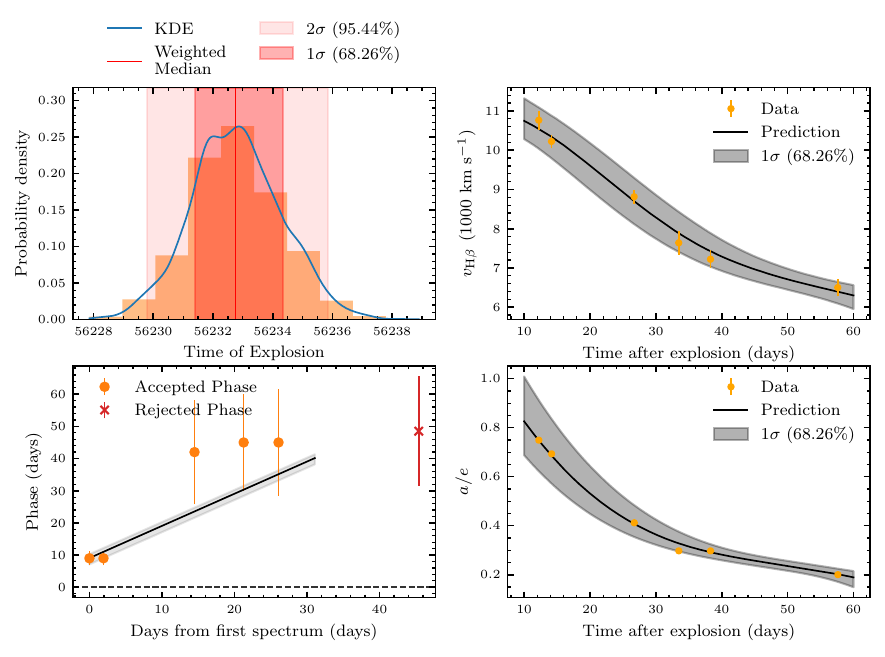}
	\caption{\gls{toe} fit and resultant KDE (left) as well as the interpolated $v_{\mathrm{H}\beta}$ and $a/e$ values (right) of \fvq.}
\end{figure}

\begin{figure}[!htb]
	\centering
    \includegraphics{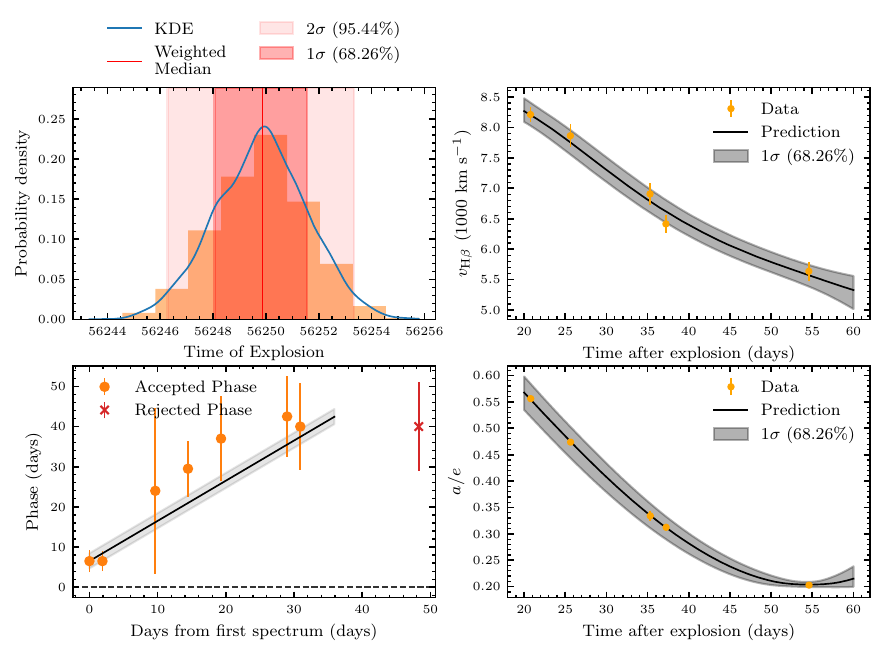}
	\caption{\gls{toe} fit and resultant KDE (left) as well as the interpolated $v_{\mathrm{H}\beta}$ and $a/e$ values (right) of \ljg.}
\end{figure}

\begin{figure}[!htb]
	\centering
    \includegraphics{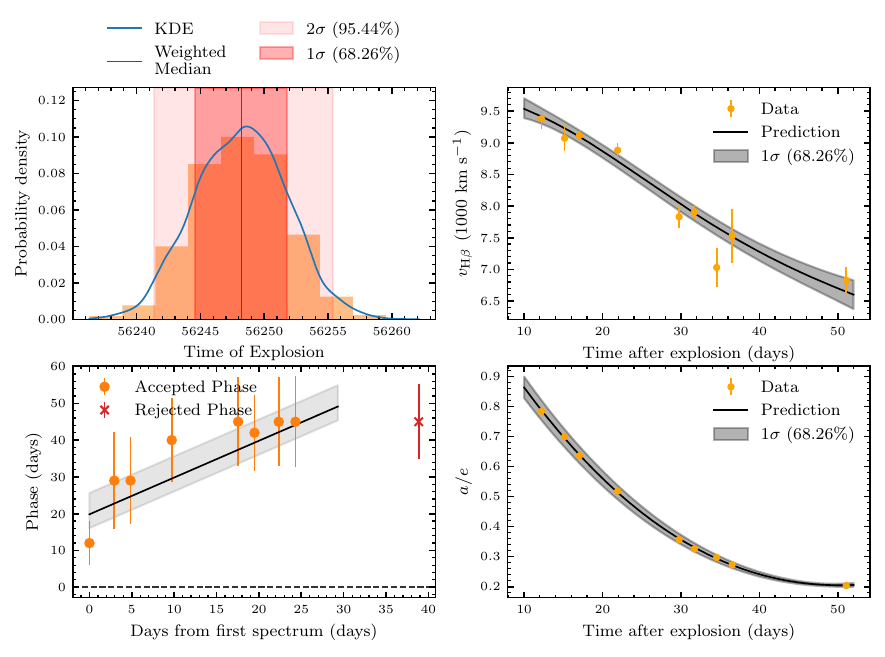}
	\caption{\gls{toe} fit and resultant KDE (left) as well as the interpolated $v_{\mathrm{H}\beta}$ and $a/e$ values (right) of \hi.}
\end{figure}

\begin{figure}[!htb]
	\centering
    \includegraphics{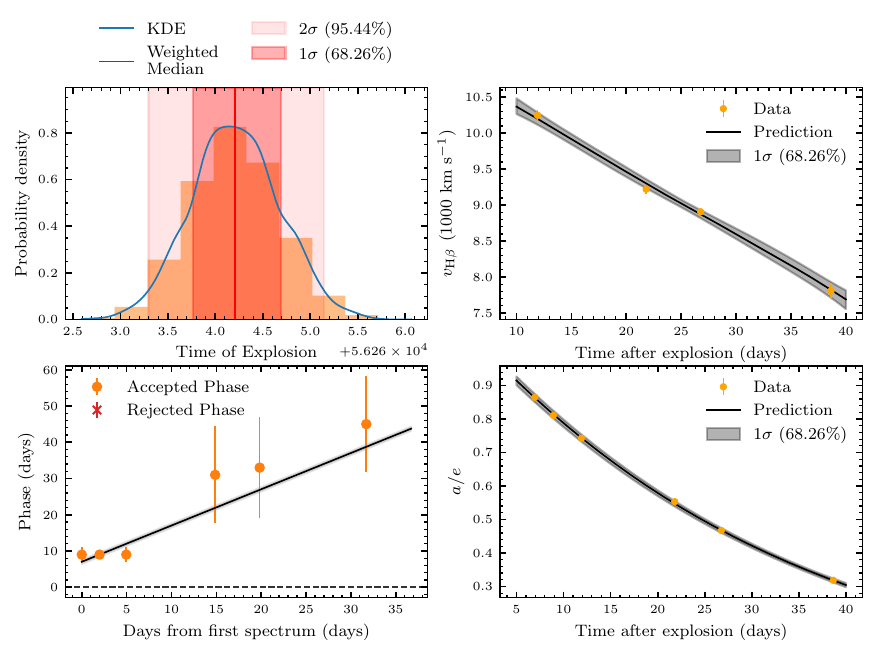}
	\caption{\gls{toe} fit and resultant KDE (left) as well as the interpolated $v_{\mathrm{H}\beta}$ and $a/e$ values (right) of \zw.}
\end{figure}

\begin{figure}[!htb]
	\centering
    \includegraphics{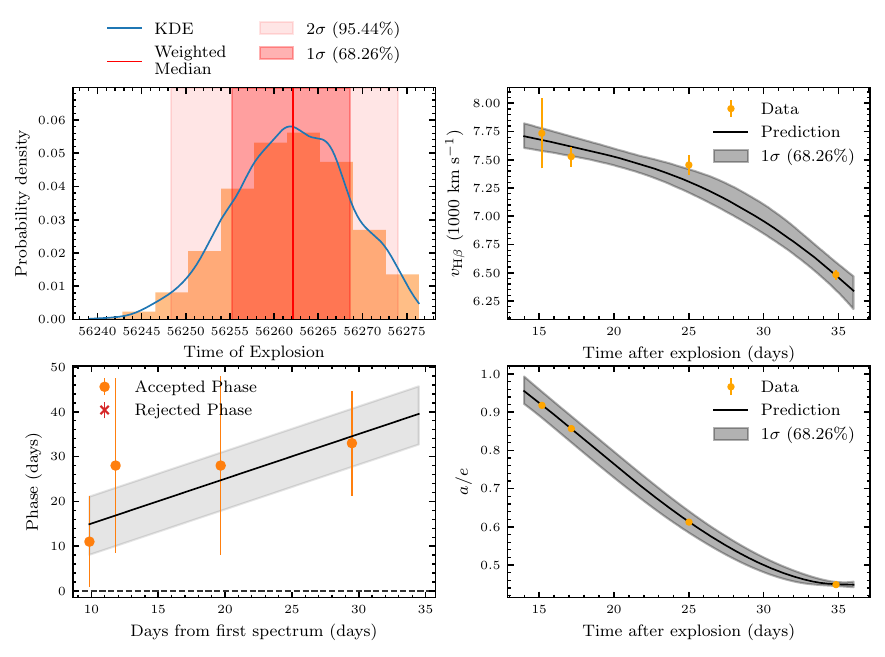}
	\caption{\gls{toe} fit and resultant KDE (left) as well as the interpolated $v_{\mathrm{H}\beta}$ and $a/e$ values (right) of \hnj.}
\end{figure}

\begin{figure}[!htb]
	\centering
    \includegraphics{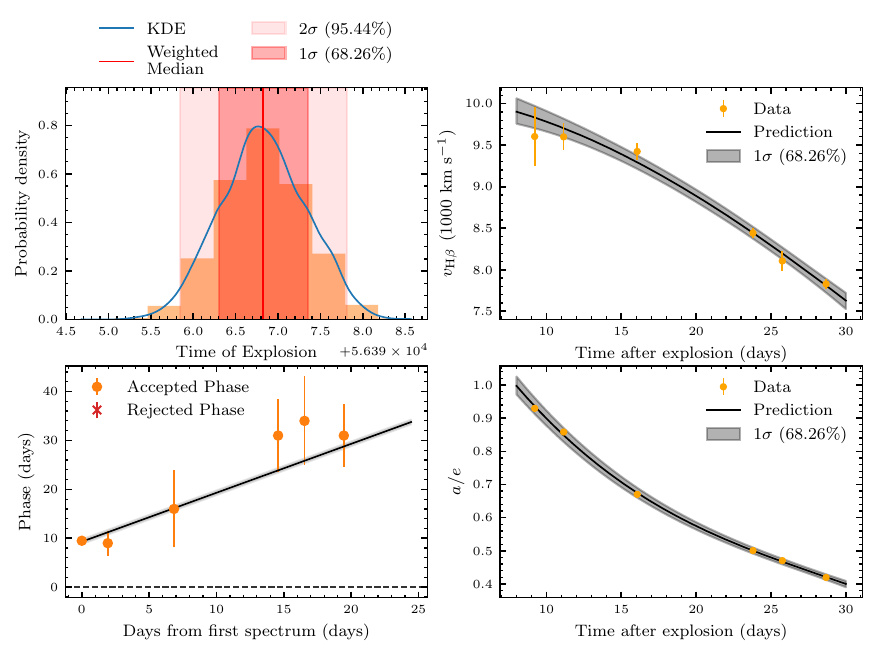}
	\caption{\gls{toe} fit and resultant KDE (left) as well as the interpolated $v_{\mathrm{H}\beta}$ and $a/e$ values (right) of \ugc.}
\end{figure}

\begin{figure}[!htb]
	\centering
    \includegraphics{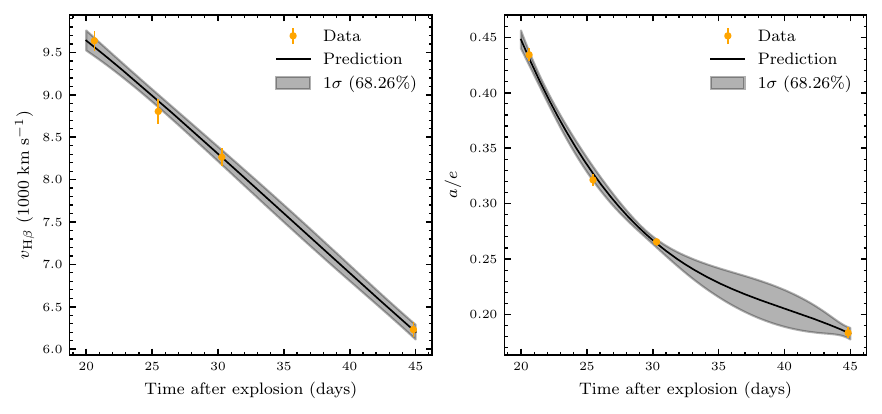}
	\caption{The interpolated $v_{\mathrm{H}\beta}$ and $a/e$ values of \bjx.}\label{fig:app_sample4}
\end{figure}

\FloatBarrier

\section{The calibrator sample}\label{sec:appendix_calib}
\begin{figure}[!htb]
	\centering
    \includegraphics{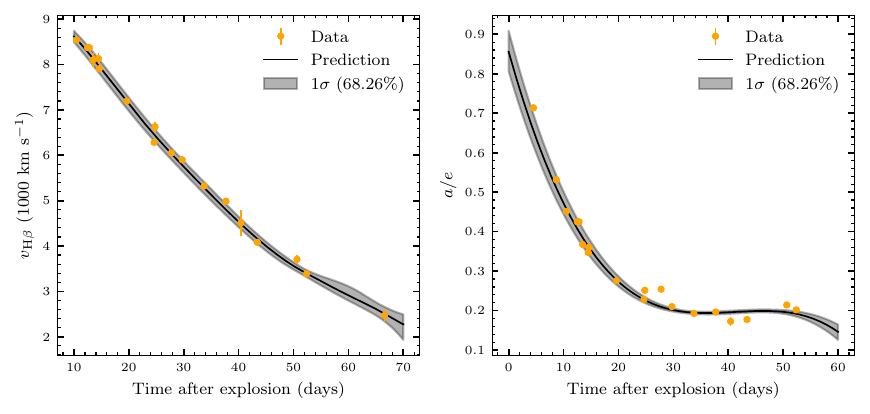}
	\caption{The interpolated $v_{\mathrm{H}\beta}$ and $a/e$ values of \sem.}\label{fig:app_sample_calib_start}
\end{figure}

This section (Tables~\ref{tab:calibphot} and \ref{tab:calibratorspec}) contains an overview of the sources of the utilized photometry and spectra, as well as detail on e.g. the used photometric system.
Figures~\ref{fig:app_sample_calib_start} to \ref{fig:app_sample_calib_end} contain illustrations of the interpolation of the $v_{\mathrm{H}\beta}$ and $a/e$ values of the calibrator \glspl{sne} used in our fiducial Hubble-flow sample.
In Figure~\ref{fig:app_03hn_toe} we additionally illustrate the \gls{toe} fit for \hn{}.

\begin{table}[!htb]
    \centering
    \caption{Overview of the calibrator sample photometry.}\label{tab:calibphot}
    \begin{tabular}{l|c|c|c}
        \hline\hline
        \gls{sne} name & Photometric system & Source & References \\
        \hline \\[-2.0ex]
        \sem & \gls{kait}-2 & T. de Jaeger (2021) priv. comm. & 1 \\
        \gi & \gls{kait}-2 & T. de Jaeger (2021) priv. comm. & 1 \\
        \hn & \acrshort{lco} & \acrshort{osc}\tablefootmark{a} & 2\\
        \fourdj & BATC & \acrshort{osc}\tablefootmark{a}  & 3 \\
        \et & \gls{kait}-3 & T. de Jaeger (2022) priv. comm. & 4 \\
		\cs & \gls{kait}-3 & T. de Jaeger (2021) priv. comm. & 1 \\
        \bk & \gls{csp} & T. de Jaeger (2022) priv. comm. & 5\\
        \ib & \acrshort{ctio} & \acrshort{osc} \tablefootmark{a} & 6 \\
        \aw & \gls{kait}-4 & \acrshort{osc} \tablefootmark{a} & 5, \acrshort{sndb} \tablefootmark{b}\\
        \ej & \gls{kait}-4 & \acrshort{osc} \tablefootmark{a} & 5, \acrshort{sndb} \tablefootmark{b}\\
        \eaw & \gls{kait}-4 & T. de Jaeger (2022) priv. comm. & 7 \\
        \aoq & \gls{kait}-4 & T. de Jaeger (2022) priv. comm. & 8 \\
        \hline
    \end{tabular}
\tablefoot{
This table indicates the utilized photometric system, the source of the used data and references to the original data reduction.
\tablefoottext{a}{The \acrlong{osc} (\cite{Guillochon2017}; \url{https://sne.space/}).}
\tablefoottext{b}{\acrlong{sndb} (\cite{Silverman2012}; \url{http://heracles.astro.berkeley.edu/sndb/}).}
}
\tablebib{
(1)~\citet{Poznanski2009};
(2)~\citet{Galbany2016a};
(3)~\citet{Zhang2006a};
(4)~\citet{Faran2014};
(5)~\citet{deJaeger2019};
(6)~\citet{Takats2015};
(7)~\citet{Dyk2019a};
(8)~\citet{Jaeger2022a};
}
\end{table}

\begin{table}[!htb]
	\centering
	\caption{Overview of the references to the original data publications of the calibrator sample spectra.}\label{tab:calibratorspec}
	\begin{tabular}{l|c}
		\hline\hline
		\gls{sne} name & References \\
		\hline \\[-2.0ex]
		\sem & \cite{Hamuy2001a,Faran2014,Leonard2002a}\\
		\gi & \cite{Shivvers2017,Leonard2002}\\
        \hn & \cite{Harutyunyan2008a,Guti_rrez_2017} \\
        \fourdj & \cite{Vinko2006a} \\
        \et & \cite{Sahu2006a,Faran2014} \\
		\cs & \cite{Pastorello2006,Faran2014,Bufano2009,Pastorello2009} \\
        \bk & \cite{Jaeger2017} \\
		\ib & \cite{Takats2015} \\
		\aw & \cite{DallOra2014,Bose2013,Jerkstrand2014} \\
		\ej & \cite{Bose2015,Valenti2014,Yuan2016}\\
        \eaw & \cite{Szalai2019a}\\
        \aoq & T. de Jaeger (2022) priv. comm.\\
		\hline
	\end{tabular}
\tablefoot{Details on the instrumentation such as the telescope they are mounted on and additional references can be found on the respective \href{https://www.wiserep.org/}{WISeREP} pages where the \gls{sne} data is publicly available (i.e. for all \glspl{sne} except \aoq{}).}
\end{table}

\begin{figure}[!htb]
	\centering
    \includegraphics{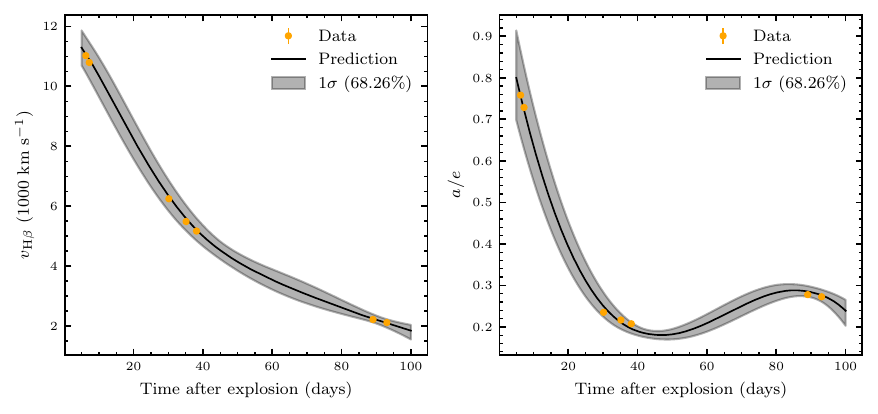}
	\caption{The interpolated $v_{\mathrm{H}\beta}$ and $a/e$ values of \gi.}
\end{figure}

\begin{figure}[!htb]
	\centering
    \includegraphics{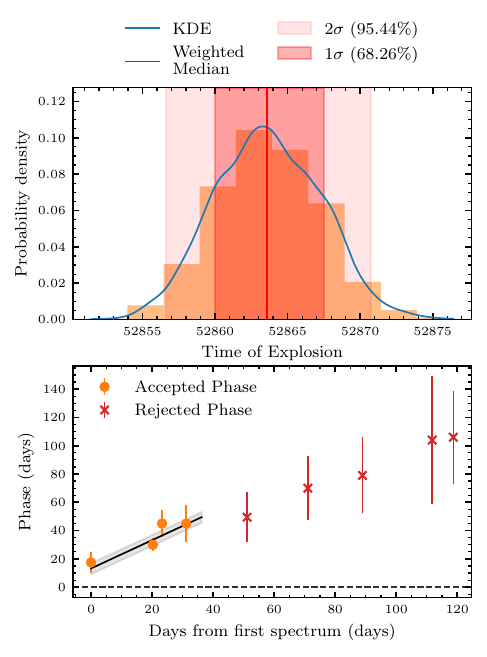}
	\caption{\gls{toe} fit and resultant KDE as well as the interpolated $v_{\mathrm{H}\beta}$ and $a/e$ values of \hn{}.}\label{fig:app_03hn_toe}
\end{figure}

\begin{figure}[!htb]
	\centering
    \includegraphics{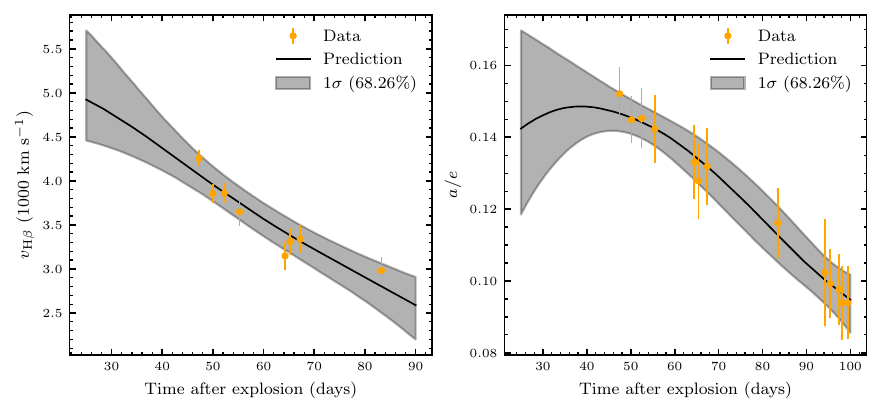}
	\caption{The interpolated $v_{\mathrm{H}\beta}$ and $a/e$ values of \fourdj.}
\end{figure}

\begin{figure}[!htb]
	\centering
    \includegraphics{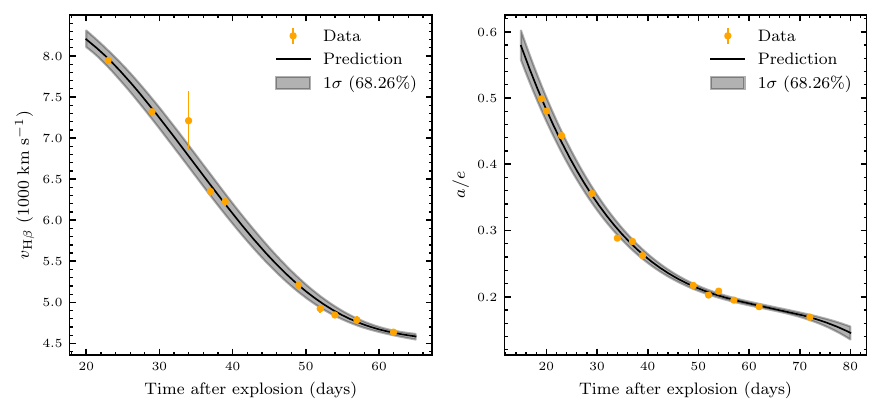}
	\caption{The interpolated $v_{\mathrm{H}\beta}$ and $a/e$ values of \et.}
\end{figure}

\begin{figure}[!htb]
	\centering
    \includegraphics{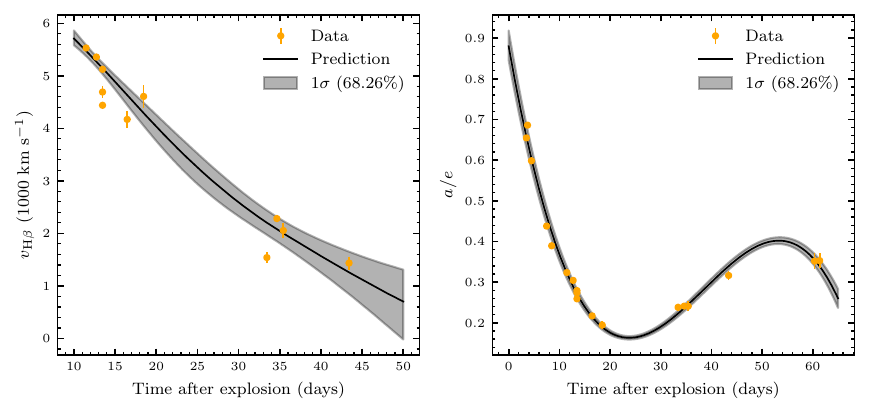}
	\caption{The interpolated $v_{\mathrm{H}\beta}$ and $a/e$ values of \cs.}
\end{figure}

\begin{figure}[!htb]
	\centering
    \includegraphics{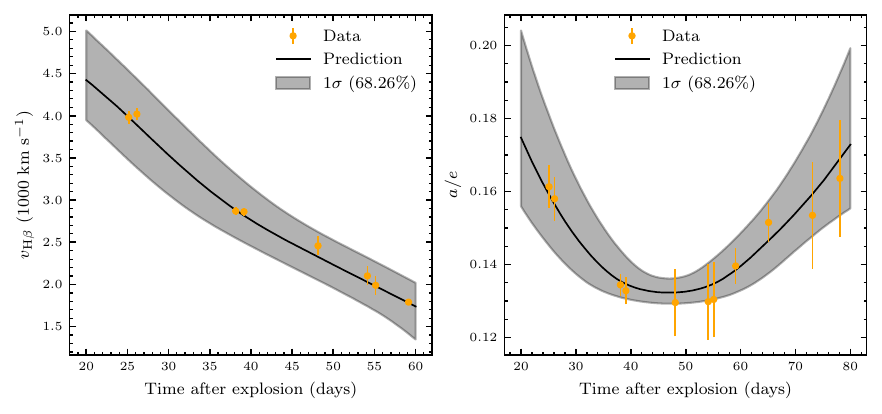}
	\caption{The interpolated $v_{\mathrm{H}\beta}$ and $a/e$ values of \bk.}
\end{figure}

\begin{figure}[!htb]
	\centering
    \includegraphics{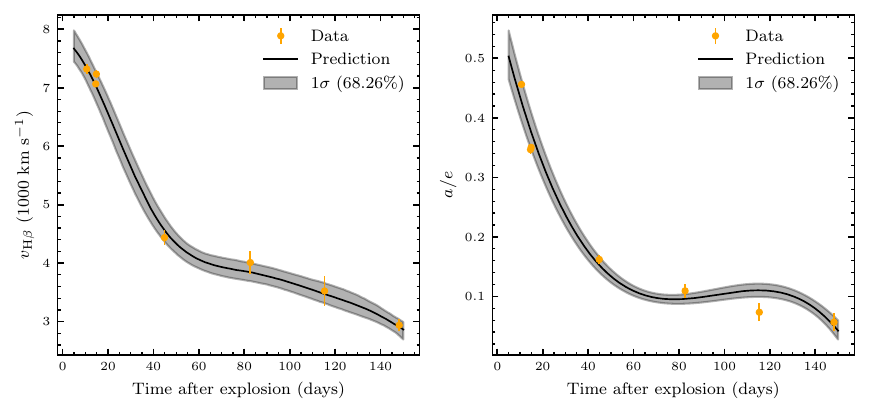}
	\caption{The interpolated $v_{\mathrm{H}\beta}$ and $a/e$ values of \ib.}\label{fig:09ib_interp}
\end{figure}

\begin{figure}[!htb]
	\centering
    \includegraphics{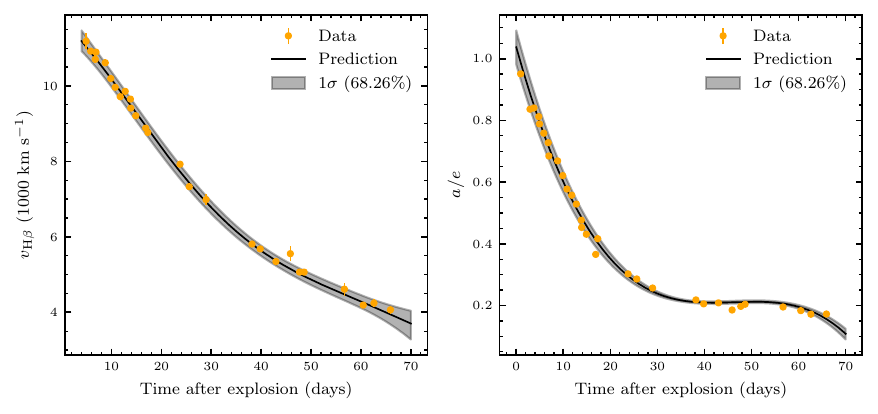}
	\caption{The interpolated $v_{\mathrm{H}\beta}$ and $a/e$ values of \aw.}
\end{figure}

\begin{figure}[!htb]
	\centering
    \includegraphics{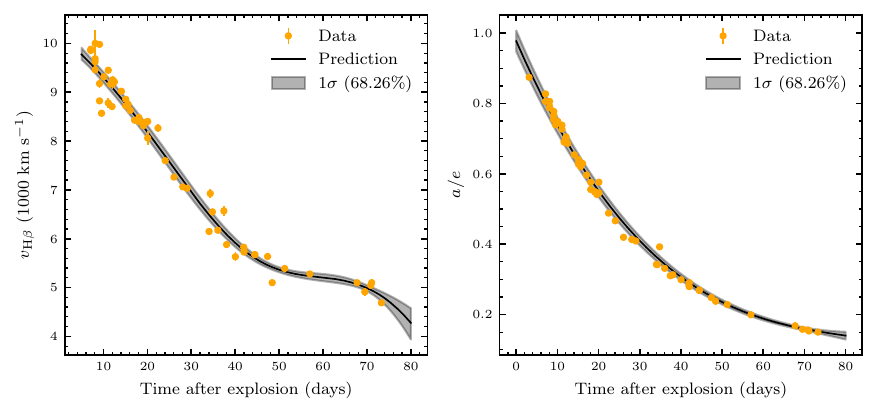}
	\caption{The interpolated $v_{\mathrm{H}\beta}$ and $a/e$ values of \ej.}
\end{figure}

\begin{figure}[!htb]
	\centering
    \includegraphics{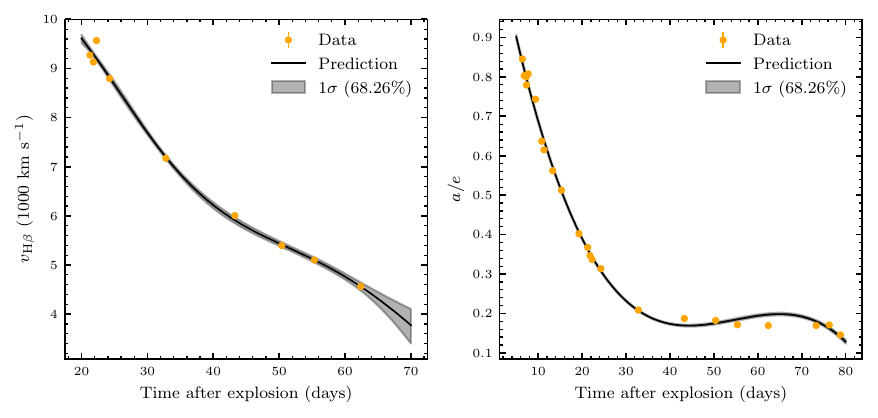}
	\caption{The interpolated $v_{\mathrm{H}\beta}$ and $a/e$ values of \eaw.}
\end{figure}

\begin{figure}[!htb]
	\centering
    \includegraphics{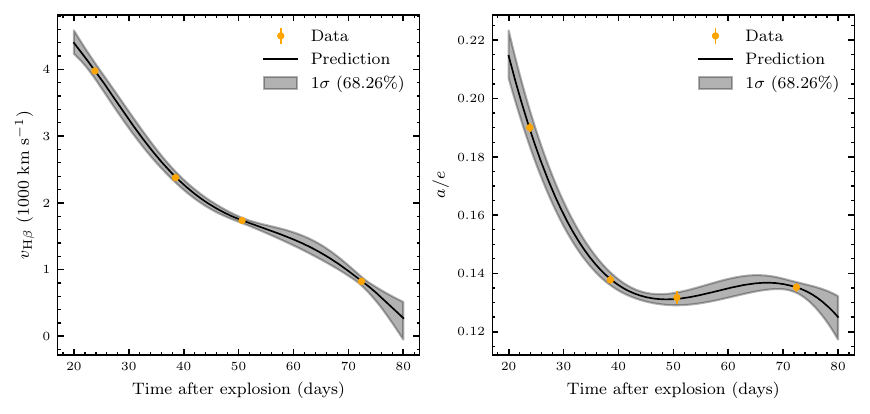}
	\caption{The interpolated $v_{\mathrm{H}\beta}$ and $a/e$ values of \aoq.}\label{fig:app_sample_calib_end}
\end{figure}
\FloatBarrier

\clearpage
\section{Hyperparameter posterior distributions}
\label{sec:app_posteriors}

This section contains the posterior distributions of the main $H_0$ \gls{scm} fit described in Section~\ref{sec:res_h0}, complementing the posterior distributions illustrated in Figure~\ref{fig:corner}.

\begin{figure}[!hbt]
    \centering
    \includegraphics[width=\linewidth]{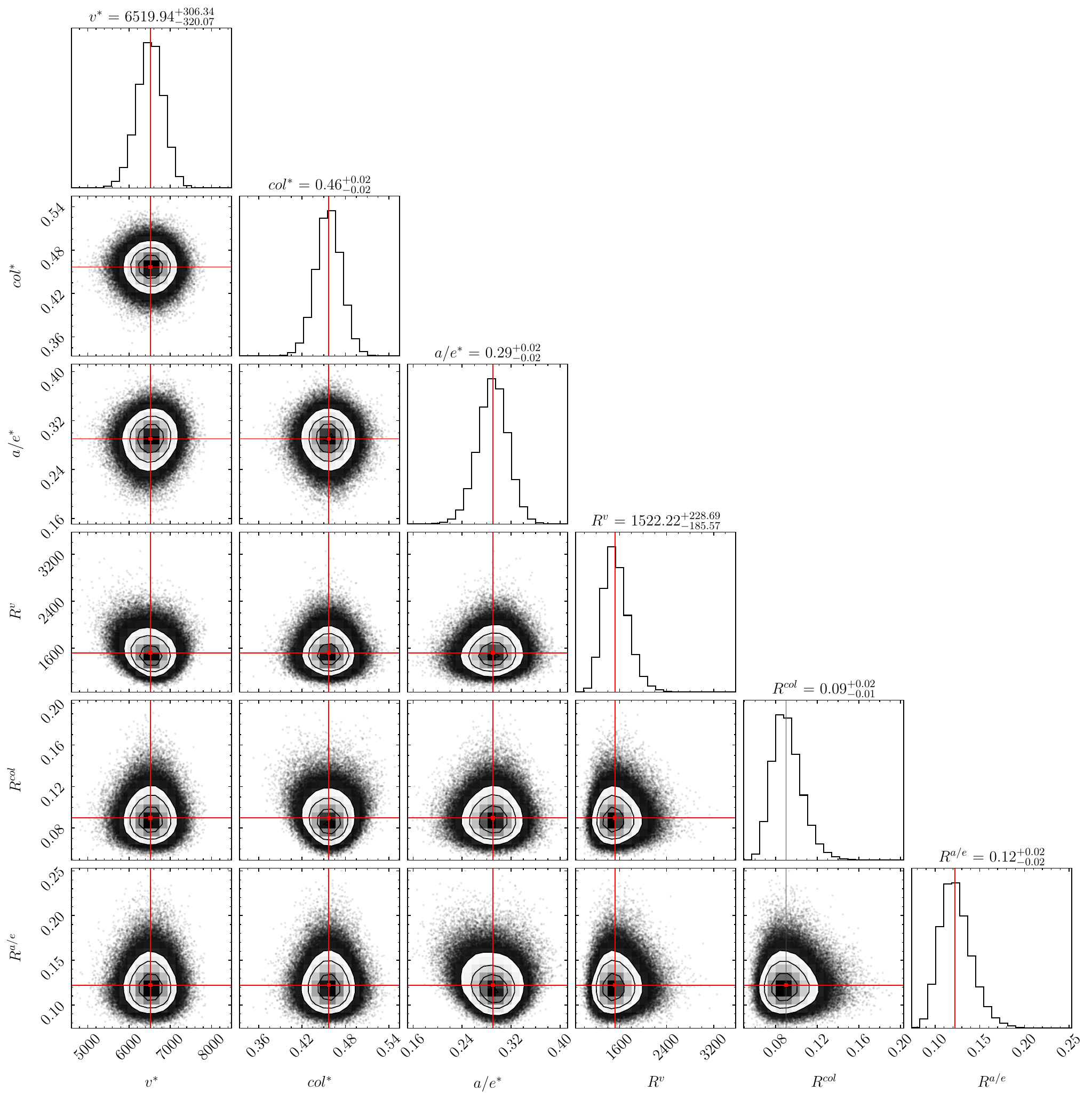}
    \caption{Corner plot of the posterior sample distributions of the hyperparameters of the \gls{snfactory} sample obtained from the \gls{scm} at $30$ days corresponding to the Hubble diagram in Figure~\ref{fig:hubble_diag}. Here, the values in the titles of the individual plots refer to the $16$th, $50$th and $84$th percentile. The red lines indicate the $50$th percentile value.}
    \label{fig:corner_snfactory}
\end{figure}

\clearpage
\section{SCM data}\label{app:data}
This section contains the data used for the results discussed in Sections~\ref{sec:results} and \ref{sec:res_h0}, listed in Table~\ref{tab:scmdata}.
\begin{table*}[!h]
	\centering
	\caption{\gls{sne} data interpolated to $30$ days after the explosion used for our fiducial analysis.}
	\label{tab:scmdata}
	\begingroup
	\setlength{\tabcolsep}{6pt}
	\renewcommand{\arraystretch}{1.25}
	\begin{tabular}{l|c|c|c|c|c|c}
		\hline\hline
		\gls{sne} name & Dataset\tablefootmark{a} & $m_I$ (mag)\tablefootmark{b} & $col$ (mag)\tablefootmark{b} & $v_{\mathrm{H}\beta}$ (km\,s$^{-1}$)\tablefootmark{b} & $a/e$\tablefootmark{b} & $\sigma_\mathrm{tot}$\tablefootmark{c} \\
		\hline
        \hb & SNfactory & $16.91\pm0.06$ & $0.447\pm0.052$ & $7355\pm554$ & $0.482\pm0.065$ & $0.23$ \\
\wmf & SNfactory & $18.07\pm0.02$ & $0.584\pm0.015$ & $7916\pm326$ & $0.235\pm0.015$ & $0.10$ \\
\xlr & SNfactory & $15.56\pm0.06$ & $0.458\pm0.002$ & $8442\pm169$ & $0.744\pm0.042$ & $0.14$ \\
\icthree & SNfactory & $16.52\pm0.01$ & $0.392\pm0.003$ & $6943\pm262$ & $0.344\pm0.017$ & $0.11$ \\
\ngctwo & SNfactory & $15.44\pm0.01$ & $0.406\pm0.009$ & $7482\pm267$ & $0.344\pm0.029$ & $0.17$ \\
\ngcfour & SNfactory & $18.39\pm0.02$ & $0.463\pm0.009$ & $8343\pm157$ & $0.372\pm0.025$ & $0.08$ \\
\pgc & SNfactory & $18.08\pm0.02$ & $0.464\pm0.003$ & $6898\pm136$ & $0.182\pm0.005$ & $0.06$ \\
\cer & SNfactory & $17.42\pm0.02$ & $0.502\pm0.007$ & $8314\pm109$ & $0.387\pm0.006$ & $0.06$ \\
\css & SNfactory & $15.58\pm0.03$ & $0.511\pm0.028$ & $8147\pm66$ & $0.306\pm0.003$ & $0.13$ \\
\icone & SNfactory & $17.80\pm0.02$ & $0.521\pm0.008$ & $7098\pm486$ & $0.201\pm0.016$ & $0.15$ \\
\icthreefive & SNfactory & $17.23\pm0.02$ & $0.661\pm0.010$ & $5807\pm193$ & $0.212\pm0.006$ & $0.10$ \\
\fvq & SNfactory & $18.29\pm0.06$ & $0.289\pm0.033$ & $8283\pm559$ & $0.364\pm0.052$ & $0.19$ \\
\ljg & SNfactory & $18.22\pm0.04$ & $0.393\pm0.007$ & $7302\pm203$ & $0.407\pm0.027$ & $0.10$ \\
\hi & SNfactory & $17.71\pm0.03$ & $0.366\pm0.009$ & $8027\pm116$ & $0.356\pm0.016$ & $0.07$ \\
\hnj & SNfactory & $16.39\pm0.05$ & $0.421\pm0.010$ & $6956\pm130$ & $0.500\pm0.019$ & $0.11$ \\
\zw & SNfactory & $14.80\pm0.04$ & $0.389\pm0.010$ & $8597\pm73$ & $0.422\pm0.007$ & $0.13$ \\
\ugc & SNfactory & $17.71\pm0.03$ & $0.499\pm0.006$ & $7629\pm97$ & $0.400\pm0.009$ & $0.06$ \\
\bjx & SNfactory & $18.07\pm0.03$ & $0.442\pm0.027$ & $8304\pm86$ & $0.267\pm0.003$ & $0.05$ \\
\sem & Cepheid & $13.38\pm0.01$ & $0.413\pm0.015$ & $5748\pm128$ & $0.201\pm0.004$ & $0.10$ \\
\gi & Cepheid & $14.06\pm0.02$ & $0.665\pm0.041$ & $6361\pm518$ & $0.252\pm0.030$ & $0.20$ \\
\hn & Cepheid & $14.14\pm0.03$ & $0.525\pm0.072$ & $6618\pm440$ & $0.310\pm0.034$ & $0.16$ \\
\fourdj & Cepheid & $11.31\pm0.14$ & $0.373\pm0.150$ & $4761\pm513$ & $0.146\pm0.019$ & $0.30$ \\
\et & TRGB & $11.46\pm0.02$ & $0.319\pm0.040$ & $7247\pm128$ & $0.342\pm0.012$ & $0.17$ \\
\cs & TRGB & $14.17\pm0.01$ & $0.557\pm0.018$ & $2592\pm258$ & $0.190\pm0.006$ & $0.21$ \\
\bk & TRGB & $12.65\pm0.02$ & $0.181\pm0.111$ & $3534\pm499$ & $0.148\pm0.013$ & $0.29$ \\
\ib & Cepheid & $15.40\pm0.01$ & $0.546\pm0.037$ & $5646\pm291$ & $0.237\pm0.020$ & $0.13$ \\
\aw & Cepheid & $12.91\pm0.03$ & $0.421\pm0.029$ & $6777\pm164$ & $0.239\pm0.007$ & $0.11$ \\
\ej & TRGB & $12.28\pm0.02$ & $0.386\pm0.031$ & $6952\pm129$ & $0.410\pm0.013$ & $0.09$ \\
\eaw & TRGB & $11.72\pm0.01$ & $0.410\pm0.013$ & $7692\pm48$ & $0.232\pm0.002$ & $0.16$ \\
\aoq & Cepheid & $15.11\pm0.02$ & $0.512\pm0.025$ & $3250\pm125$ & $0.160\pm0.004$ & $0.10$ \\
        \hline
	\end{tabular}
	\endgroup
 \tablefoot{
 \tablefoottext{a}{\glspl{sne} labeled with \gls{trgb} and Cepheid make up the combined calibrator dataset and not two individual datasets. We provide separate labels only for convenience.}
 \tablefoottext{b}{The given uncertainties refer to the total statistical uncertainty.}
 \tablefoottext{c}{The value of $\sigma_\mathrm{tot}$ depends on the value of the model coefficients, for example $\alpha$, see Eq.~\ref{eq:sigma_tot_hf} and \ref{eq:sigma_tot_calib}. The value here is calculated for our fiducial result, i.e., calculated with the coefficients given in Table~\ref{tab:h0vals}.}
 }
\end{table*}

\end{appendix}
\twocolumn

\end{document}